# Power System Studies Using Open-Access Software

**Juan A. Martinez-Velasco[1], Pau Casals-Torrens[2], Ricard Bosch-Tous[3], Alexandre Serrano-Fontova[4]**

1 Retired (formerly with Universitat Politècnica de Catalunya); juan-antonio.martinez@upc.edu
2 Retired (formerly with Universitat Politècnica de Catalunya); p.casals@upc.edu
3 Retired (formerly with Universitat Politècnica de Catalunya); ricard.bosch@upc.edu
4 Private Consultant; advancedconsulting1990@gmail.com

**Abstract:** The use of open-access software is an option that can be considered by those interested in power system studies. In addition, the combination of two or more of these tools can expand the capabilities and the fields of application of each tool. This paper proposes the implementation of a flexible and powerful simulation environment based on **R/Rstudio** for carrying out power system studies. Several simple case studies are presented aimed at showing how the combination of either **EMTP/ATP** or **OpenDSS** with **R/RStudio** can expand the capabilities of each of these tools for performing either steady-state or transient power system studies. Basically, the proposed environment uses **RStudio** as control center from which each simulation tool (e.g., **R**, **ATP**, **OpenDSS**) can be run. Some procedures for generating information that must be exchanged between **RStudio** and **ATP** or **RStudio** and **OpenDSS** have been implemented. Such exchanges are bidirectional: **ATP** and **OpenDSS** produce simulation results that can be read by **RStudio** (text files in the case of **ATP**, comma separated value (CSV) and text files in the case of **OpenDSS**), while **RStudio** capabilities are used to generate files that are embedded into the input file to be read by either **ATP** or **OpenDSS**. This late option can be used to change either the configuration or some parameters of the test system under study. Finally, one very interesting option illustrated in this paper is the possibility of using machine learning algorithms to predict the performance of the test system.

## 1 Introduction

*Open-access software*, *open-source software* and *free software* are distinct concepts often used interchangeably. In addition, these concepts have also been used as synonymous of *public-domain software*. *Open access* refers to a set of principles and a range of practices through which nominally copyrightable publications are delivered to readers free of access charges or other barriers. *Free software* is software that grants the user the freedom to run, copy, distribute, study, change, and improve the software. The term *free* refers to freedom, not price: even if the user pays for a copy, the user retains the right to share and modify it. *Open-source software* is software usually released under a license in which the copyright holder grants users the right to use, study, change, and distribute the software and its source code to anyone and for any purpose. *Public-domain software* is software for which there is no ownership such as copyright, trademark, or patent; it can be modified, distributed, or sold even without any attribution by anyone. Wikipedia provides very useful information and documentation on these concepts; see [1]-[7].

The goal of this paper is to propose a simulation environment to perform power system studies using open-access and free software. Strictly speaking, the tools used in this work are open-access tools available to everybody at no cost, although a license might be required for some tools.

The proposed environment uses **Rstudio** [8] as control center and combines capabilities of **R** [9], a popular tool for statistical studies, and capabilities of two popular simulation tools for power system studies: **ATP** [10] and **OpenDSS** [11]. Some simple case studies have been selected to illustrate the scope of the studies that can be performed using the proposed environment.

The paper has been organized as follows. Section 2 provides a short summary of the capabilities available in the simulation tools: **R** environment (including obviously **RStudio**), **ATP** and **OpenDSS**. An important aspect of the section is to clarify the type of studies and/or simulations that can be performed using every tool alone. Section 3 includes four case studies, two using each power system simulation tool (i.e., **ATP** and **OpenDSS**). Since the number of studies that can be carried out using either **ATP** or **OpenDSS** is basically endless, it is obvious that the selected case studies are just a small sample of the studies that can be performed using the proposed environment. A common aspect to most selected case studies is the application of one or several supervised machine learning (ML) approaches taking advantages of **R** capabilities to show how to predict the performance of the test system; to achieve such a goal, a parametric study aimed at generating the data frames needed to train and test a ML algorithm has to be implemented in each case study. The last two sections add some discussion comments and a summary of the main conclusions.
The test systems used in this paper are very simple; the goal is to illustrate how the combination of power system simulation tools and the **R** environment can be used to expand the capabilities of these tools taking advantage of **RStudio** to show simulation results.

## 2 An overview of simulation tools

This section provides a short summary of capabilities and applications of the simulation tools used in the proposed simulation environment. The information used to prepare the next subsections comes from reference [10] for **ATP**, from the Manual and the official web site for **OpenDSS** (see [12] and [13]), and from several sites for **R/RStudio** (see, for instance, [14] and [15]).

### 2.1 R environment

**R** is a free and open-source programming language for data science and statistics. **R** allows users to handle data, generate high quality visualizations, and perform a range of statistical and analytic computing tasks. In addition to the base built-in computing and graphical functions, users can incorporate additional functions by installing and loading packages available in specific repositories. **R** has been open-source for most of its lifespan, the language itself has undergone some changes and the fields in what it can be used have also significantly expanded. The development of **R** was started in 1991 by Ross Ihaka and Robert Gentleman as a research project for the Department of Statistics at the University of Auckland (New Zealand). **R** was released under a GNU general public license, making this tool both free to use and open-source, in 1995 [16]. The R Core Team, the group that writes, reviews and enacts changes to the language, was formed in 1997. The *Comprehensive R Archive Network* (CRAN), a repository of open-source **R** software packages, was formed the same year. The *R Foundation* was formed in 2003 to hold and administer the software copyright and to provide support for the language project. The *R Journal*, an open-access journal for statistical computing and research, was established in 2009. This tool is supported by a very active community; in addition, several **R** conferences are held on a regular basis.
The most popular way of using **R** is perhaps to run it from **RStudio**, an integrated development environment specifically designed for the **R** programming language, although the current version can also be used with other simulation tools (e.g., **Python**). **RStudio** is currently developed by Posit PBC (formerly RStudio PBC, formerly RStudio Inc.).
The main features of **RStudio** can be summarized as follows:

- It is a friendly graphical user interface (GUI) that simplifies the process of writing and executing **R** code. The interface is divided into multiple panels, including the Console, Environment, History, Files, Plots, Packages, Help, and Viewer tabs (see Figure 1).
- **RStudio** allows users to write **R** scripts and run them or save them for future use; install and load **R** packages; import and explore data from various sources; create high quality visualizations.

- **RStudio** was primarily designed for **R**, but it also supports other programming languages such as **Python**, **Bash**, or **SQL**.
- **RStudio** supports reproducible analyses through *RMarkdown*, which allows users to combine text, code, and output in a single document. These documents can be exported in various formats, including HTML, PDF, and Word.
- **RStudio** provides extensive help documentation for **R** functions and packages, making it easier for users to write and debug code.

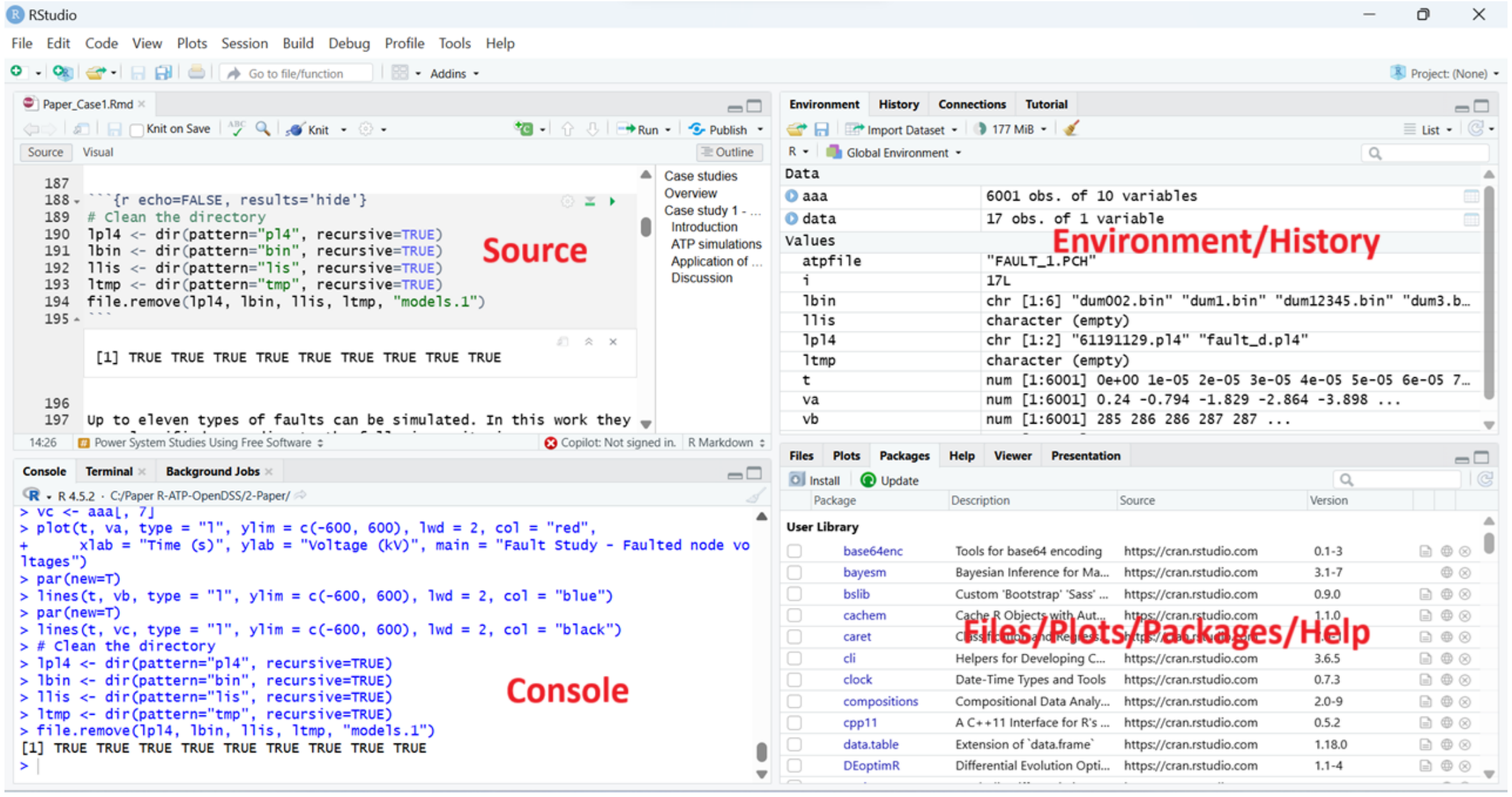

***Figure 1***. *RStudio environment.*

**R** packages are extensions to the programming language, containing code, data, and documentation in a standardized format. They can be installed via CRAN. It contains an archive of the latest versions of the language, documentation, and contributed packages. Other repositories are the *Posit Package Manager* (developed and maintained by the developers of **RStudio**), *Bioconductor* (which provides **R** packages for the analysis of genomic data), and *R-Forge* (a platform for the collaborative development of the **R** environment). In some cases, the developers of a specific package decide to make it available using other options; **Github** is one of the most popular. The large number of packages available for **R** and the ease of installing and using them is one of the major factors that justify the adoption of this language. At the time of writing this paper, **R** users could benefit from more than 15000 packages.

**R** is widely used for statistical computing, data analysis, and graphical representation of data. Its advantages include:

1. **Open source**: **R** is a free open-source language, which means that anyone can use it, modify it, and distribute it.
2. **Extensive libraries**: **R** boasts a vast collection of packages tailored for statistical analyses, machine learning applications, data visualization, numerical computing, time-series analysis, and more.
3. **Statistical analysis**: **R** was specifically designed for statistical analysis. It offers a wide range of statistical functions and methods, making it well-suited for tasks such as hypothesis testing, regression analysis, time-series analysis, clustering, and more.
4. **Data visualization**: **R** provides powerful tools for data visualization, so users can easily generate high-quality plots to explore data effectively.

5. **Integration with other languages**: **R** can be integrated with other programming languages such as **Python**, **Java**, or **C++**, allowing users to leverage the strengths of each language within a single workflow.
6. **Community support**: **R** has a large and active community of users and developers, who share their knowledge and best practices through forums and online resources.
7. **Reproducibility and documentation**: **R** promotes reproducible research through its support for literate programming, which allows users to embed code, documentation, and visualizations in a single document.

**R** can be applied into a myriad of fields: finance, banking, supply chain, retail, marketing, e-commerce, manufacturing, data sciences, environmental and climate sciences, astronomy, chemistry, or genomics. Users can take advantage of capabilities implemented in a vast collection of packages that were developed for machine learning and deep learning analyses [17-21]. The **R** environment exhibits an almost endless list of features and capabilities: to all those mentioned above add parallel computing [22], animation [23], and capabilities for editing any type of document (e.g., books, slides, papers, etc.) [24-26].

## 2.2 The ATP package

### 2.2.1 Introduction

**ATP** is an acronym that stands for Alternative Transients Program. It is a non-commercial simulation program based on the original **EMTP** (ElectroMagnetic Transients Program) developed by Bonneville Power Administration. **ATP** was originally developed for simulation of transients in power systems; current **ATP** capabilities allow user to apply it to the analysis of electromagnetic and electromechanical transients (i.e., overvoltages, subsynchronous resonance, wind power generation), the development of models for power electronics devices (FACTS, Custom Power), or the design of protective relays. This section is a short version of Chapter 4 of reference [27]. Presently, the acronym **ATP** is used to denote a package that consists of at least three tools: (i) the graphical user interface **ATPDraw**; (ii) the simulation tool **TPBIG**; (iii) a postprocessor for graphical output.

**ATPDraw** is an interactive Windows-based graphical user interface (GUI) that can act as a shell for the whole package; that is, users can control the execution of all programs integrated in the package from **ATPDraw**. The ATP package can include many tools depending on the case studies and results in which a user is interested. Several tools are currently available to **ATP** users for postprocessing the simulation results generated by **TPBIG** (e.g., **PCPlot**, **TPPLOT**, **GTPPLOT**, **PlotXY**, **ATP Analyzer**, **TOP**). Other tools, such as **ATPDesigner** and **ATP Control Center**, have been developed to work as *control center*. Figure 2 depicts the most common tasks that can be carried out in typical studies by the tools usually integrated in the ATP package. Although other files can be generated and manipulated by the package tools, and more interactions between programs and files can be activated, the figure shows the most important connections between tools and files.

The capabilities of the **ATP** package are the combination of capabilities provided by each tool. Given that users can expand the package by adding more tools and/or creating custom-made environments, the **ATP** applications are almost unlimited. However, when used here in combination with **RStudio** only the simulation engine (i.e., **TPBIG**) will be used, therefore only a summary of **TPBIG** capabilities is provided below.

**TPBIG** is a tool for simulation of transients in power systems. In addition to the modelling capabilities, several non-simulation supporting routines are available to calculate the parameters of lines, cables and transformers, or create their models. The program can also represent control systems and interface them with an electric network. Although this program is mainly intended for transients simulation, it can also be used to perform ac steady state calculations, to obtain system impedance as a function of the frequency, and to calculate harmonic power flows. A schematic overview of available simulation modules and supporting routines, as well as their interaction, is shown in Figure 3.

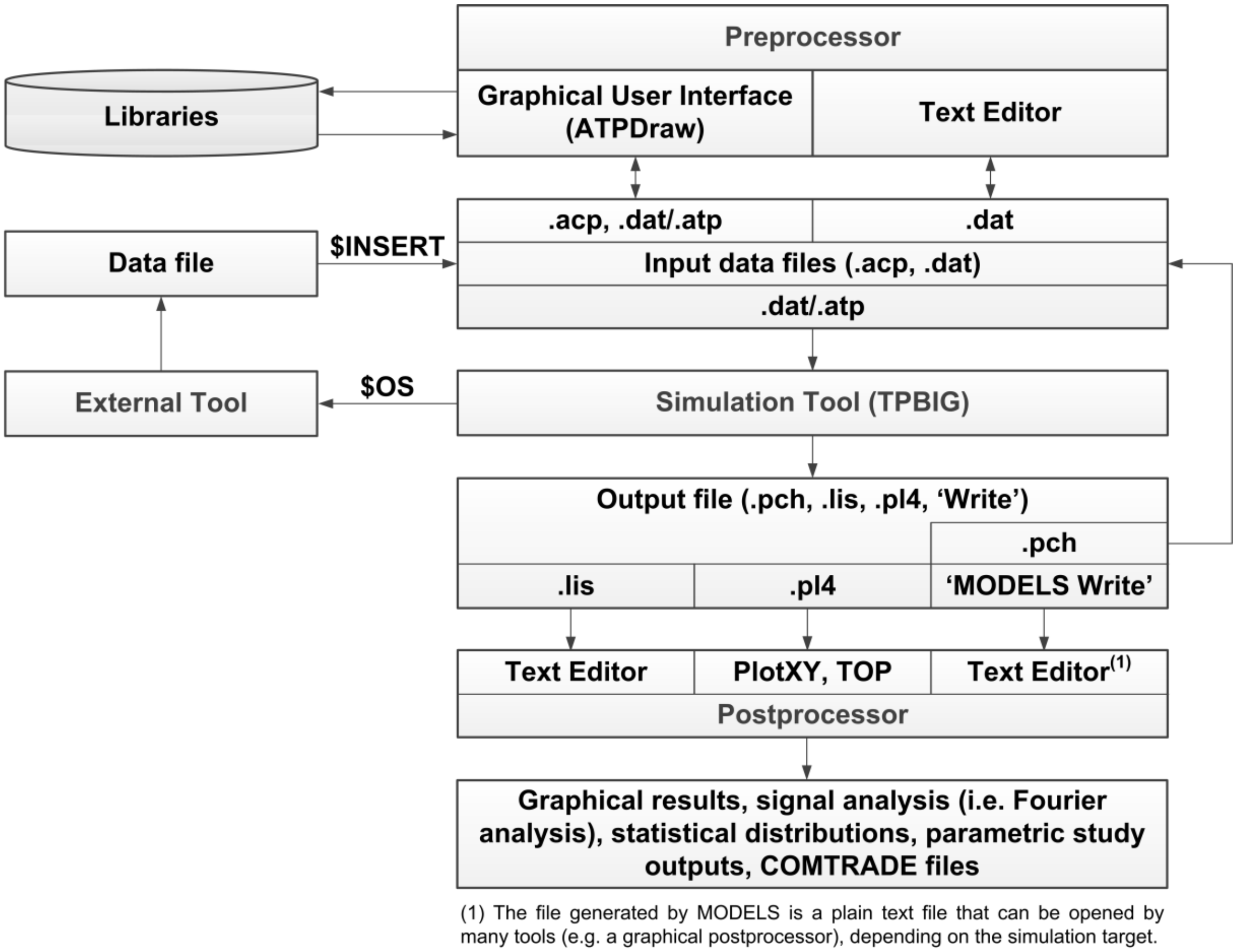

***Figure 2**. Tools and tasks of the ATP package [10].*

**TPBIG** can represent a high number of electrical network components, from simple passive linear and non-linear branches, switches, and static sources to sophisticated models of transmission lines, power transformers, and rotating machines. Its basic capabilities include built-in models that cover multi-phase line and cable models with either lumped or distributed parameters, ideal and saturable multi-phase transformer models, rotating machine models, ideal time- and voltage-controlled switches, ideal current and voltage sources of various shapes. **TPBIG** also supports two options for representing control systems:

1. TACS (Transient Analysis of Control Systems) is a simulation module for time-domain analysis of control systems based on a block diagram representation. A TACS section within an input data file consists of predefined blocks categorized into sources, transfer functions, devices, and FORTRAN statements. One time-step delay exists in the interface between a TACS-based model and the electrical network. An extra time delay can also occur due to nonlinearities in the control system; the number of this type of delay can be reduced by optimal sorting of the control blocks. Input variables can be node voltages, switch currents, internal machine variables, and switch statuses. Special internal sources can also be defined in TACS. FORTRAN statements can be used within TACS to define variables with a FORTRAN-like syntax.
2. MODELS is a symbolic language that provides an option for numerical and logical manipulation of variables. The main MODELS features are: (a) a distinction between model description and model use; (b) the decomposition of a large model into submodels with a hierarchical organization; (c) the self-documenting nature of a model description, which can be easily understood and can serve as a reference document.

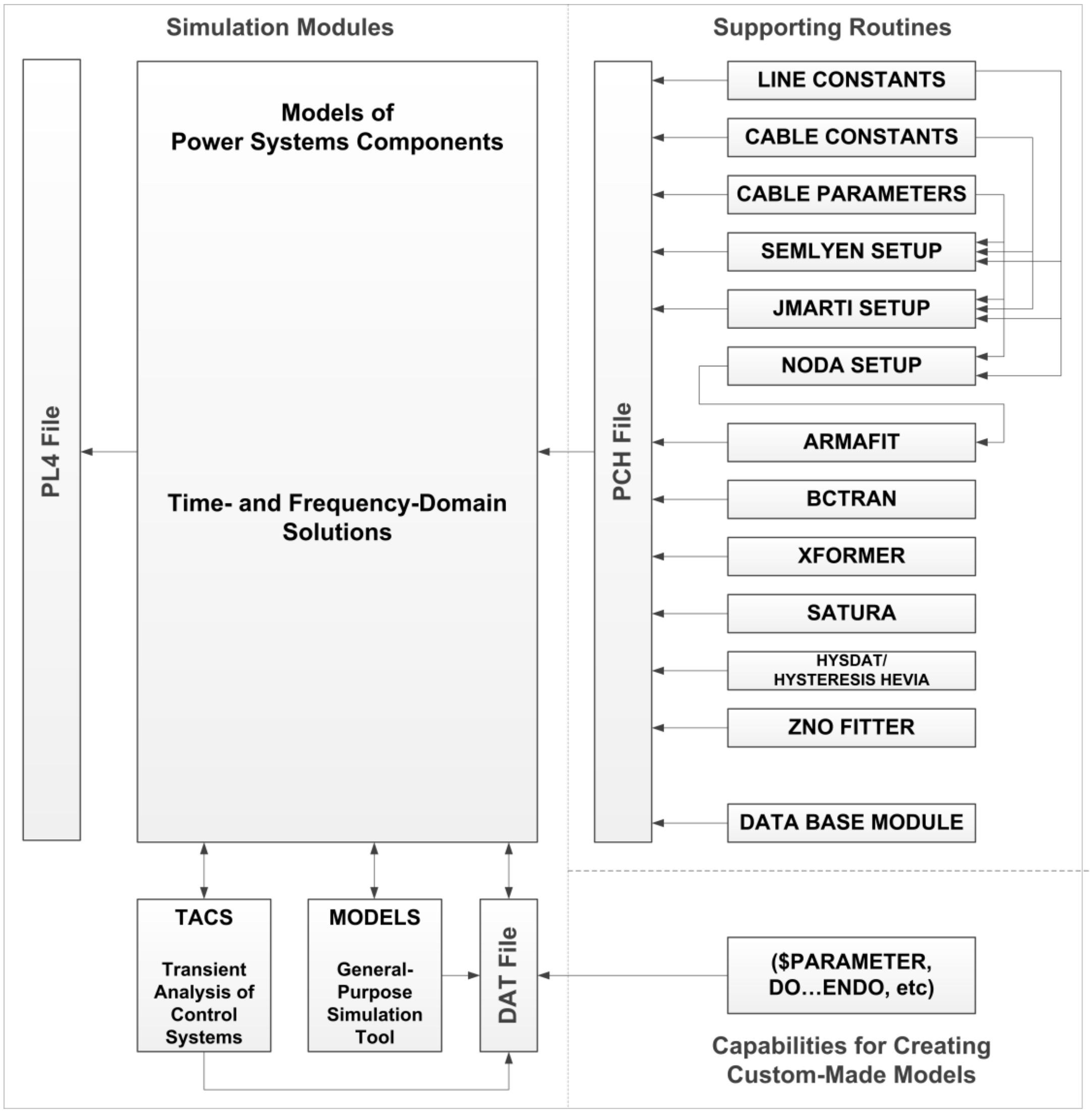

***Figure 3***. *TPBIG capabilities [10].*

### 2.2.2 Creation of custom-made models

Several **TPBIG** capabilities can be used to create custom-made models not available in the program (e.g., relay models) or to facilitate the usage of other components (e.g., a converter for representing a solid-state switch). In addition to built-in components, the options to be considered for this type of task are the supporting routine DATA BASE MODULE and $PARAMETER.

### 2.2.3 Solution Techniques

**Time-domain solution techniques**: **TPBIG** uses the Dommel scheme to convert the differential equations of the network components into algebraic equations [28]: the discrete companion model of a lumped parameter component is obtained from its differential equation using the trapezoidal integration; the discrete companion model of a component with distributed parameters (e.g., a lossless transmission line) is deduced from the application of the Bergeron method. With the Dommel scheme, the discretized equations of network components are assembled using the nodal admittance matrix of the equivalent companion circuit. Since this scheme can only be used to solve linear networks, some approaches have been implemented to cope with nonlinear and time-varying elements; they can be based on a piecewise-linear representation or the compensation method. The trapezoidal rule is simple and numerically stable; however, it has some drawbacks: it uses a fixed time-step size and can originate numerical oscillations. In switching operations or transitions between segments in piecewise-linear inductances, the trapezoidal rule acts as a differentiator, and introduces sustained numerical oscillations. Several techniques have been proposed to control or reduce these oscillations.

**Frequency-domain solution techniques**: The incorporation of frequency-domain solution techniques in a transients tool can be very useful; they can be used to obtain the steady-state solution that exists prior to the disturbance, detect resonance conditions, or analyze harmonic propagation. When initial operating conditions are specified as power constraints, the available power flow option available in ATP, known as FIX SOURCE, is used to obtain the initial conditions for the subsequent transient simulation. The FREQUENCY SCAN (FS) option performs repetitive steady-state phasor solutions, as the frequency of sinusoidal sources is incremented between a lower and upper value, a frequency-response of node voltages, branch currents or driving-point impedances/admittances is obtained. Typical applications of FS are the analysis and identification of resonant frequencies, and the frequency response of driving-point network impedances or admittances seen from a busbar (e.g., positive-, negative- or zero-sequence impedance). HARMONIC FREQUENCY SCAN (HFS) is a companion to FS: FS solves the network for the specified sources, incrementing in each subsequent step the frequency of the sources but not their amplitudes, HFS performs harmonic analysis by executing a string of phasor solutions determined by a list of sinusoidal sources entered by the user.

**Control systems dynamics**: The solution method for control systems based on the TACS option is also based on the trapezoidal rule: transfer functions are converted into algebraic equations. Since the resulting matrices are unsymmetrical by nature, the electric network and the control system are solved separately. The network solution is first advanced, network variables are next passed to the control section, and control equations are solved. Finally, the network receives control commands. The whole procedure introduces a time-step delay. When a nonlinear block is inside a closed-loop configuration, a true simultaneous solution is not possible. The procedure is simultaneous only for linear blocks, and sequential for nonlinear blocks. Therefore, the loop is broken and the system is solved by inserting a time delay [29].

### 2.2.4 Applications

The capabilities implemented in **TPBIG** can be used to perform many studies in power systems. Table 1 shows a summary of the most important applications.

ATP applications can also be classified as follows:

- *Time-domain simulations*: They are generally used for simulation of transients, such as switching or lightning overvoltages. They can also be used for analyzing harmonic distortion created by power electronics devices.
- *Frequency-domain simulations*: **TPBIG** capabilities can be used to obtain the driving point impedance at a particular node versus frequency, detect resonance conditions, design filter banks, or analyze harmonic propagation.
- *Parametric studies*: They are usually performed to evaluate the relationship between variables and parameters; when one or more parameters cannot be accurately specified, this analysis will determine the range of values for which they are of concern.
- *Statistical studies*: In addition to statistical switches, users can combine some **ATP** capabilities to perform all types of Monte Carlo simulations, not covered by statistical switches.

***Table 1*** *- ATP Applications [10]*

| Study | Requirements |
|---|---|
| Ferroresonance | Basic power components |
| Parallel line resonance | Steady-state initialization (SSI) |
| | Basic power components |
| Harmonics | Frequency scan/Harmonic power flow |
| | Basic power components |
| | Frequency-dependent models |
| Subsynchronous resonance | SSI + Basic power components |
| | Three-phase synchronous machine |
| Transient stability | SSI + Basic power components |
| | Three-phase synchronous machine |
| | Control system dynamics |
| Switching studies | SSI + Basic power components |
| | Statistical switches/tabulation |
| | Frequency-dependent line model |
| Lightning studies | Basic power components |
| | Frequency-dependent line model |
| HVDC | Basic power components |
| | Semiconductor devices |
| | Control system dynamics |
| FACTS devices | Basic power components |
| | Semiconductor devices |
| | Control system dynamics |
| Adjustable speed drives | Basic power components |
| | Semiconductor devices + TACS |
| | Rotating machine models |
| System protection | SSI + Basic power components |
| | Control system dynamics |
| | Frequency-dependent line model |
| Secondary arc | SSI + Basic power components |
| | Control system dynamics |
| | Frequency-dependent line model |

**TPBIG** capabilities can also be used to expand the application fields; for instance, parameters of the system under simulation can be changed according to a given law, some components can be either disconnected or activated, and some calculations can be carried out by external programs. In addition, it is possible, if required, to modify online the simulation time or the number of runs. Other concepts (e.g., multiple run option, data symbol replacement, data module) can be of paramount importance for expanding **ATP** applications; for details about them see [27].

## 2.3 OpenDSS

### 2.3.1 Introduction

The Open Distribution System Simulator (**OpenDSS**) is an open-source simulation tool specifically designed for distribution system studies [11-13,]. It has been implemented as both a stand-alone executable program and a component object-model (COM) dynamic-link library (DLL) designed to be driven from a variety of existing software platforms. The executable version adds a basic user interface on to the solution engine to assist users in developing scripts and viewing solutions.

**OpenDSS** represents distribution circuits using nodal admittance equations. The nonlinear behavior of some devices (e.g., some load models) is modeled by current source injections using *compensation*: the current predicted from the linear portion of the model that resides in the admittance matrix is compensated by an external injection to iteratively obtain the correct current.

**OpenDSS** may be scripted via text commands, text files, or through calls to functions in the COM interface from user-written programs. Thus, every single element is required to be connected to the model through the command line.

**OpenDSS** has special features for creating models of electric power distribution systems and performing many types of analysis related to distribution planning and power quality. Although it does not perform time domain simulations, it can perform transient stability studies. All types of analysis are currently in the frequency domain.

**OpenDSS** can be executed as a stand-alone program or integrated into other software platforms (e.g., **MATLAB**, **Excel**, **Python**) by using the MS Windows COM interface, which allows interaction among different objects without dictating the methods of implementation. This tool is windows based and is not compatible with other operating systems. Through the COM interface, the user can add other solution modes and features externally and perform the functions of the simulator, including definition of the model data. **OpenDSS** can be driven entirely from a MS Office tool through VBA, or from any other 3rd party analysis program that can handle COM. Users can drive **OpenDSS** from **MATLAB**; this approach provides powerful external analytical capabilities as well as excellent graphics for displaying results.

The COM interface also provides direct access to the text-based command interface as well as numerous methods and properties for accessing many of the parameters and functions of the simulator's models. Through the command interface, user-written programs can generate scripts to do several desired functions in sequence. The input may be redirected to a text file to accomplish the same effect as macros and provide some database-like characteristics. Many of the results can be retrieved through the COM interface as well as from various output files. Output files are typically written in CSV format that imports easily into other tools for post processing.

Advanced users have two additional options for using **OpenDSS**: (i) downloading the source code and modifying it to suit special needs; (ii) developing DLLs that plug into generic containers the **OpenDSS** provides. This allows developers to concentrate on the model of the device of interest while letting the DSS executive take care of all other aspects of the distribution system model. Such DLLs can be written in most common programming languages.

Figure 4 illustrates the structure of **OpenDSS**. The user can interact with the simulation engine by using text scripts, COM interface calls, and user-written DLLs. Output scripts can be saved and used by other software programs that can import common CSV formats.

**OpenDSS** can analyze unbalanced, n-phase power utility distribution systems and distributed generation (DG) interconnections to the distribution system. Combined with transmission system studies, these analyses can provide a holistic assessment of power system response.

**Open-DSS** provides different generator models, inverters included. PVSystem objects can be controlled by advanced inverter controls to achieve volt-VAR, volt-watt and other advanced inverter control characteristics. **OpenDSS** also provides models for fuses, reclosers, and relays whose parameters and technical specifications can be changed by the user.

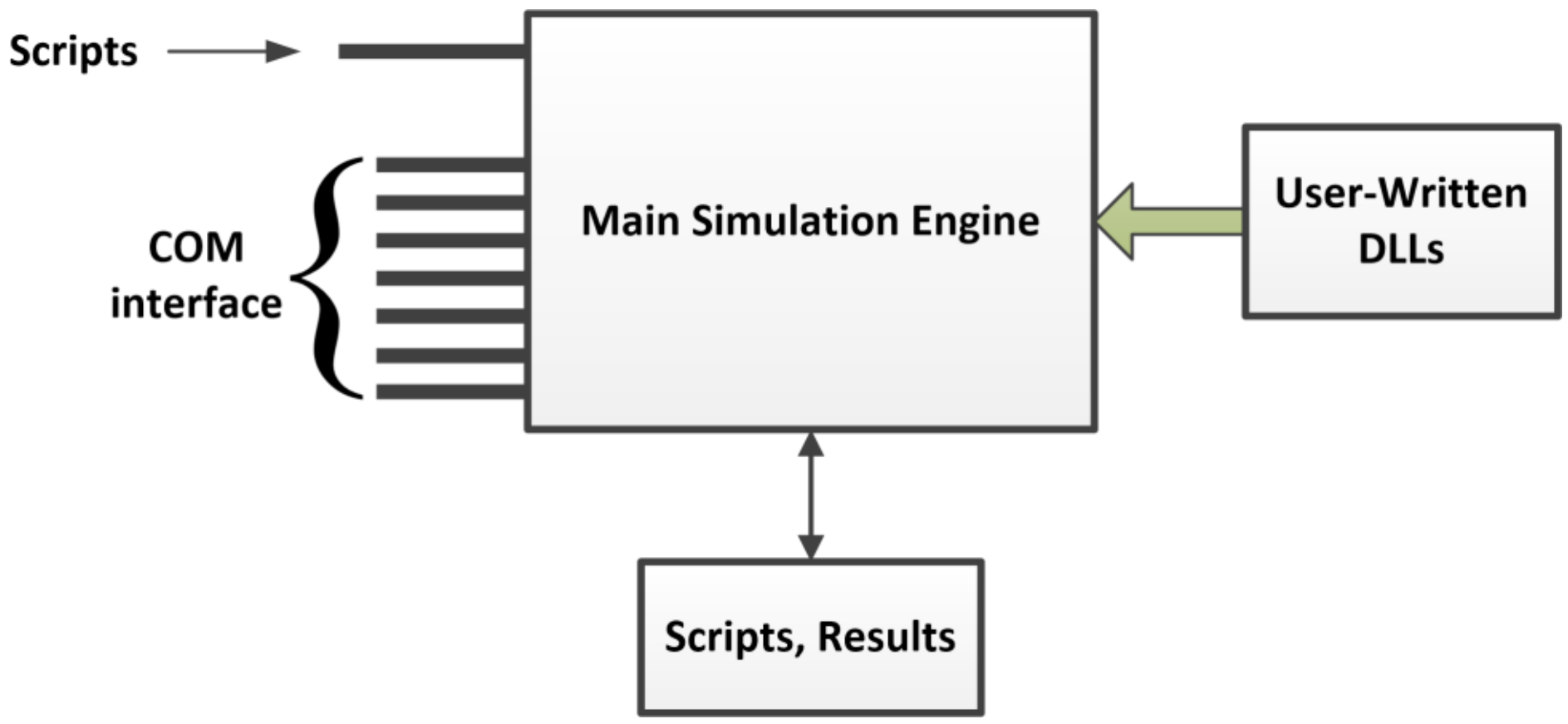

***Figure 4***. *OpenDSS Structure.*

Sequential-time simulations are essential for modeling variable elements, such as renewable generators, electric vehicles, and storage [11-13]. Sequential-time simulations, built-in solution modes, and separate models for controllers and circuit elements allow **OpenDSS** to investigate the fluctuation of renewable resources and assess the impact of interconnecting high penetrations of variable generation.

**OpenDSS** has several built-in solution modes: snapshot power flow, daily mode (24 hours,1-hour increments), yearly mode (8760 hours, 1-hour increments), and duty cycle (time steps ranging from 1 seconds to 5 seconds). This facilitates distribution planning studies and interconnection studies with high penetration of variable generation. Figure 5 shows the detailed components of the main simulation engine of **OpenDSS**.

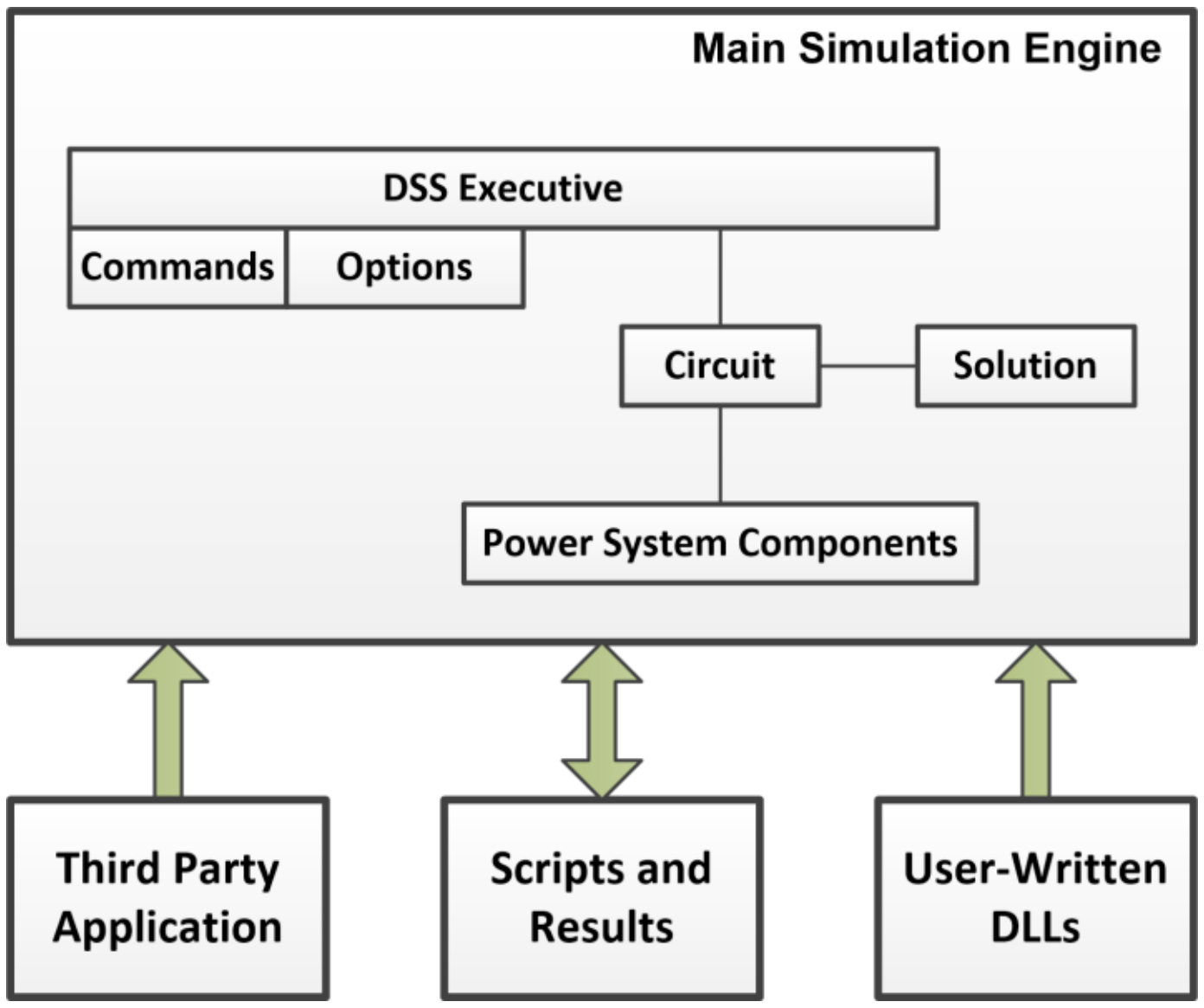

***Figure 5***. *Detailed simulation engine of OpenDSS.*

### 2.3.2 Capabilities and Applications

The list of **OpenDSS** features includes among others [11-13]:

- **Grid elements**. **OpenDSS** can model distribution network elements related to delivery (e.g., line, cable, capacitor, reactor, and transformer), conversion (e.g., generator, motor, PV system, storage, and load), control (e.g., controllers of capacitor banks, storage units, DG), and protection (e.g., overcurrent devices). These models enable the package to support elements of the analysis of the distribution power grid with variable generation, such as power loss calculation, short-circuit currents, unbalanced loads, and storage management.

- **Preprocessing and post processing**. **OpenDSS** supports CSV files, **MATLAB**, and **Excel** to preprocess or post process data.
- **Compatibility with other applications**. **OpenDSS** can be easily integrated with other software platforms through its COM interface, which allows users to control the simulation from their own software. **OpenDSS** is compatible with **MATLAB**, **Excel**, and **VBA**.

**OpenDSS** can be used for:

- **Power flow analysis**. Multi-phase models allow **OpenDSS** to run power flow analysis for radial networks and meshed systems. The software can also perform power flow for unbalanced loads. Currently, **OpenDSS** employs three iterative power flow algorithms: normal current injection mode (the default method), Newton mode, and NCIM (Node Current Injection Method) mode.
- **Harmonic analysis**. **OpenDSS** has a harmonics solution mode to perform simulations of harmonic current and voltage distortion in networks.
- **Capacitor bank control**. **OpenDSS** can model standard utility capacitor bank control modes, such as voltage, current, kVAR, and time. Voltage override can be applied to any mode; capacitors can be controlled remotely and by dynamic-linked libraries written by the user.
- **Short-circuit analysis**. **OpenDSS** has an existing command that can perform short-circuit analysis. It can also solve for a short circuit while executing a power flow study.
- **Transient stability solution**. **OpenDSS** can perform transient stability analysis, allowing for islanding studies related to distributed generation and microgrids.
- **Voltage regulator models**. **OpenDSS** has a realistic regulator model with several options, such as regulator controls (with consideration for line drop compensation), remote bus regulation, and automatic reversing.

The list of applications is much longer since it can also be used for distribution planning and analysis, open conductor fault analysis, distribution state estimation, simulation of wind and photovoltaic power generation, analysis of distributed generation impact, analysis of unusual transformer configurations, or development of test systems.

# 3 Case studies

## 3.1 Overview

This section shows how to proceed with the proposed simulation environment and what results can be derived. Up to four case studies using a combination of either **R** and **ATP** or **R** and **OpenDSS** are presented below. The first two cases use **ATP** capabilities, the last two cases use **OpenDSS** capabilities. In three case studies the capabilities of either **ATP** or **OpenDSS** are used to obtain results that will be later used to apply machine learning (ML) algorithms available in **R** libraries. The goal of every case study is summarized below:

- *Case study 1*: Fault location using **ATP**. The case study uses the nearest neighbor algorithm to predict both the type and location of the fault from power system measurements.
- *Case study 2*: Lightning performance of a transmission line using **ATP**. The case study applies the support vector machine algorithm to predict the lightning flashover rate of the test line from the geometry of the line and the lightning activity in the area where the line is operating.
- *Case study 3*: Distribution system analysis using **OpenDSS** considering time-varying and random loads. The case study also analyzes the impact of local distribution and energy storage.
- *Case study 4*: Introduction to transient stability of power systems using **OpenDSS**. This case study applies several machine learning algorithms to predict the stability of the test system.

The procedure established for all case studies may be summarized as follows:

1. Edit the input file for **ATP** or **OpenDSS** as usual. Since the goal is to apply **RStudio** in parametric studies and **R** capabilities for ML studies, the edition of the input files for the two simulation tools should consider the possibility of modifying the files as simulations progress. Fortunately, both simulation tools, **ATP** and **OpenDSS**, allow users to include/insert scripts that can be created within **RStudio**.
2. Edit an *RMarkdown* for every case study. The nature of this type of file allows users to create very flexible structures that can be used to run a simple simulation, a parametric study or a predictive study based on an ML algorithm. Consider that this late option will be based on training and testing data sets that could be obtained by implementing a parametric study in the *RMarkdown*.
3. Create the figures that will be used in the final version of the paper. Remember that some **R** packages can be used to generate high quality figures and even some animation, which could be an option in transient studies.
4. Create a single *RMarkdown* by joining all case studies and generate a notebook using the formats that can be used with **RStudio**: word, pdf, and html. This is an interesting option that adds value to the combination of **RStudio** and a power simulation tool. As mentioned in the Introduction, the draft of this paper has been created using a single *RMarkdown* with which all illustrative case studies included in this section were run.

A notebook generated from an *RMarkdown* benefits from the capabilities available in **R** packages: high-quality visualization, statistical studies, flow control to adapt the input file to the requirements of the study, implementation of ML algorithms to predict test system performance.

## 3.2 Case study 1 - Fault location

### 3.2.1 Introduction

The goal is to apply the k-nearest neighbors (kNN) algorithm to locate and classify faults using the RMS voltage values that correspond to several fault locations along an overhead test line. That is, the recorded information will only use steady state RMS values of line voltages without considering the voltages during any transient period. The kNN algorithm is a non-parametric supervised ML method that can be used for both classification and regression. It predicts the label or value of a sample by looking at the $k$ closest points from the training set in feature space and using their majority vote or average. It was first developed by E. Fix and J. Hodges in 1951 [30] and later expanded by T. Cover and P.E. Hart [31]; see also [32].

Figure 6 shows the configuration of the test line. It is a 400 kV, 100 km transmission line shielded by two ground wires.

The simulations aimed at obtaining the input from which train and test data sets will be derived are controlled by *RMarkdown*. Since **TPBIG** will be the only tool of the **ATP** package to be used here, some previous editing is required: a preprocessing step is needed to create the input files in text code. The procedure is detailed below.

1. Use **ATPDraw** to create the input file. The test line is represented as a two-section line, being the middle point the node at which the fault is located. Figure 7 shows the decoration used in **ATPDraw** for creating the test system model. Note that the test system includes an ideal three-phase 400 kV (phase-to-phase RMS value) voltage source, an equivalent impedance (needed to represent the system to which the line is connected), measuring switches, and the various system elements required for representing a fault. Voltages are measured at the three-phases of the ideal source voltage terminals, the two line ends and the fault location (the node between the line sections).

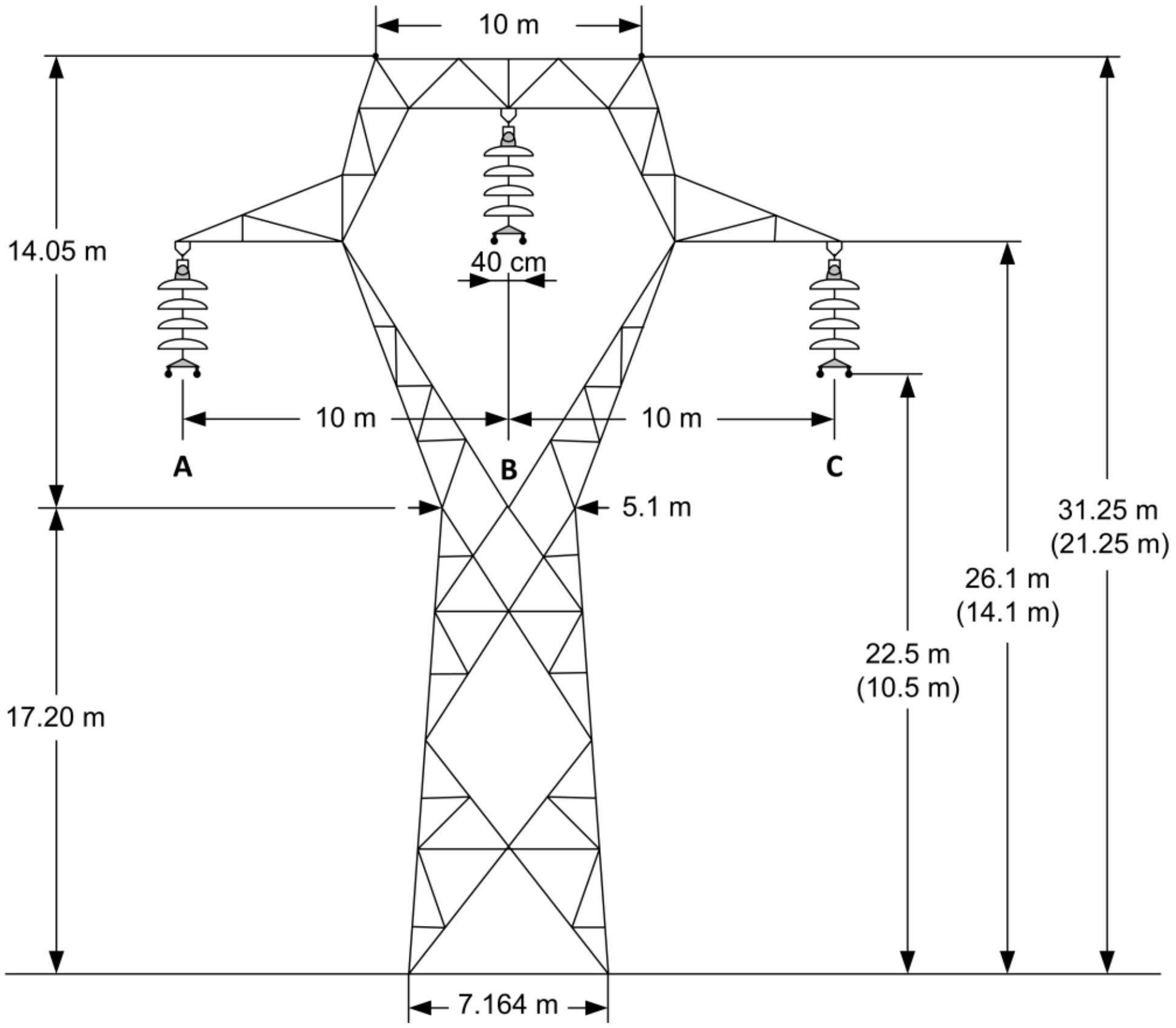

***Figure 6***. *Case study 1: Test line configuration.*

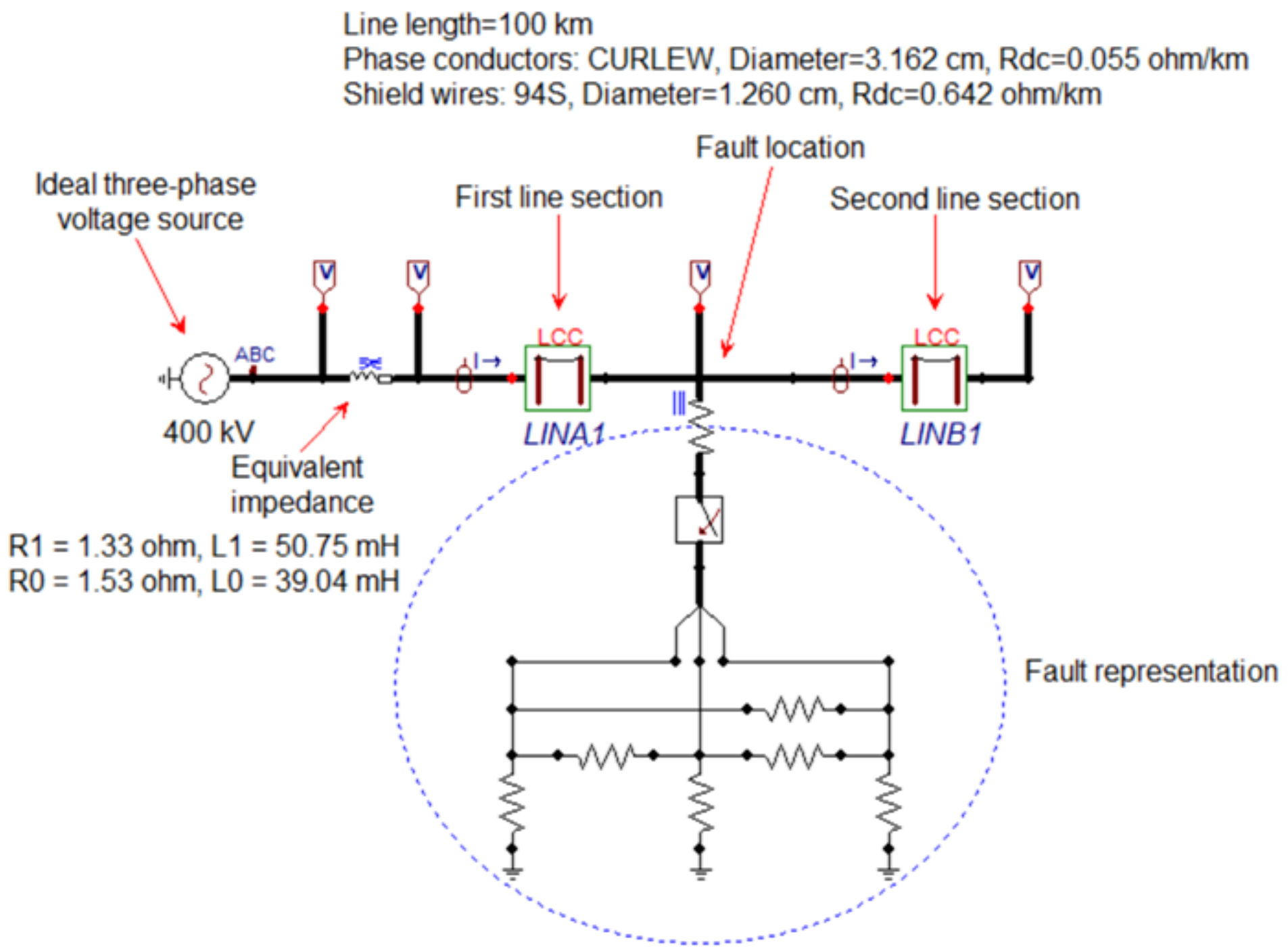

***Figure 7***. *Case study 1: APTDraw file for the test line under a fault condition.*

2. To obtain a data set adequate for application of the kNN algorithm, a reasonable number of **TPBIG** simulations will be executed by changing the type and location of the fault along the test line. The text data files required to run the cases are generated within the *Rmarkdown* and included into the input data files to be used by **TPBIG** by means of a $INCLUDE statement.

### 3.2.2 ATP simulation

The test system is simulated under fault condition and voltage values are passed to a MODELS section for further manipulation and visualization using **R** capabilities. The simulation results shown in Figure 8 correspond to a two-phase-to-ground fault (ACG fault) located 30 km from the beginning of the test line. Note that an actual steady state is not reached during the simulations presented in Figure 8. Since the fault condition appears 10 milliseconds after starting the simulation, the voltage of the unfaulted phase exhibits some ripple. To obtain an actual steady state (without ripple in voltage waveforms), the test system can be simulated by assuming that the fault condition is already at the beginning of the simulation (see Figure 9).

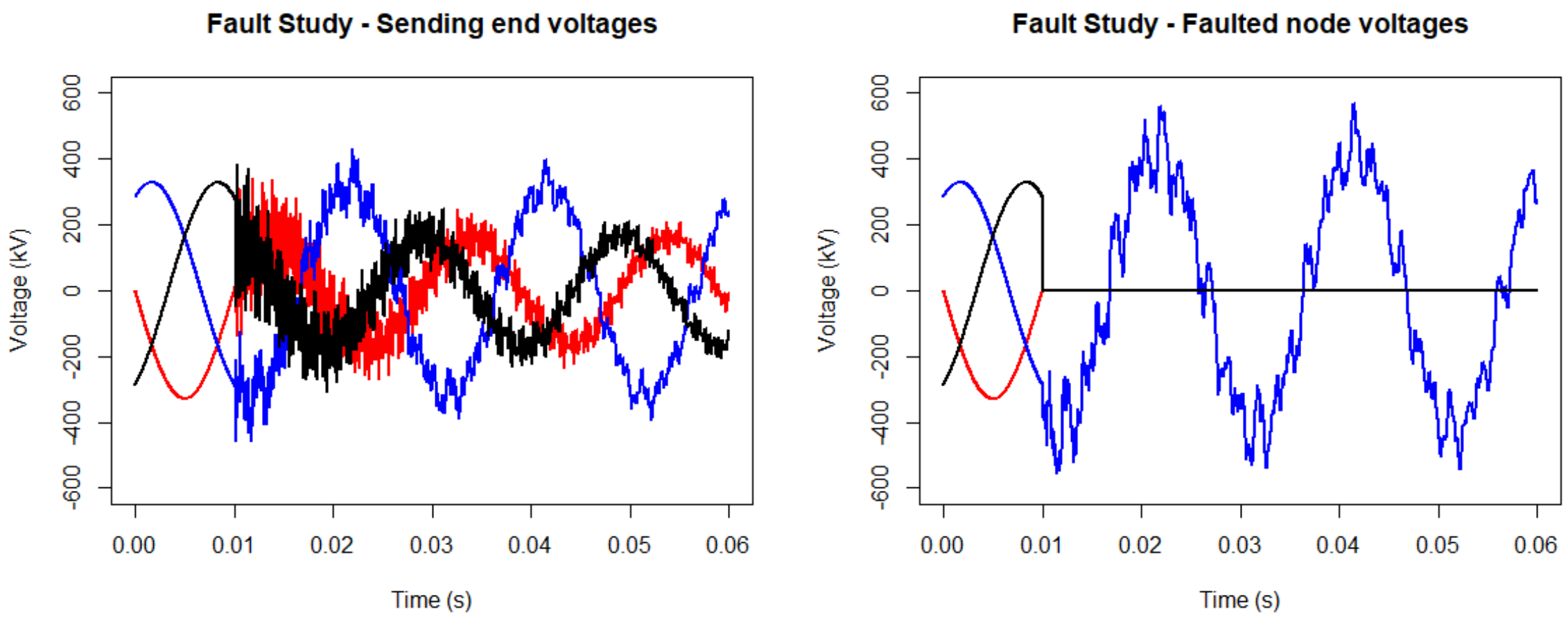

***Figure 8**. Case study 1: Simulation results for a double-phase-to-ground (ACG) at 30 km from the sending end of the test line.*

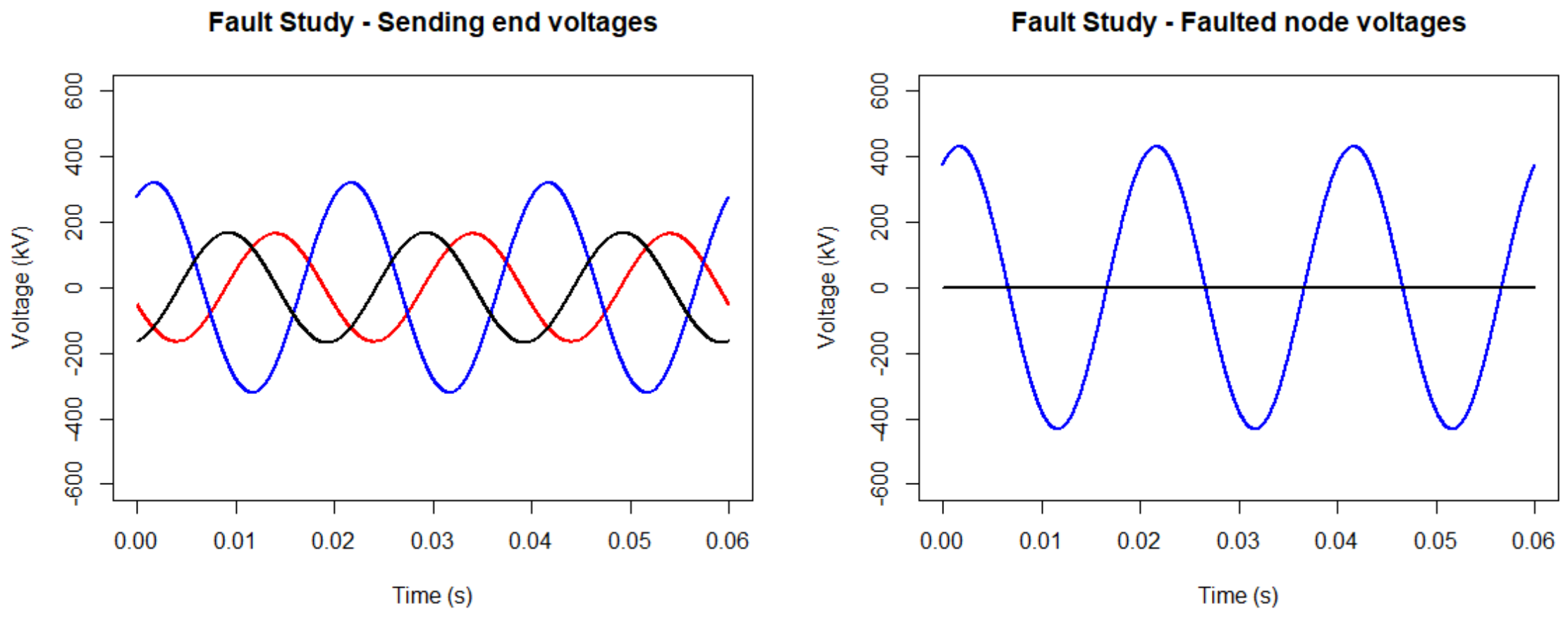

***Figure 9**. Case study 1: Steady state simulation results for a double-phase-to-ground (ACG) at 30 km from the sending end of the test line.*

Up to eleven types of faults can be simulated. In this work they are classified according to the following criterium:

- $j = 1$, Three-phase-to-ground (ABCG) fault

- $j = 2$, Three-phase (ABC) fault
- $j = 3$, Double-phase-to-ground (ABG) fault
- $j = 4$, Double-phase-to-ground (BCG) fault
- $j = 5$, Double-phase-to-ground (ACG) fault
- $j = 6$, Double-phase (AB) fault
- $j = 7$, Double-phase (BC) fault
- $j = 8$, Double-phase (AC) fault
- $j = 9$, Single-phase-to-ground (AG) fault
- $j = 10$, Single-phase-to-ground (BG) fault
- $j = 11$, Single-phase-to-ground (CG) fault

The next subsection shows how to obtain data for training and testing the algorithm and how to apply it to predict the type of fault and its location.

### 3.2.3 Application of the nearest neighbor algorithm

A parametric approach is used now to generate both the train and test data sets. In both cases, all types of faults (remember there are eleven types) are simulated every 5 km, beginning at 5 km from the first line terminal and ending at 5 km from the second line terminal, with a total of 19 fault locations. That is, two data frames with $19 \times 11 = 209$ rows are created. The distance between the fault location and the type of fault (from 1 to 11) are systematically generated to create the train data set and randomly generated to create the test data set. All fault models generated for the training data set are bold (fault resistances are zero), while resistance values for the testing data set are randomly generated with values below 1 ohm.

Figure 10 depicts a scheme of the procedure implemented here for applying the kNN algorithm to fault classification and location.

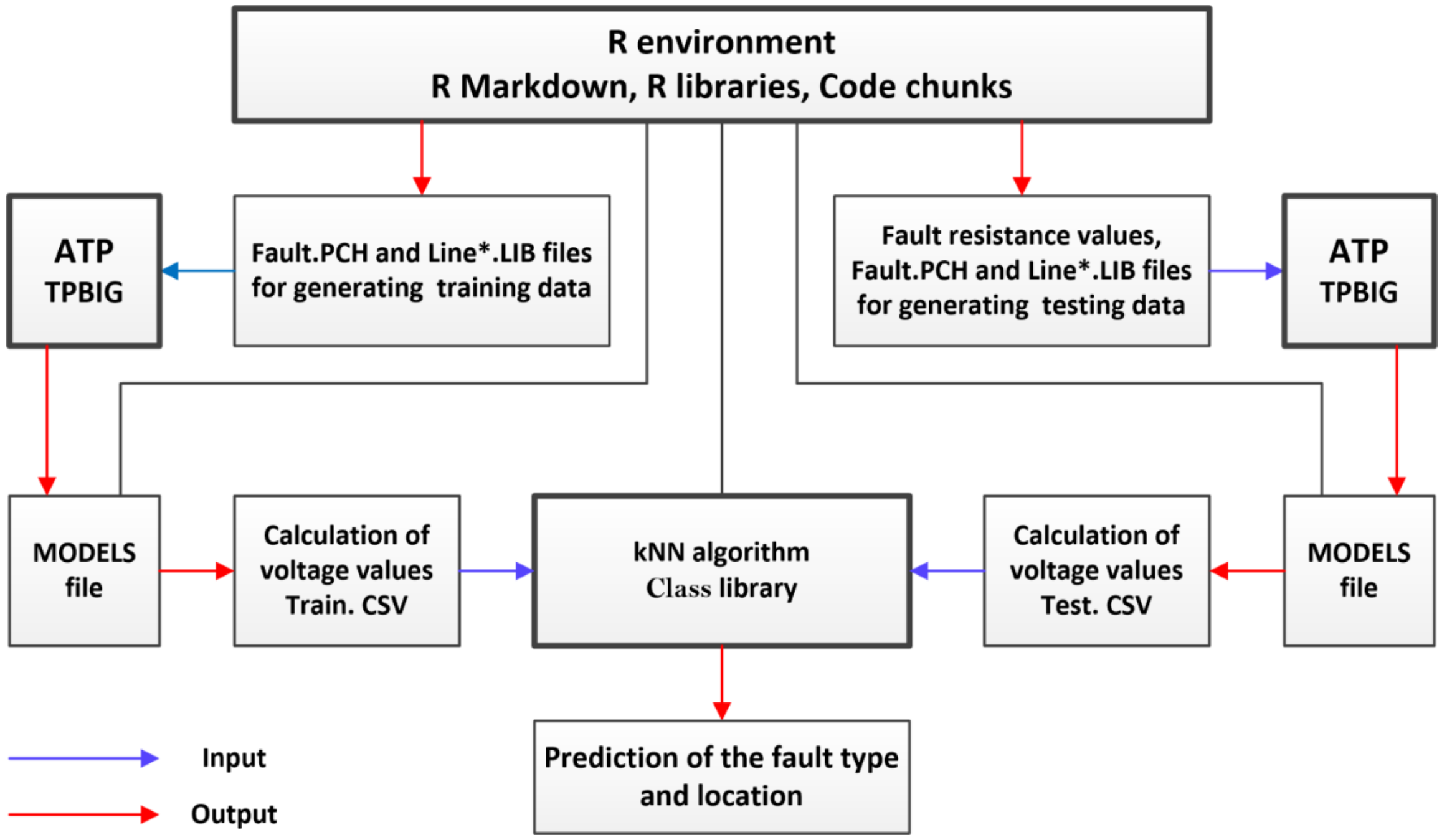

***Figure 10**. Case study 1: Scheme implemented for classification and location of faults in overhead transmission lines.*

For each row, the RMS values of the three voltages at the two line terminals and the fault position are saved. Two additional columns are used to save the codes of the fault position (distance between the fault and the line beginning) and the fault type.

1. *Generation of the train data set*: The generated values are saved in a CSV file. The values saved in each of the $19 \times 11 = 209$ rows are the rms values of voltages at the two line terminals and at the fault location, as well as the code of distance from the fault location to the sending end terminal of the line and the fault type code. Remember that the line length is 100 km and the considered fault locations are homogeneously distributed along this distance, beginning at 5 km and ending at 95 from the sending end of the test line.
2. *Generation of the test data set*: As for the training data set, the generated values are saved in a CSV file. The values saved in each row are, as for the train data set, the rms values of voltages at the two line terminals and at the fault location, as well as the code of distance from the fault location to the sending end terminal of the line and the fault type code. Remember that the values used for representing the fault are all zero in the training step and randomly generated, with values below 1 ohm, for the testing step.

Figure 11 shows some features of the train and test data sets. Observe the different configuration of both data sets: the values used for training were systematically generated, while the values for testing were randomly generated.

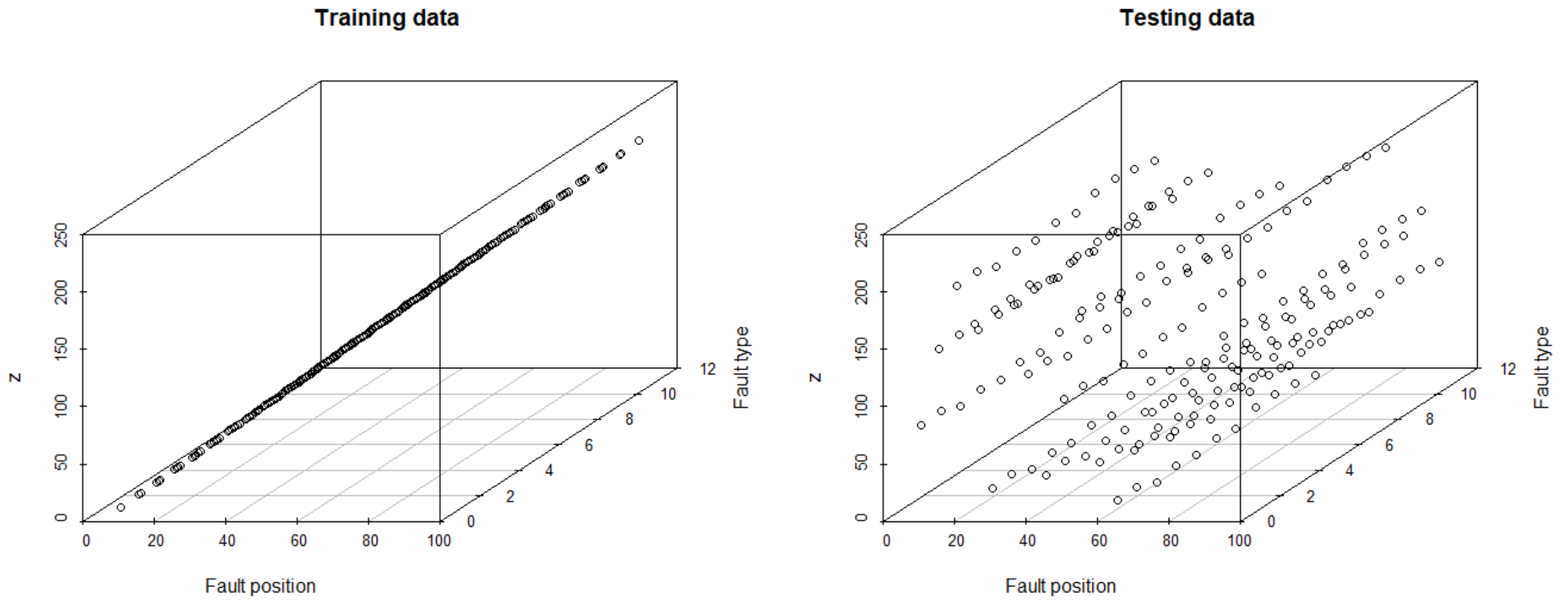

***Figure11***. *Case study 1: Fault location and fault type values for training and testing.*

The kNN algorithm is applied using the `knn` function available in the `class` library. First, both (train and test) CSV files are read and prepared for the application of the kNN algorithm. Next, they are manipulated for further application of the algorithm: all voltage values are passed to per unit values (using the rated phase-to-ground RMS voltage value of the test line) and a single code is generated to simultaneously cope with fault type and location.

One example is shown below: some of the values corresponding to the row 200 of the training data set before and after the manipulation mentioned above. This example corresponds to three-phase fault without contact to ground (the fault type is 02, according to the classification detailed above) at 95 (= 19*5) km from the beginning line terminal.

```
##       VloadA   VloadB   VloadC Distance
## 200 43555.65 43555.99 43555.66       95

##        VloadA   VloadB    VloadC Code
## 200 0.1886015 0.188603 0.1886015 1902

## Training data set sample

##          VbusA      VbusB      VbusC      VfaultA      VfaultB      VfaultC Code
## 1  0.09282529 0.10423275 0.09123346 1.137875e-07 1.114481e-07 1.141726e-07  101
## 2  0.09828552 0.09084722 0.09765437 2.576860e-02 2.576867e-02 2.576873e-02  102
## 3  0.18691577 0.20490622 0.94775462 1.110021e-07 1.072618e-07 1.076148e+00  103
## 4  0.94660499 0.21363364 0.17838497 1.072257e+00 1.066317e-07 1.115582e-07  104
## 5  0.18106608 0.94484202 0.18925699 1.095352e-07 1.091459e+00 1.092735e-07  105
```

```
## 6  0.50916220 0.50918251 1.00557354 5.017877e-01 5.017876e-01 1.008690e+00  106
## 7  1.00545375 0.50913666 0.50907598 1.004120e+00 5.050531e-01 5.050529e-01  107
## 8  0.51103646 1.00603795 0.51009852 5.033373e-01 1.006726e+00 5.033374e-01  108
## 9  0.22044650 0.96729175 0.98283929 1.056353e-07 1.057640e+00 1.037007e+00  109
## 10 0.98318844 0.25267109 0.96846130 1.049062e+00 1.013168e-07 1.054587e+00  110

## Testing data set sample

##        VbusA     VbusB     VbusC     VfaultA      VfaultB     VfaultC Code
## 1  0.6349413 0.6569411 0.9945020 0.006371069 0.0234308888 1.322766323 1203
## 2  0.6486202 0.9933925 0.6634977 0.016496933 1.3847766571 0.017230645 1205
## 3  1.0054691 0.6968155 0.6921027 1.008016856 0.5376208548 0.523229051 1207
## 4  0.9936852 0.6739742 0.6194204 1.301114295 0.0225393108 0.009341099 1204
## 5  0.9947481 0.9989213 0.7756749 1.295118697 1.3423794902 0.003238305 1211
## 6  0.6944012 0.6941526 1.0055583 0.510400708 0.5091797066 1.036118224 1206
## 7  0.5596489 0.5533650 0.5584582 0.021902750 0.0008864356 0.006119956 1201
## 8  0.5673703 0.5447769 0.5661413 0.151398215 0.1595082371 0.161124745 1202
## 9  0.9991649 0.8058251 0.9962832 1.291881735 0.0029831701 1.317720201 1210
## 10 0.7058951 1.0060379 0.7068156 0.500005491 1.0119736127 0.509591795 1208
```

The algorithm is applied considering different values for the nearest neighbors.

```
## Number of nearest neighbors = 1
## agreement1
##       FALSE        TRUE
## 0.004784689 0.995215311

## Number of nearest neighbors = 2
## agreement2
##     FALSE      TRUE
## 0.5023923 0.4976077

## Number of nearest neighbors = 3
## agreement3
##     FALSE      TRUE
## 0.6985646 0.3014354

## Number of nearest neighbors = 4
## agreement4
##     FALSE      TRUE
## 0.7751196 0.2248804
```

Figure 12 shows the accuracy in percent achieved with the kNN algorithm considering the number of nearest neighbors. The procedure is repeated with fault resistance values randomly generated and values below 5 ohm. Figure 13 compares the accuracy achieved with different fault resistance values. It is evident that the best accuracy in both scenarios is achieved when $k = 1$.

### 3.2.4 Discussion

The approach presented here is just a theoretical work without any practical application since it has some significant limitations: it is based on steady state values during the fault condition, and the range of values used for training the model is rather limited. The main goal of the case study is to illustrate how the combination of the **R** environment and **TPBIG** can be used for studies related to fault diagnosis of power system studies. For a deeper analysis of this type of study see [33].

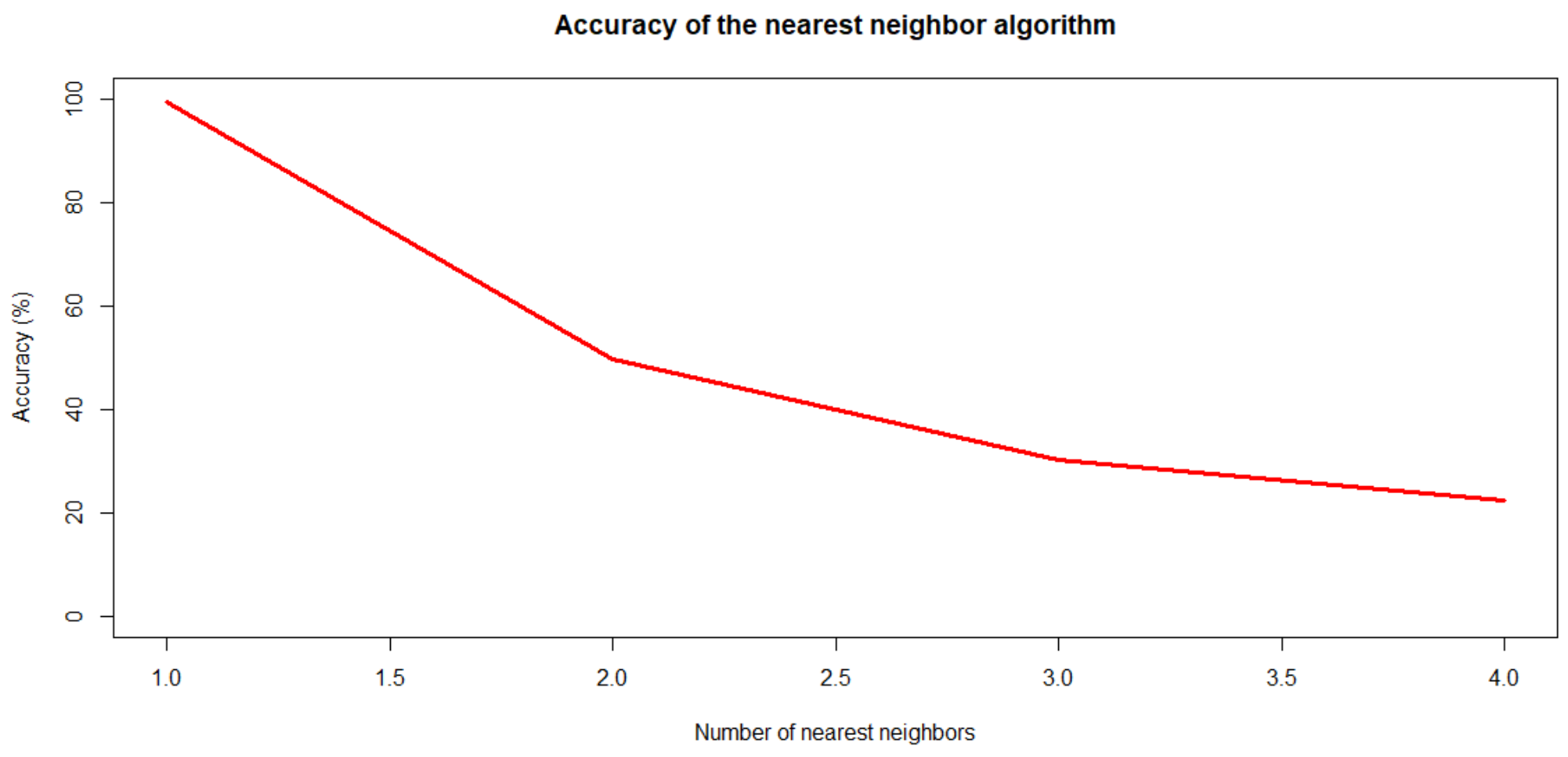

***Figure 12***. *Accuracy of the nearest neighbor algorithm - Fault resistances < 1 ohm.*

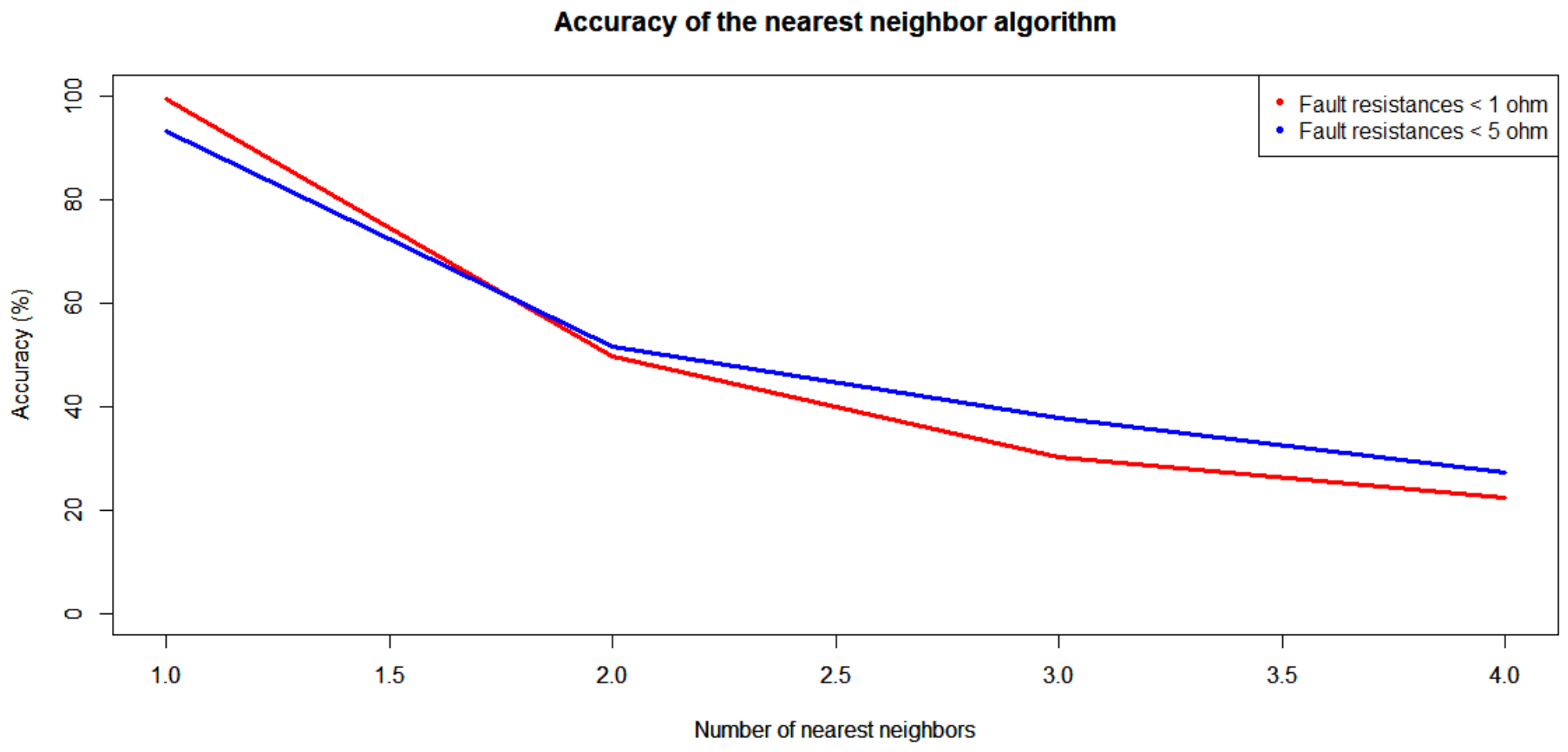

***Figure 13***. *Accuracy of the nearest neighbor algorithm - Fault resistances < 5 ohm.*

## 3.3 Case study 2 - Lightning performance of an overhead line

### 3.3.1 Introduction

The goal of this case study is to implement a procedure for estimating the lightning performance of an overhead transmission line using rather simple models for representing both the line and the lightning stroke, and then use the generated information to train a support vector machine (SVM) algorithm that could predict the response of the test line to a given lightning stroke discharge [34], [35]. The lightning performance of an overhead line depends on its design (e.g., shielding, grounding) and the atmospheric activity in the region where the line is located. An unacceptable flashover rate can be due either to a very high backflashover rate (BFOR), a very high shielding failure flashover rate (SFFOR), or both. A reason for a very high BFOR is a poor grounding, while the main reason for a very high SSFOR is a poor shielding. An additional reason that can affect both rates is the strike distance of insulator strings.

This case study is based on the overhead transmission line analyzed in Section 4 of the EMTP Primer edited by EPRI (EL-4202 Research Project 2149-1, September 1985) [36]. Figure 14 shows the geometry of the 230 kV test line; it is a single-circuit line with one conductor per phase and shielded by two ground wires.

The next subsections present the guidelines used here to implement the test line model, a short description of the incident model needed to estimate the lightning performance of the test line, and the **ATP** implementation of the line model. A statistical study, based on the Monte Carlo, method, aimed at estimating the lightning performance of the test line will follow. The results derived from the application of the statistical study will be used to train a model that could predict the lightning performance of the test line by means of the SVM algorithm.

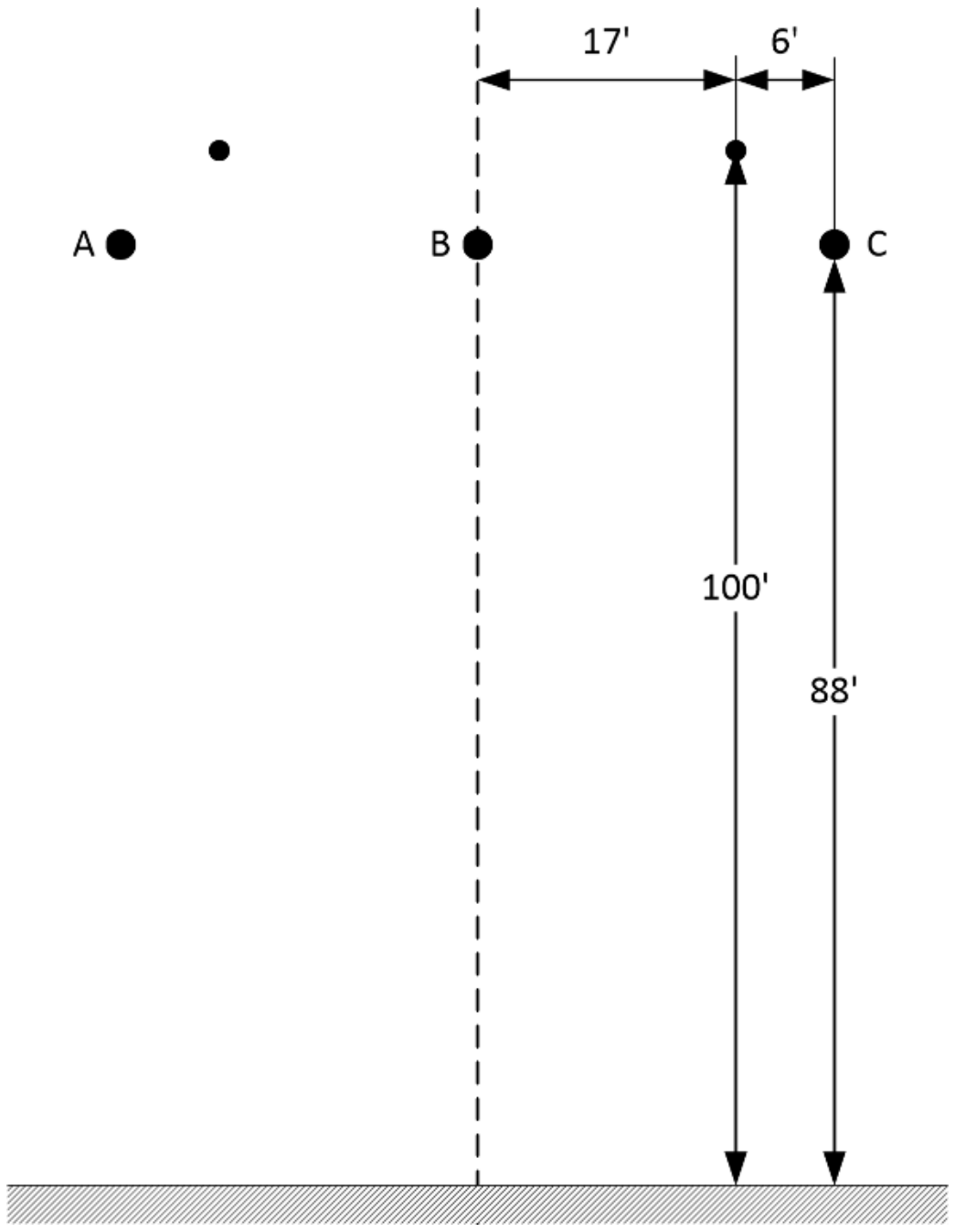

***Figure 14**. Test line configuration.*

### 3.3.2 Modelling guidelines

The following paragraphs detail the models used to represent the different parts of the transmission test line analyzed in this case study [37].

1) The line is modelled by means of several spans at each side of the point of impact. Each span is represented as a multi-phase untransposed constant and distributed-parameter line section. Figure 15 shows the right side of the model implemented in **ATP**: the model includes four spans at each side of the point of impact. The depicted model assumes the lightning discharge reaches a ground wire at the tower top.
2) The line termination at each side of the above model is represented by means of a long enough section, whose parameters are calculated as for the line spans.
3) A tower is represented as an ideal single-conductor distributed-parameter line with surge impedance is calculated according to CIGRE recommendations [38].
4) Insulator strings are represented as voltage-controlled switches whose flashover voltage value follows a normal distribution characterized by a mean and a standard deviation value.
5) Tower footing impedances are represented as resistances whose values are determined assuming they are uniformly distributed.
6) A lightning stroke is represented as an ideal current source with a double ramp waveform defined by three parameters (see Figure 16): the peak current magnitude, $I_p$, the front time, $t_f$, and the tail time, $t_h$, that is the time interval between the start of the wave and the 50% of peak current on tail.

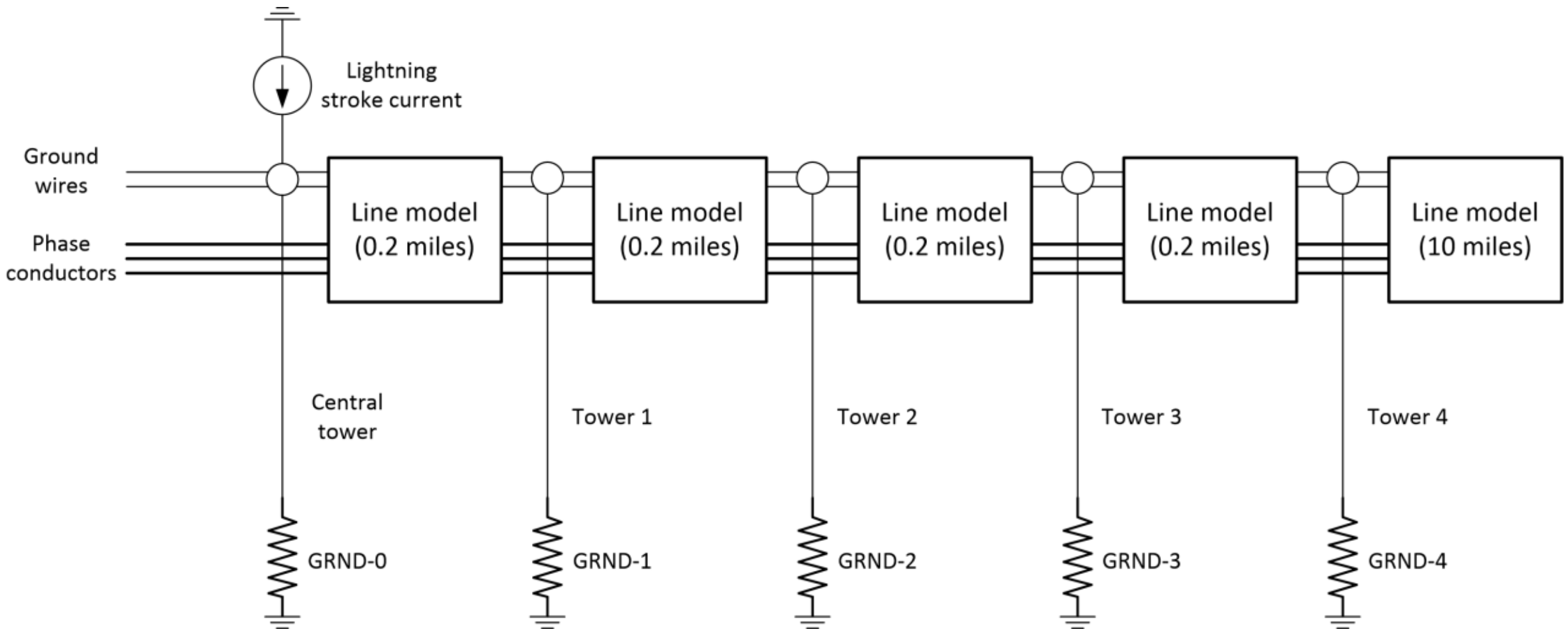

***Figure 15****. Test line model for lightning overvoltage calculations.*

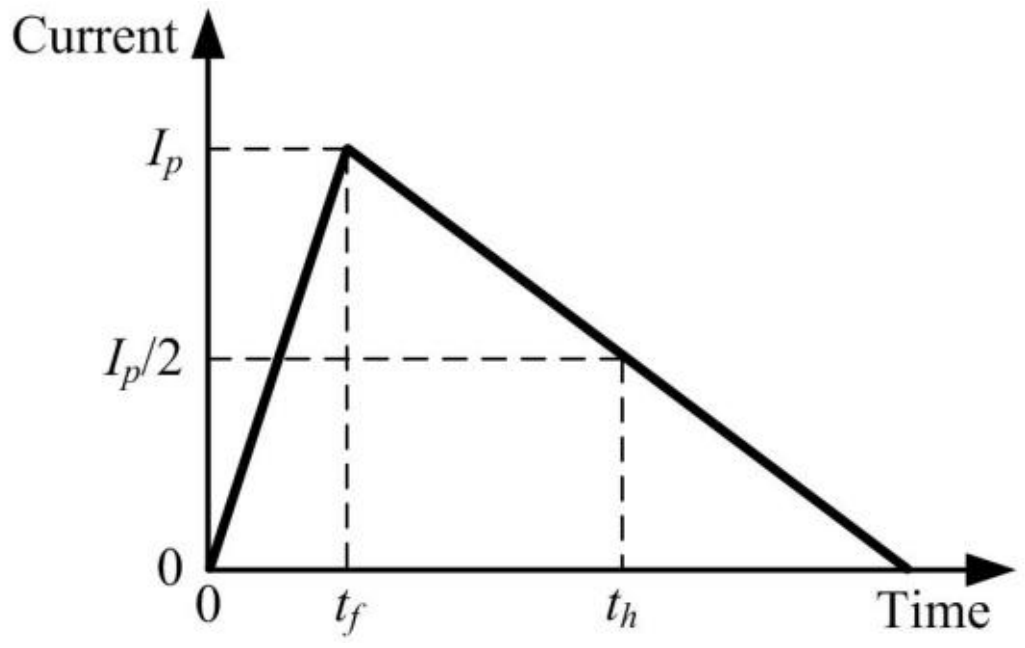

***Figure 16****. Double ramp waveform of the lightning stroke.*

### 3.3.3 The incidence model

The incidence model used in this example is the electrogeometric model (EGM) proposed by Brown and Whitehead [39]. According to this model, the striking distances are calculated as follows:

$$r_c = 7.1 I_p^{0.75} \qquad r_g = 6.4 I_p^{0.75}$$

where $r_c$, is the striking distance to both phase conductors and shield wires, $r_g$ is the striking distance to earth, and $I_p$ is the current peak of the return stroke current.

Figure 17 shows how to apply the EGM to a shielded line. First, calculate the values of $r_c$ and $r_g$ for a given current peak value and determine the location of points A and B. Any stroke with a vertical channel within A and B will impact the phase conductor, any stroke with a vertical channel at the left of point B will impact the shield wire, any stroke with a vertical channel at the right of point A will go to ground. Figure 18 shows how determine whether a discharge will go a ground wire or a phase conductor.

After implementing the code to obtain the point of impact, the following results are derived:

```
## Critical current to shield wires     = 17.62 (A)
## Critical current that shields a span = 64.15 (A)
```

That is, every discharge with a peak value above 17.62 (kA) will go to a ground wire (although it can also go to ground if the distance of the vertical channel with respect to the line is long enough). For every discharge with a peak value above 64.15 (kA) a span is fully shielded.

**R** capabilities are used to illustrate the application of the electrogeometric model.

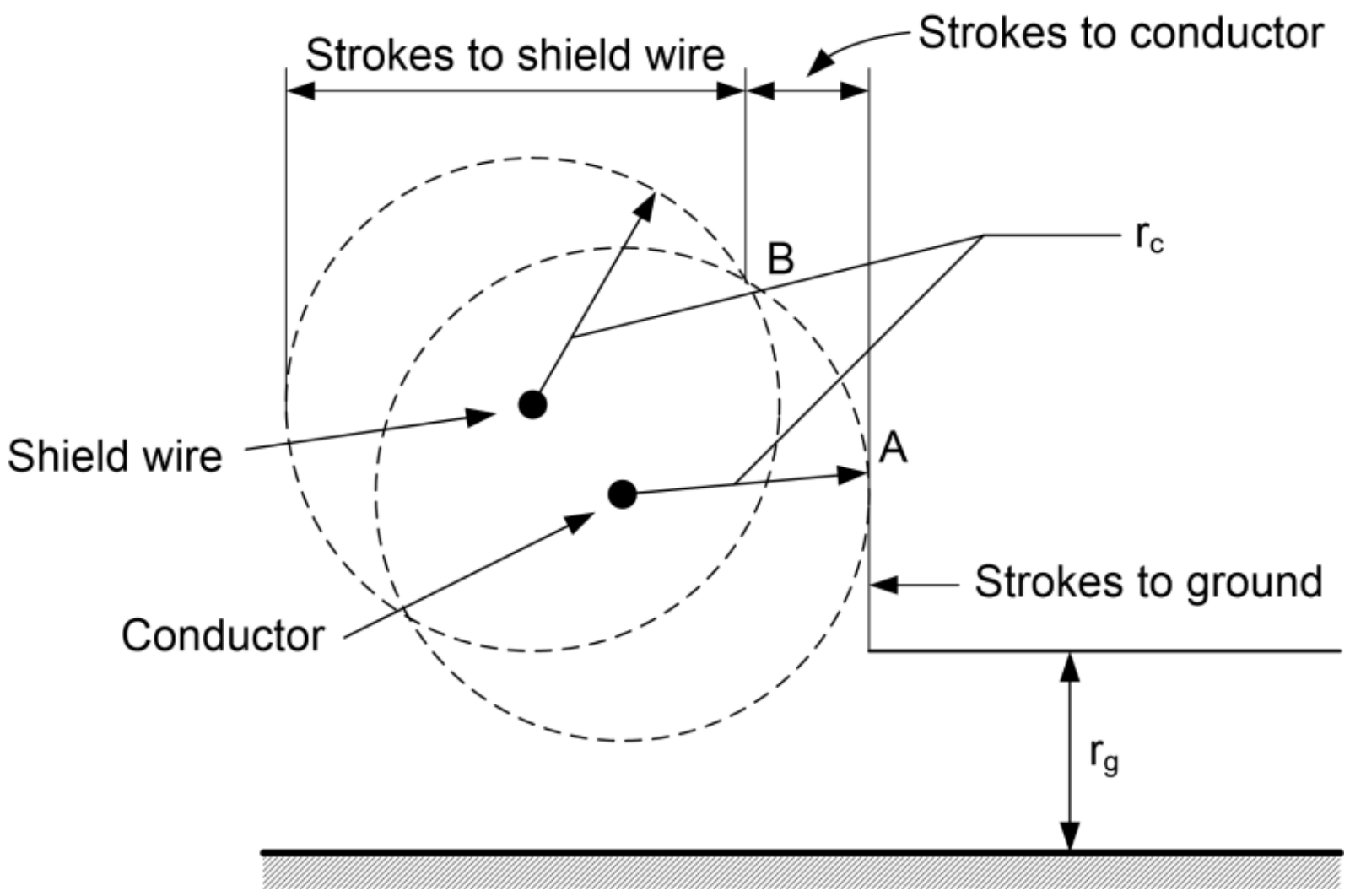

***Figure 17**. Aplication of the electrogeometric model.*

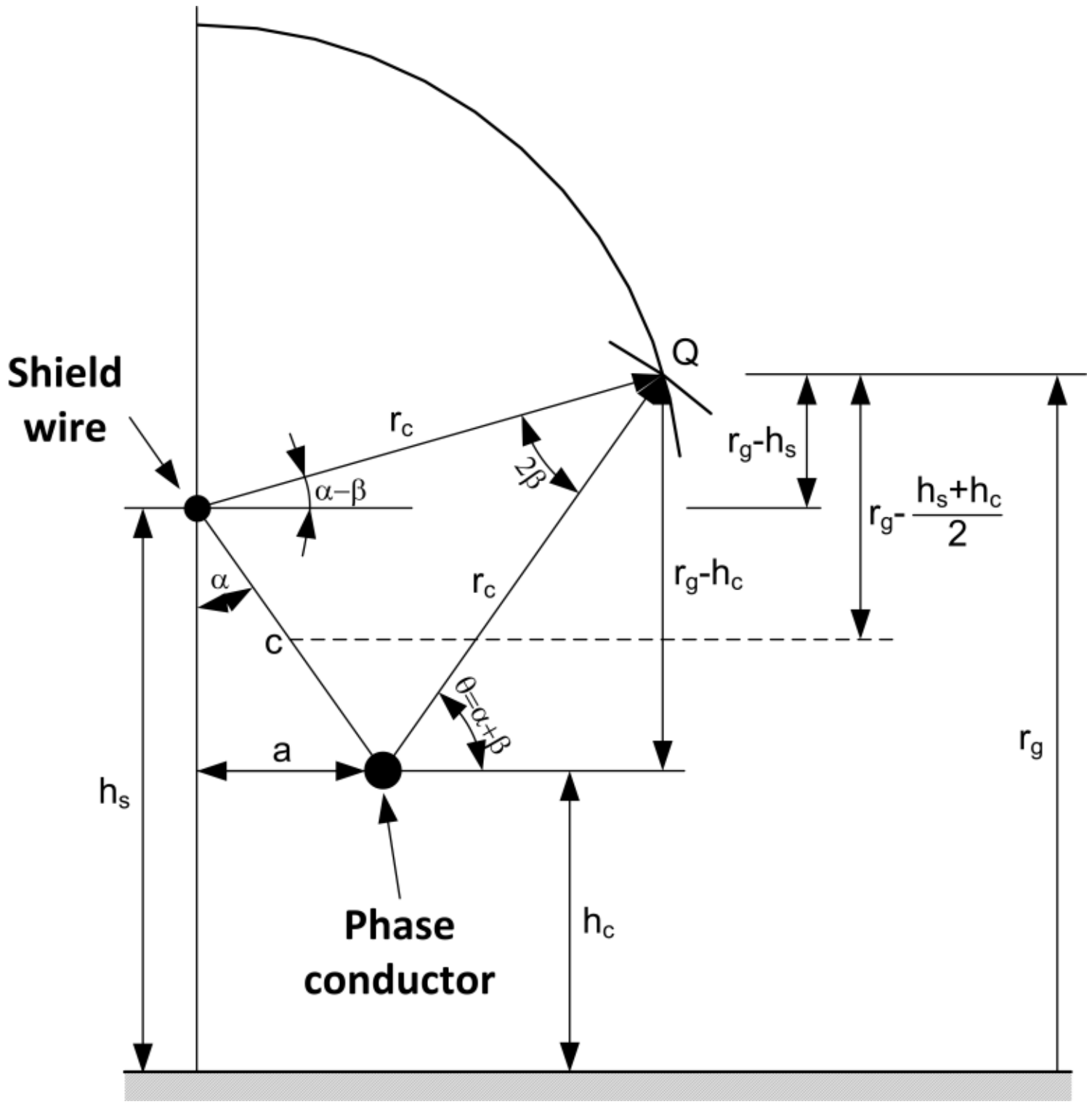

***Figure 18**. Calculations related to the application of the electrogeometric model.*

Figure 19 shows the application of the EGM to check whether the lightning stroke will go to a ground wire or to an outer phase conductor. The figure shows the circles corresponding to three striking distances (i.e., the peak values). According to the figure, for peak values above 17.62 kA, phases A and C are shielded, and the lightning discharge will go to a ground wire. It can be assumed that the middle phase (i.e., phase B) is fully shielded by the two ground wires.

Figure 20 shows the application of the EGM to check whether the lightning discharge will impact a tower or somewhere in a span. The geometry used in the figure corresponds to a ground wire.

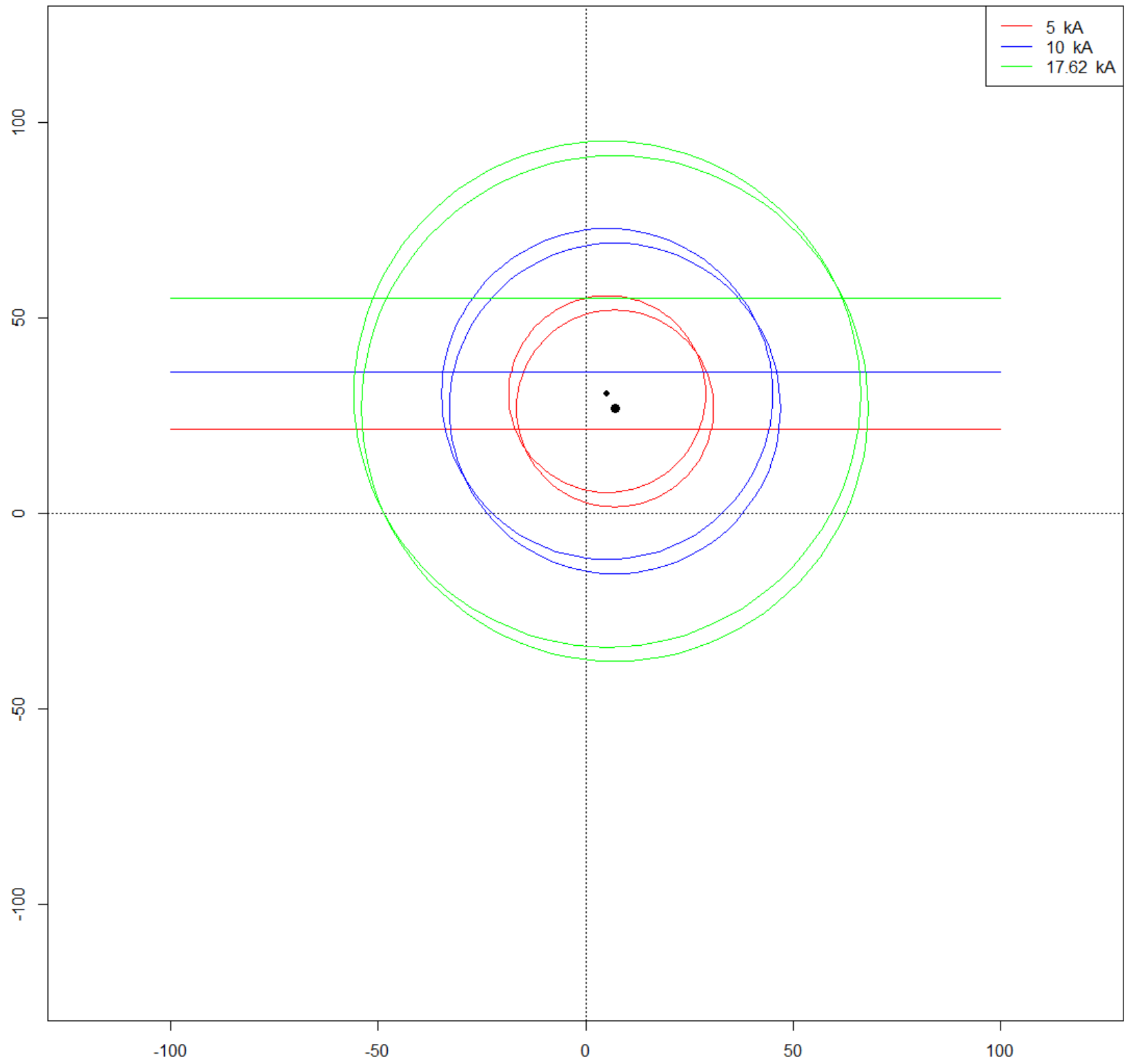

***Figure 19**. Application of the electrogeometric model - Impact on a tower (right side).*

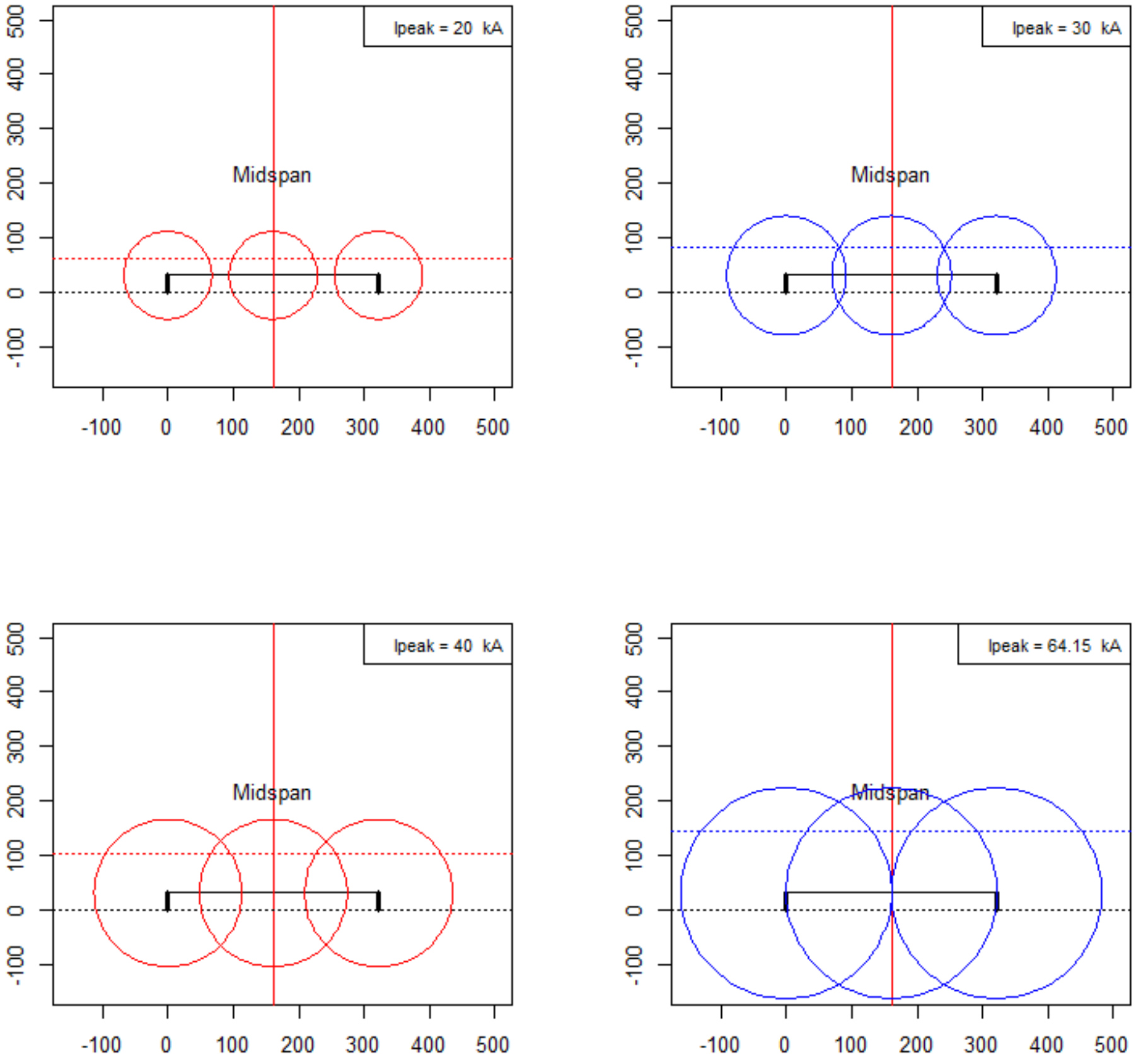

***Figure 20**. Application of the electrogeometric model - Impact on the span.*

As expected, the figure shows that as the current peak value increases, the length of the span to which the stroke can impact decreases. It is worth mentioning that, for a rigorous application of the EGM, a 3D geometry should be used, and calculations should be carried out using spheres instead of circles. In other words, the rigorous application of the EGM to a geometry like that of the test line is very difficult, due, among other aspects, to the catenary shape of all line wires (ground wires and phase conductors). A simplified approach is used in this case study to estimate the final step of the lightning discharge; that is, to distinguish whether the stroke will go to a tower or to a span, to a ground wire or to a phase conductor. In addition, it will be assumed that if the discharge goes to a span it will always impact the midspan point, either at a ground wire or a phase conductor.
Another aspect to consider in this study is that the conductor of the middle phase (phase B) is fully shielded by the ground wires. According to the EGM model used here, only discharges with a very small current peak value will reach that conductor.

### 3.3.4 Lightning performance study

To estimate the lightning flashover rate of an overhead line, a procedure based on the Monte Carlo method with the following features will be applied [40]:

a) Random values are calculated. The list includes the parameters of the lightning stroke (peak current, rise time, tail time, and location of the vertical leader channel), phase conductor voltages, the grounding resistance and the insulator strength.
b) The last step of a return stroke is determined by means of the electrogeometric model. The model used in this study is that proposed by Brown and Whitehead [39]. See above.
c) Once the point of impact has been determined, the overvoltage calculation is performed. Only those strokes to the line are simulated considering two options: stroke to ground wire, stroke to phase conductor.
d) The flashover rate is determined once all strokes to the line have been simulated considering the statistics of lightning flashes to ground in the area where the line is located.

The following probability distributions are used:

- Lightning flashes are assumed to be of negative polarity, with a single stroke and parameters independently distributed, being their statistical behavior approximated by a log-normal distribution, whose probability density function is:

$$p(x) = \frac{1}{\sqrt{2\pi} x \sigma_{\ln x}} \exp\left[-\frac{1}{2}\left(\frac{\ln x - \ln x_m}{\sigma \ln x}\right)^2\right]$$

  where $\sigma_{\ln x}$ is the standard deviation of $\ln x$, and $x_m$ is the median value of $x$.
- The power-frequency reference angle of phase conductors is uniformly distributed between 0 and 360 degrees.
- The insulator strength values follow a normal distribution for which a mean and a standard deviation have to be specified. In this study, the critical flashover voltage (CFO), used as mean value, is 977.5 kV, the standard deviation being a 5% of the CFO value.
- The tower footing resistances have a constant value, being the resistance value the same for all line towers. The resistance value for each run is randomly determined by using a uniform distribution within a selected range of values. In this case study, the resistance values are equal or above 10 ohms, and equal or below 100 ohms.
- Before the application of the electrogeometric model, the stroke location is estimated by assuming a vertical path and a uniform ground distribution of the leader. Only flashovers across insulator strings are assumed.

### 3.3.5 Generation of random values

#### *3.3.5.1 Introduction*

To apply the Monte Carlo method the values of variables and parameters of random nature have to be generated. In this study these values are:

- the parameters that characterize the lightning stroke current ($I_p, t_f, t_h$, see Figure 16);
- the coordinates of the lightning discharges, for which a vertical channel is assumed;
- the reference angle for voltages of phase conductors;
- the tower footing resistances;
- the strength of insulator strings.

Note that the geometry of the test line is needed to obtain the coordinates of the stroke vertical channel and to apply the electrogeometric model. In this case, the parameters to be specified include the number of line spans and the length of each span; two aspects that need to be considered for building the test line model.

The first value to be specified when applying the Monte Carlo method is the number of random values to be generated for variable/parameter. In this study this number is $n = 50000$. Remember that to apply **TPBIG**, the information to be specified for each lightning stoke current in the data input file is the point of impact, the three parameters of the stroke current, the reference phase angle and the tower footing resistance values.

The rest of this section is dedicated to obtaining the coordinates of the lightning stroke vertical channel, the parameters of the stroke current, and the phase conductor voltage angles. The point of impact will be later estimated by applying the EGM once the coordinates of the vertical channel and the peak current value of the lightning stroke are known. The random values of tower footing resistances will be obtained in the subsequent section.

#### *3.3.5.2 Estimation of the lightning discharge vertical channels*

The coordinates of the vertical channel of the lightning strokes are obtained by assuming a uniform distribution along the section length represented in the test line model (here 4 spans; that is 0.8 miles or 1287.4752 meters) and within a perpendicular distance of 500 meters to both sides of the line. Figure 21 shows the point of impact of the first 1000 vertical channels. It is evident that these points are more or less uniformly distributed in the area of concern.

#### *3.3.5.3 Parameters of lightning strokes*

Only negative polarity and single-stroke flashes are assumed for this study. The stroke current of lightning discharges exhibits a double-ramp waveform characterized by three parameters (see Figure 16): the current peak, $I_p$ (in kA), the from time, $t_f$ (in $\mu$segundos), and the tail time, $t_h$ (in $\mu$seconds). It is assumed that the three parameters follow an uncorrelated log-normal distribution.

The values of means and variances assumed for the three distributions are as follows:

- Current peak: $x = 34.0$, $\sigma_{\ln x} = 0.740$ kA
- Front time: $x = 2.0$, $\sigma_{\ln x} = 0.494$ $\mu$s
- Tail time: $x = 77.5$, $\sigma_{\ln x} = 0.577$ $\mu$s.

Figures 22 to 24 show the results obtained for the three parameters that characterize the lightning stroke current.

#### *3.3.5.4 Phase angle of line voltages*

The power-frequency reference angle of phase conductors is uniformly distributed between 0 and 360º. Figure 25 shows the generated values for the test line studied here. Although the distribution depicted in the figure does not exhibit a perfect uniform distribution, it can be accepted for the present study. Obviously, the more random values the better the practical distribution.

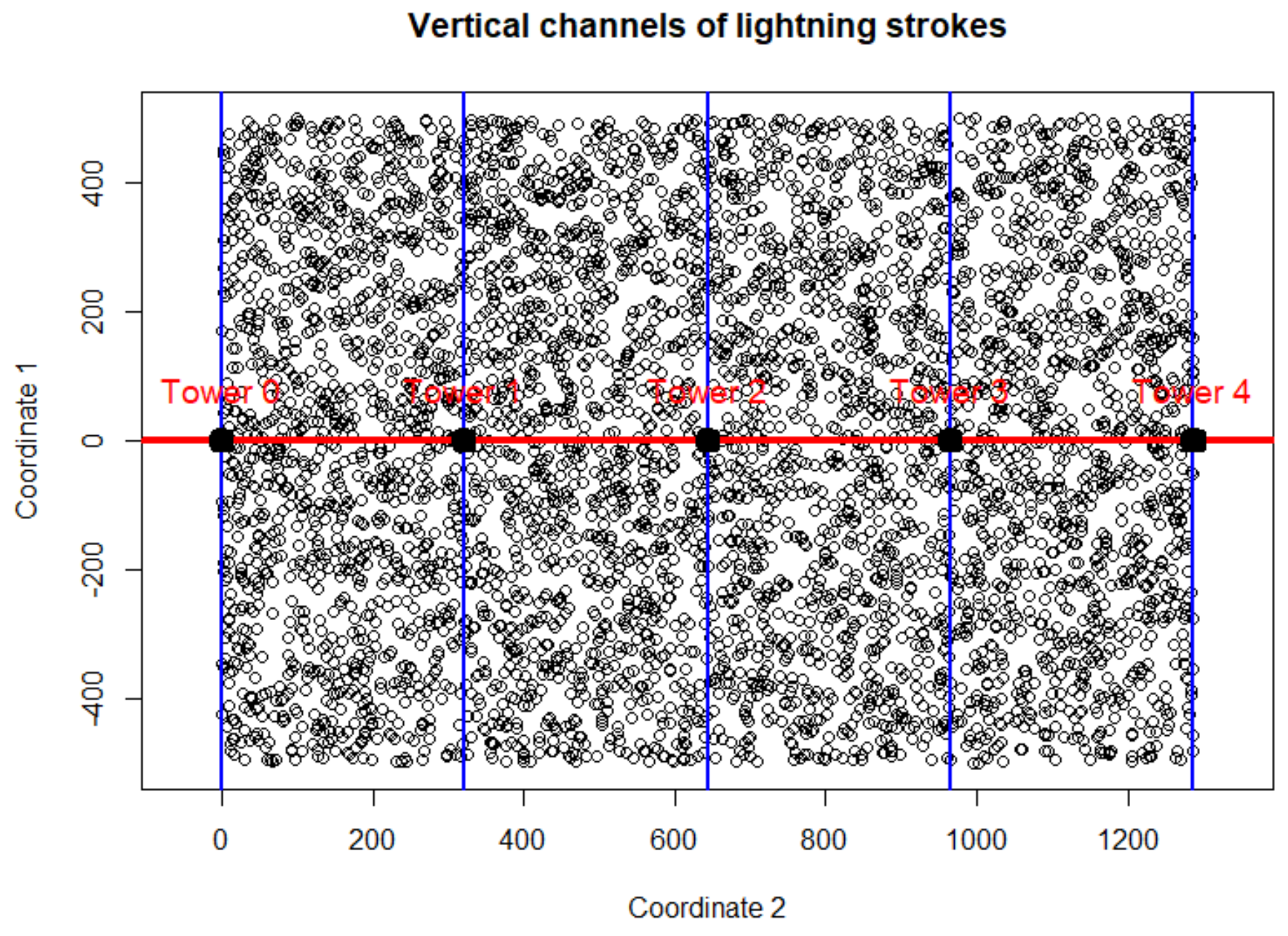

*Figure 21*. *Points of impact of the lightning stroke vertical channels.*

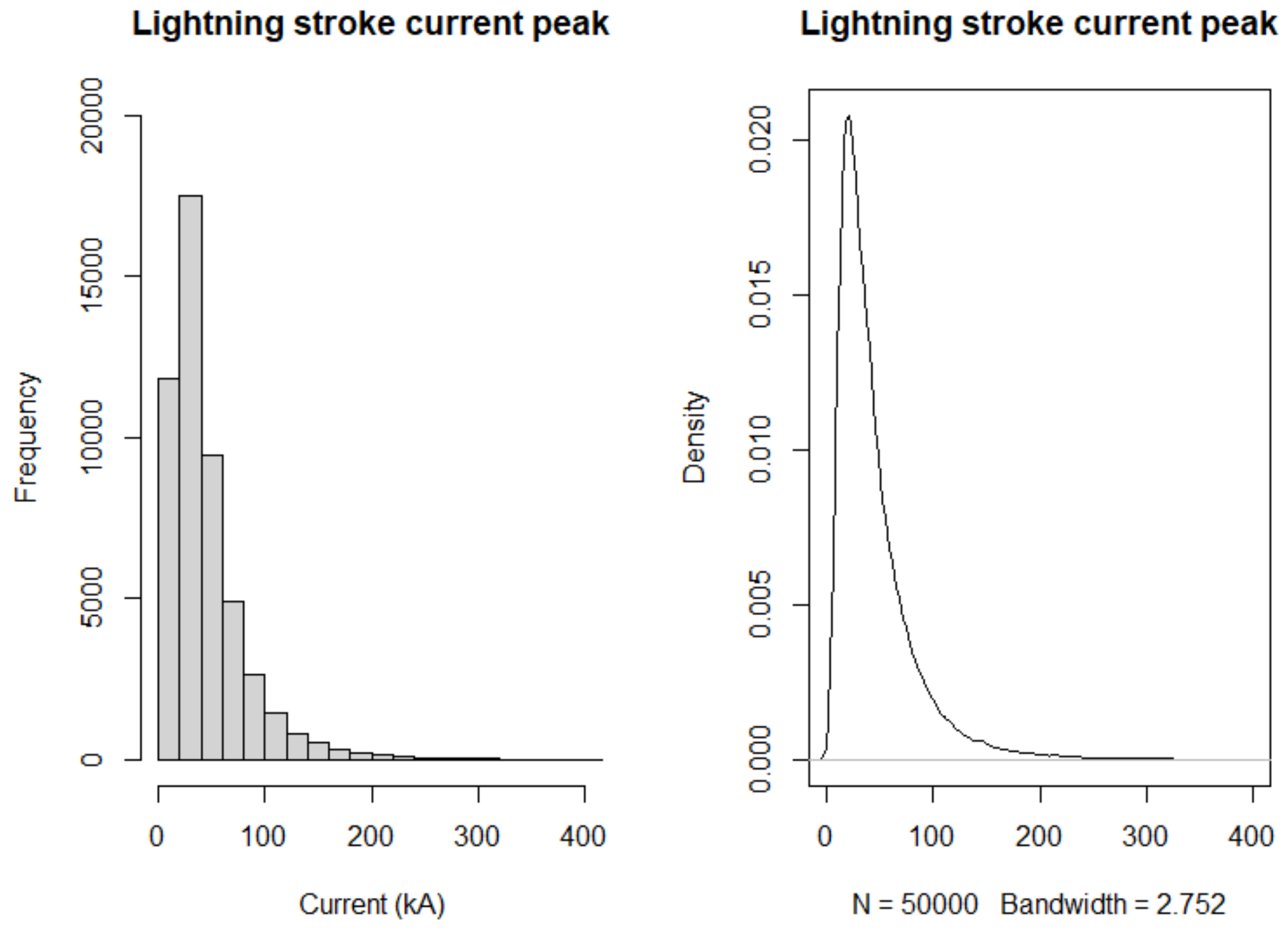

*Figure 22*. *Distribution of the stroke peak current.*

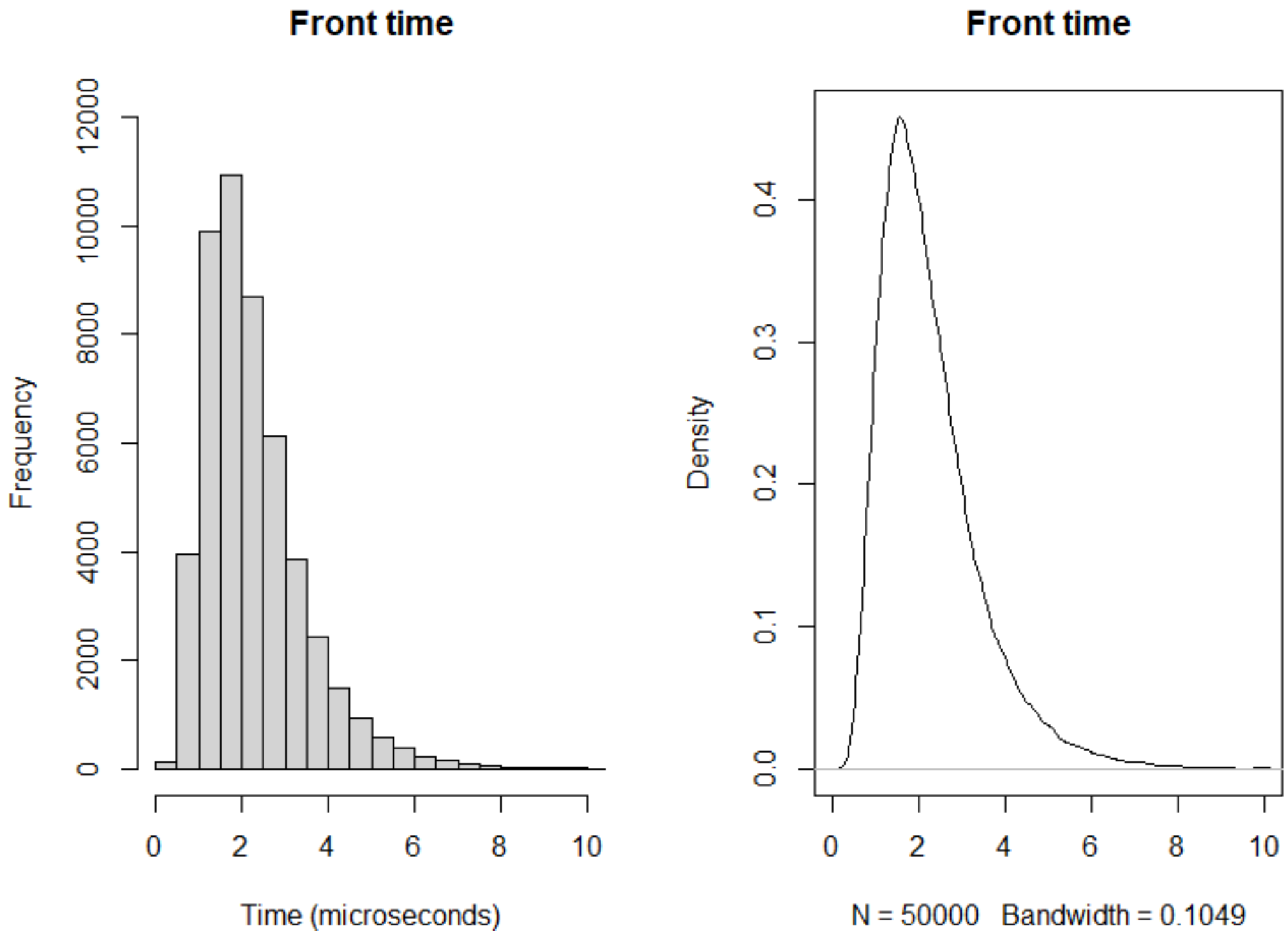

***Figure 23***. *Distribution of the stroke front time.*

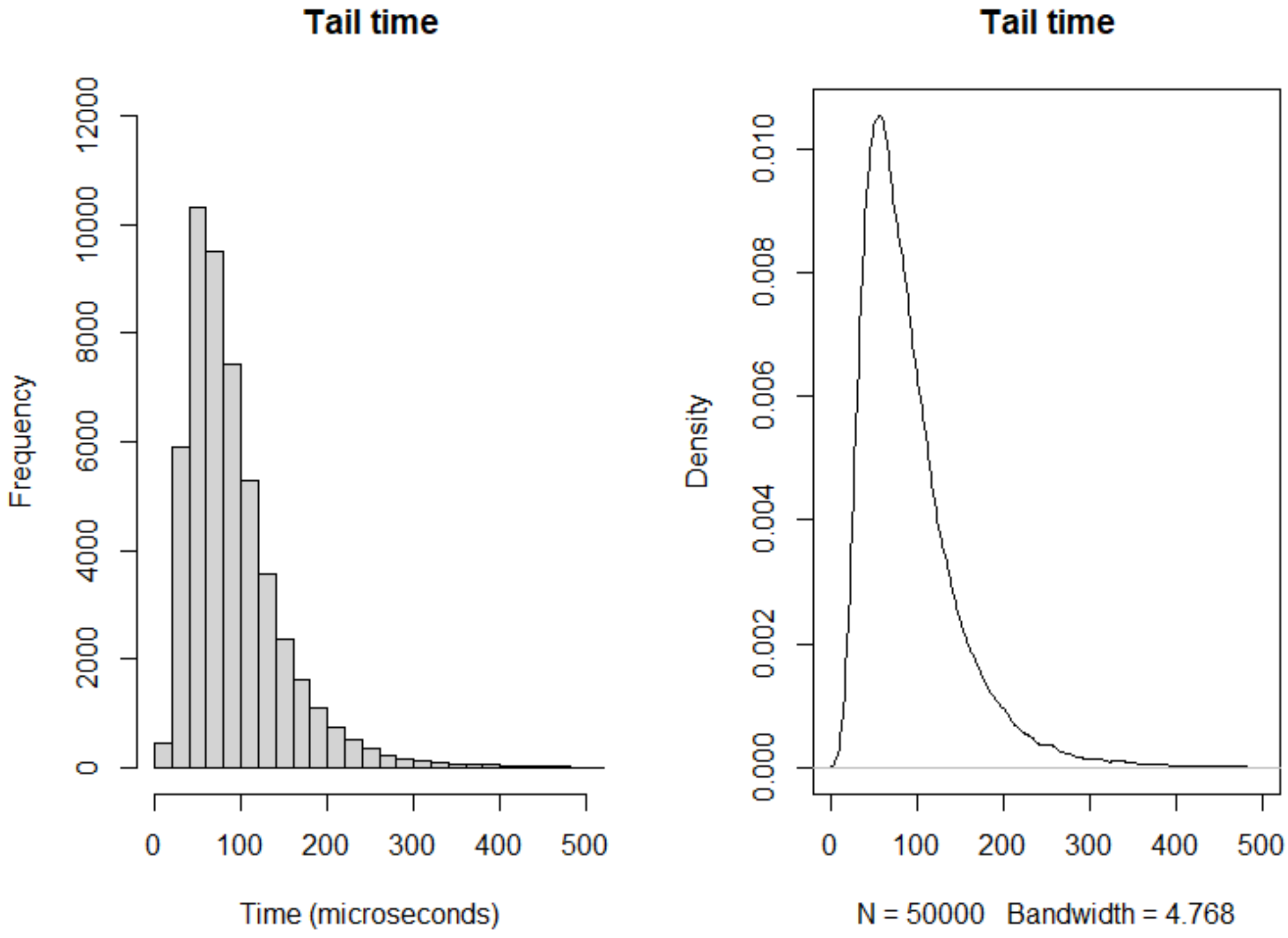

***Figure 24***. *Distribution of the stroke tail time.*

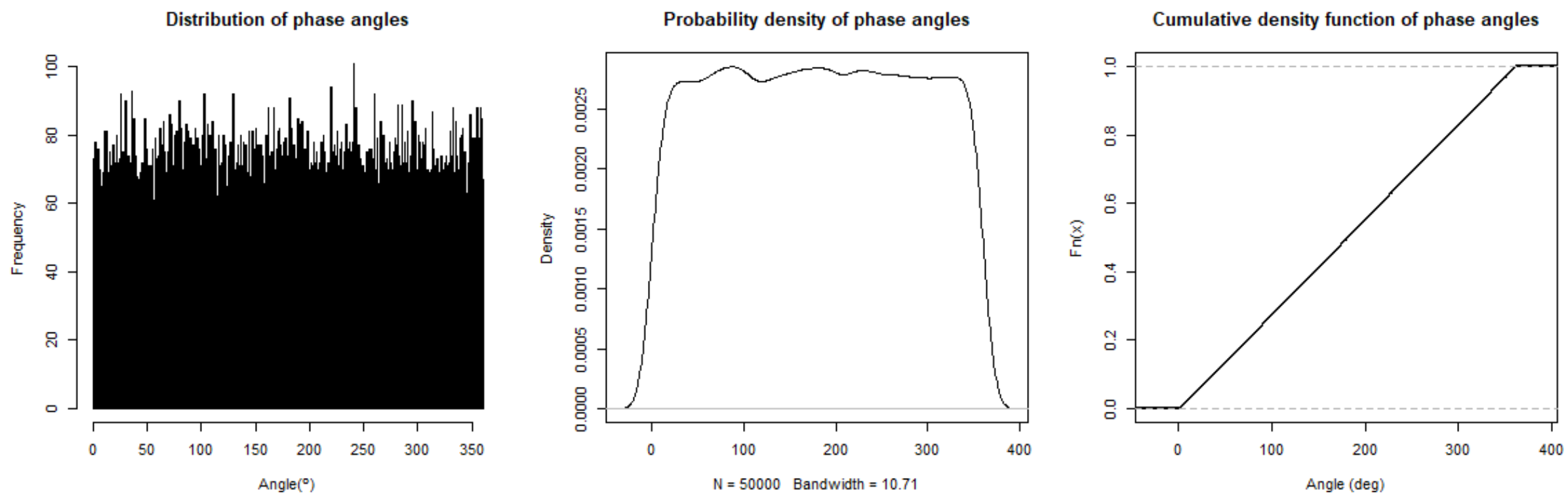

***Figure 25***. *Distribution of the phase angles.*

#### *3.3.5.5 Generation of a data frame*

The values generated above are grouped in a data frame. A sample of these values is presented below.

```
##    Coord1(m) Coord2(m) Angle(º) Peak(kA) Tail(microS) Front(microS)
## 1    232.950   968.570  310.837   48.110      120.683         1.865
## 2   -428.643    81.235   29.167   17.539       97.966         1.541
## 3    153.383   605.509   93.244   46.299       62.575         1.754
## 4   -242.656   162.841  334.192   32.851      116.471         2.067
## 5     45.874    15.694  198.319   19.248       44.164         2.573
## 6   -361.910   140.251  140.930   72.014       58.298         1.896
## 7   -217.309   634.362  281.897   23.912       80.901         1.162
## 8    147.232   816.162   10.703   11.505      105.879         0.813
```

### **3.3.6 Application of the electrogeometric model**

The next step is to estimate the point of impact of the lightning discharge. The following two options are considered: impact to ground or impact to the line. If the discharge impacts the line a distinction is to be made between the point of impact (tower, span) and the conductor (ground wire, phase conductor).

A procedure for applying the electrogeometric model can be summarized in two steps:

1. Generate random numbers to obtain the lightning current peak values and the coordinates of the vertical channels
2. Apply the electrogeometric model to distinguish between strokes to ground, to phase conductors and to shield wires.

To simplify the implementation of the electrogeometric model, the following rules will be applied:

1. Once it is known that a given discharge impacts the line, it will be assumed that the point of impact is a tower or a ground wire if the peak current value is above the critical value (see below).
2. To estimate whether the discharge goes to ground or to the line, only the geometry corresponding to towers is considered (i.e., it is assumed that the height above ground remains constant for all wires); see Figure 14.
3. To determine whether a discharge will go to a span or a tower, the following rules are applied:
    - if the peak value is above 64 kA, the discharge will go to a tower if the distance of the vertical channel with respect to the nearest tower is less than 1/4 of the span length;
    - if the peak value is above or equal to 25 kA and below or equal to 64 kA, the discharge will go to a tower if the distance of the vertical channel with respect to the nearest tower is less than 1/8 of the span length;

- if the peak value is below 25 kA, the discharge will go to a tower if the distance of the vertical channel with respect to the nearest tower is less than 1/16 of the span length.

4. If the discharge impacts somewhere in the span, it will be assumed by default that the point of impact is at midspan. This will be used for both ground wires and phase conductors.
5. It is assumed that the conductor of phase B (middle phase) is fully shielded by the two ground wires, so no discharge will reach this conductor.

First, data must be prepared for the application of the electrogeometric model. A summary of the data frame to be used for the test line is presented below after the application of the electrogeometric model.

```
##   Coord1(m) Coord2(m) Angle(º) Peak(kA) Tail(microS) Front(microS)        Wire
## 1   232.950   968.570  310.837   48.110      120.683         1.865      Ground
## 2  -428.643    81.235   29.167   17.539       97.966         1.541      Ground
## 3   153.383   605.509   93.244   46.299       62.575         1.754      Ground
## 4  -242.656   162.841  334.192   32.851      116.471         2.067      Ground
## 5    45.874    15.694  198.319   19.248       44.164         2.573 Shield wire
## 6  -361.910   140.251  140.930   72.014       58.298         1.896      Ground
## 7  -217.309   634.362  281.897   23.912       80.901         1.162      Ground
## 8   147.232   816.162   10.703   11.505      105.879         0.813      Ground
```

The characteristics of strokes to ground and to the line are analyzed. There is a distinction between strokes to towers and strokes to spans, and also between strokes to shield wires and strokes to phase conductors (shielding failures).

```
## Number of strokes to ground                 = 44082
## Number of strokes to the line               = 5918
## Number of strokes to towers                 = 1663
## Number of strokes to spans                  = 4255
## Number of strokes to shield wires           = 5869
## Number of strokes to shield wires at towers = 1658
## Number of strokes to shield wires at spans  = 4211
## Number of strokes to conductors             = 49
## Number of strokes to conductors at towers   = 5
## Number of strokes to conductors at spans    = 44
```

Figure 26 shows the location of the vertical channel of those strokes that impact the line, either at a ground wire or a phase conductor, at a tower or a span. It is evident that only strokes with a vertical channel within a certain distance to the line will impact either a tower, a shield wire or a phase conductor. Figure 27 shows the distribution of current peak values of strokes that impact the line.

### 3.3.7 Lightning performance of the test line

To obtain the lightning flashover rate of the test line only the discharges that impact the line need to be simulated (by default, it is assumed that discharges to ground do not cause flashover in transmission lines). Figure 28 depicts a flow diagram of the procedure implemented here to estimate the lightning performance of the test line.

Note that the first step is to obtain the values of the tower footing resistances, which are randomly generated assuming a uniform distribution between two specified resistance values. Remember that the line model used here assumes a fixed value for the footing resistances and the same value for all resistances. Next, the random values generated in the previous step are read to create two .PCH files that are inserted in the data input file. These files specify the parameters of the lightning stroke current waveform, the point of impact of the lightning discharge, the angle of the phase voltages, and also the tower footing resistance values. At the end of the procedure a CSV file with this information plus the voltages generated across the insulator strings is created. These voltages will be used to estimate the response of the test line; that is, will it flashover or not? At the end of the procedure a short summary is presented.

```
## Number of flashovers                                = 1103
## Number of flashovers caused by strokes to spans  = 546
## Number of flashovers caused by strokes to towers = 557
```

According to this result, about 1103 strokes will cause insulation (back)flashover. Note, however, that about one half of the flashovers were caused by strokes that impacted the span (not a tower). This is a conservative result since neither corona effect nor predischarge voltage phenomena were included in the simulations.

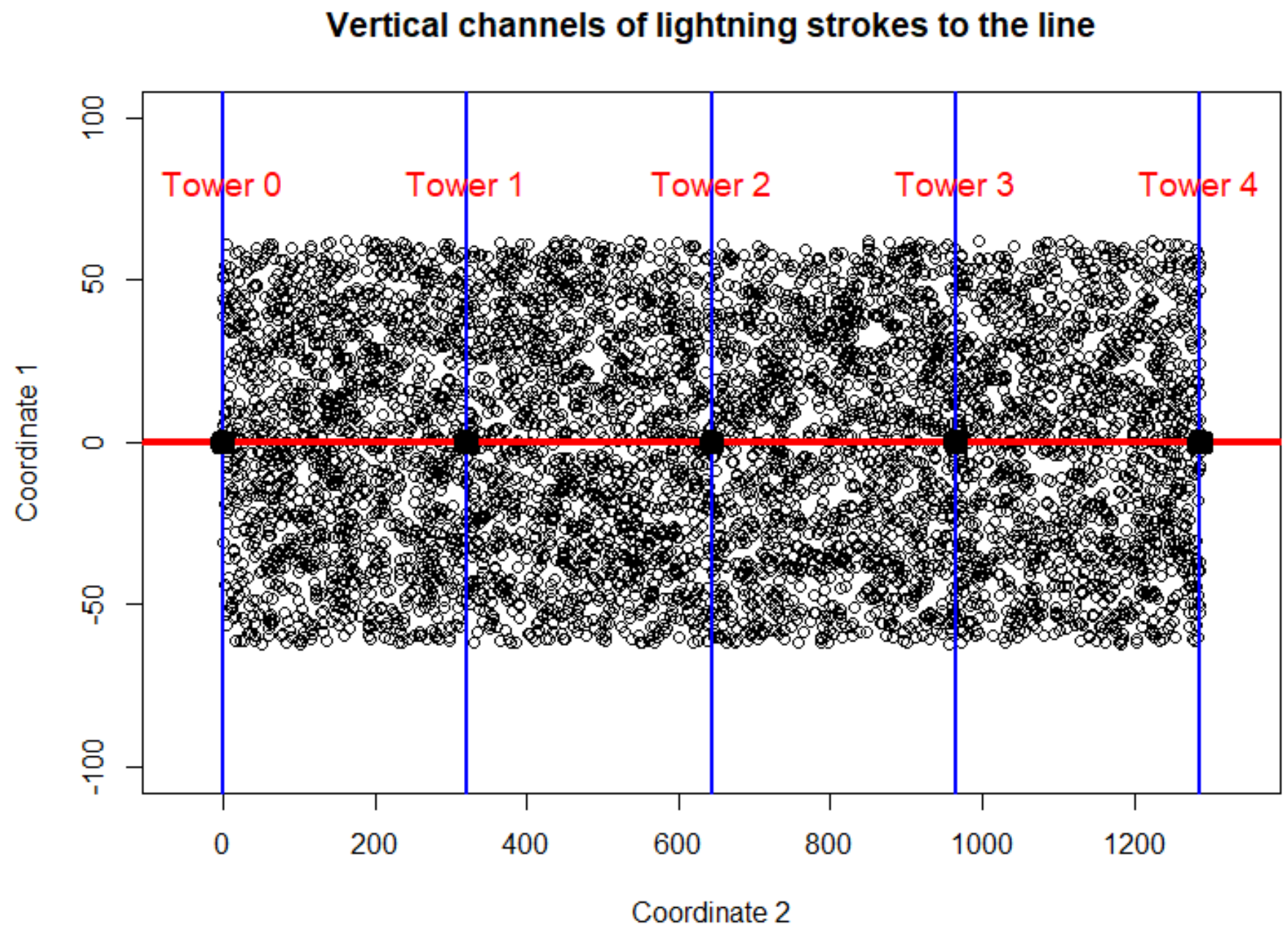

***Figure 26***. *Vertical channels of lightning strokes to the line.*

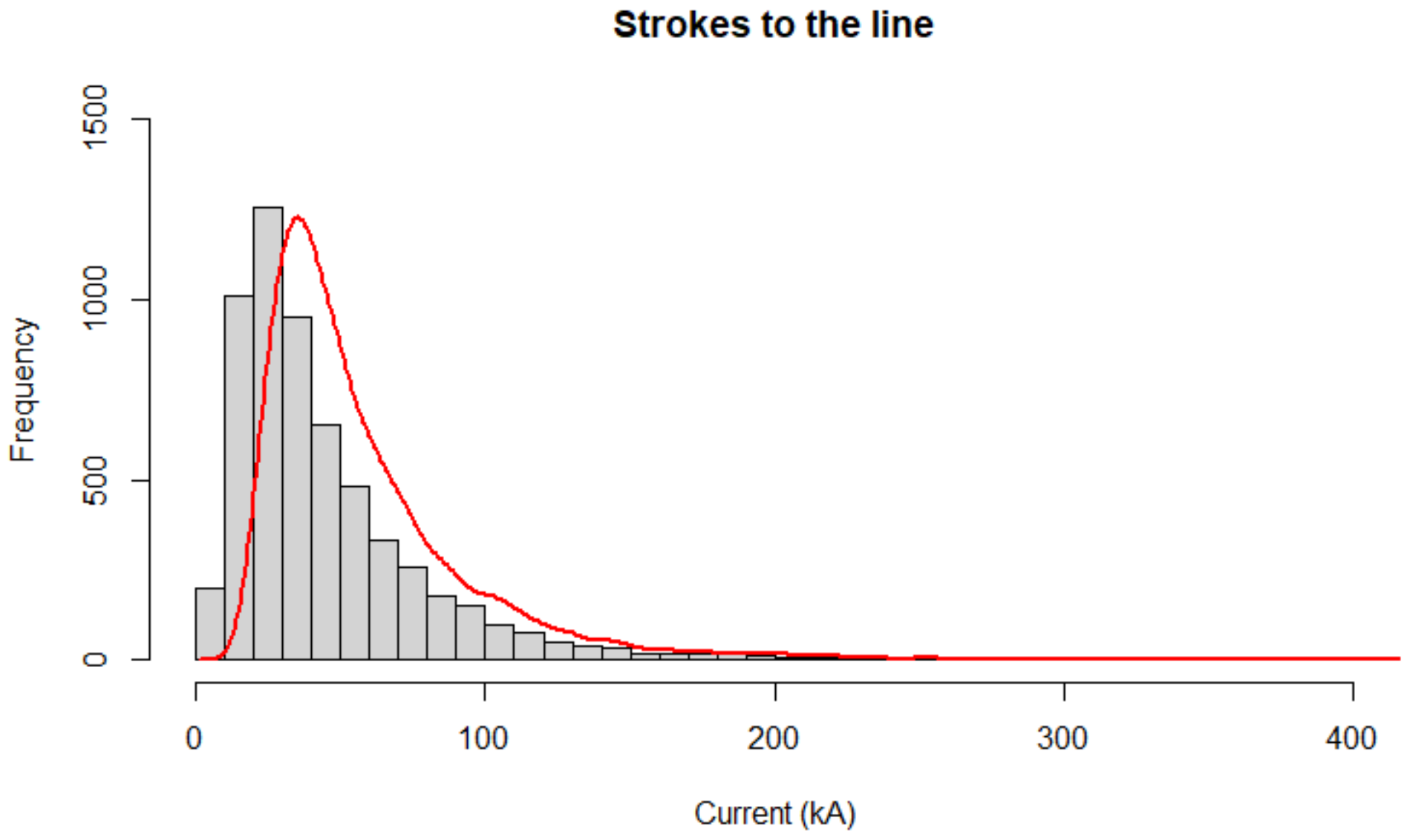

***Figure 27***. *Distribution of current peaks of strokes that impact to the line.*

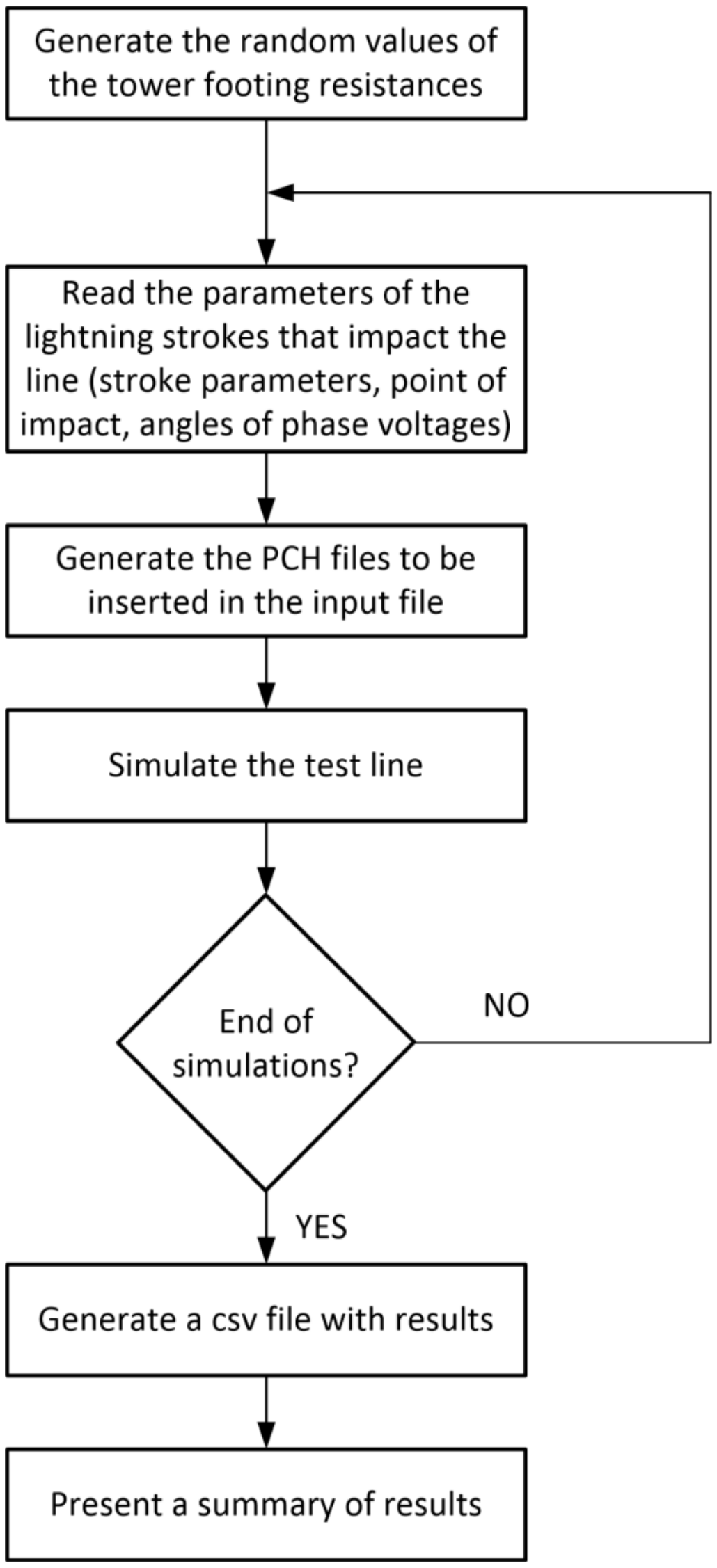

***Figure 28**. Procedure for estimating the lightning performance of the test line.*

### *3.3.7.1 Flashover rate calculation*

The flashover rate of the test line, measured in flashovers per 100 km and year, can be determined as follows:

1. Obtain the number of years during which the line has been simulated

$$N_y = \frac{n}{(l_1 \times l_2) * N_c}$$

   where

   - $n$ is the number of generated random values
   - $l_1 \times l_2$ is the area of impact of lightning strokes, in km$^2$
   - $N_c$ is density of strokes, in strokes per km$^2$ and year

2. Calculate the number of flashovers per year

$$n_f = \frac{N_f}{N_y}$$

where

- $N_f$ is the number of flashovers calculated above
- $N_y$ is the number of years during which the line has been simulated

3. Calculate the flashover rate

$$f_r = n_f \times \frac{100}{l_2}$$

Given that $n = 50000$, $l_1 = 1$ km, $l_2 = 1.287$ km and $N_c = 2.2$ strokes per km$^2$, the following results are derived

```
## Number of random generated strokes = 50000
## Length of the simulated test line  = 1.287 (km)
## Stroke density                     = 2.2 (strokes per km2 and year)
## Number of total flashovers         = 1103
## Number of simulated years          = 17653
## Flashover rate                     = 4.85 (flashovers per 100 km and year)
```

### 3.3.8 Application of the Support Vector Machine algorithm

This section shows the application of the SVM algorithm to predict the lightning performance of the test line; that is, to estimate the response of the test line (flashover or no flashover) to a given lightning stroke discharge. SVM is a supervised learning algorithm that can be used for classification and regression. It finds the optimal decision boundary that best separates data points of different classes by maximizing the margin between them [34], [35].

To predict the test line performance from the selected training data, the capabilities of the `kernlab` library are used. First, the data frames for training and testing are generated from the CSV file obtained in the previous steps: 50% of the file is randomly selected for training, the rest is selected for testing.

```
## Data set for training algorithms

##      PhaseAngle StrokePeak FrontTime  HalfPeak        Wire Tower
## 3621  336.83602   16.65463   2.66136 168.41801 Shield wire  Span
## 2987  155.65655   28.38136   3.00222  77.82828 Shield wire Tower
## 3739  147.35915   64.07407   4.52905  73.67957 Shield wire  Span
## 4778  283.72305   23.63762   2.43532 141.86153 Shield wire  Span
## 1680  143.95818   37.16293   1.70385  71.97909 Shield wire Tower
## 2538  254.24828   47.37703   2.02668 127.12414 Shield wire  Span
## 1466    1.23426  111.53214   2.59812   0.61713 Shield wire Tower
## 4916   79.84099   20.14373   1.33298  39.92050 Shield wire  Span
## 1098  276.26185   17.10230   5.54495 138.13092 Shield wire  Span
## 323    96.31580   48.65204   1.84837  48.15790 Shield wire  Span
## 2837  131.15704   18.21564   1.31883  65.57852 Shield wire Tower

## Data set for testing algorithms

##    PhaseAngle StrokePeak FrontTime   HalfPeak        Wire Tower
## 50  188.59065   56.91706   0.93517  94.295325 Shield wire Tower
## 52  188.99759   17.48473   1.12723  94.498795     Phase C  Span
## 53  296.31561   16.29315   3.03533 148.157805 Shield wire Tower
## 56   14.29547  213.06095   1.54062   7.147735 Shield wire  Span
## 57   70.01077   26.39586   6.03215  35.005385 Shield wire Tower
## 60   10.35720   31.18829   9.19364   5.178600 Shield wire  Span
## 62  212.94701   19.70671   3.66787 106.473505 Shield wire  Span
## 64   37.04510   49.94877   0.76174  18.522550 Shield wire  Span
```

```
## 65    3.73730   34.82385   2.34635   1.868650 Shield wire  Span
## 67  327.34314   44.71377   1.98780 163.671570 Shield wire Tower
## 68   43.21566   14.76824   0.75229  21.607830 Shield wire  Span
```

The SVM algorithm is now applied as classifier.

```
##  Setting default kernel parameters

## agreement
##       FALSE        TRUE
## 0.009577279 0.990422721
```

Observe that the prediction accuracy of the SVM algorithm is rather high, 99.04%.
A sample of actual and predicted performance of the test line (0 = no flashover, 1 = flashover) is shown below.

```
##      Actual Predicted
## 918       0         0
## 1027      1         1
## 1186      0         0
## 1192      0         0
## 1711      0         0
## 1901      0         0
## 2425      0         0
## 2564      0         0
## 2873      0         0
## 3027      0         0
```

### 3.3.9 Discussion

No part of the implemented test line model has been rigorously modeled: the line parameters are frequency-dependent, the corona effect was not included in the model, actual tower footing resistances exhibit a nonlinear behavior, insulator string flashover should be based on a more advanced representation (e.g., the leader progression model), most lightning flashes are multi-stroke and can exhibit positive polarity too, the double-ramp waveform is a very simplified approach for representing a lightning stroke current. The main goal of this case study was to illustrate how **R** capabilities can be used in combination with **TPBIG** capabilities for implementing a Monte Carlo study and applying a machine learning algorithm that could predict the lightning performance of an overhead transmission line.

## 3.4 Case study 3 - Distribution system analysis

### 3.4.1 Introduction

The goal of this case study is to show how **OpenDSS** can be used to simulate distribution systems with distributed generation and energy storage considering different approaches for representing loads; namely time-varying and random loads. A very simple test system configuration is simulated to illustrate the results that can be generated by **OpenDSS** and manipulated by **RStudio** to obtain high-quality visualization. Figure 29 shows the scheme of the test system to be simulated: it is a radial distribution-level system. Table 2 lists the cases to be simulated and analyzed here. Figure 30 shows a scheme of the procedure implemented in **RStudio** to visualize and analyze the simulation results generated by **OpenDSS**. Note that after running input data files, **OpenDSS** generates some CSV files that are read and manipulated by **RStudio** to obtain and present graphical results.

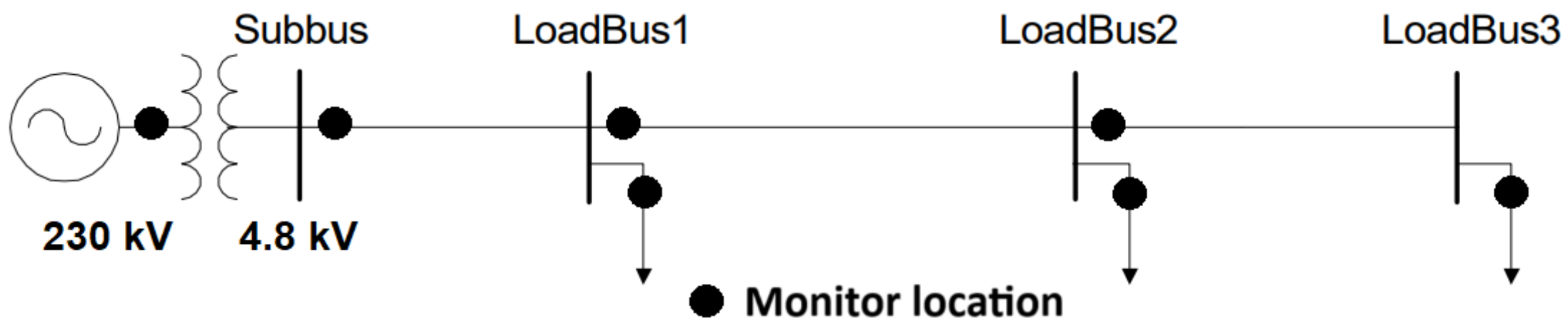

*Figure 29. Case study 3: Test system configuration.*

*Table 2 - List and Description of Studies*

| Case Name | Mode-Description |
|---|---|
| Case3A1 | Time mode - Study 1 - Without generation |
| Case3A2 | Time mode - Study 2 - With sun generation |
| Case3A3 | Time mode - Study 3 - With sun generation and storage 1 |
| Case3A4 | Time mode - Study 4 - With sun generation and storage 2 |
| Case3B1 | Monte Carlo - Study 1 - Without generation - Internally generated load distributions |
| Case3B2 | Monte Carlo - Study 2 - Constant generation - Internally generated load distributions |
| Case3B3 | Monte Carlo - Study 3 - Without generation - Externally generated load distributions |
| Case3B4 | Monte Carlo - Study 4 - Random generation - Externally generated load distributions |

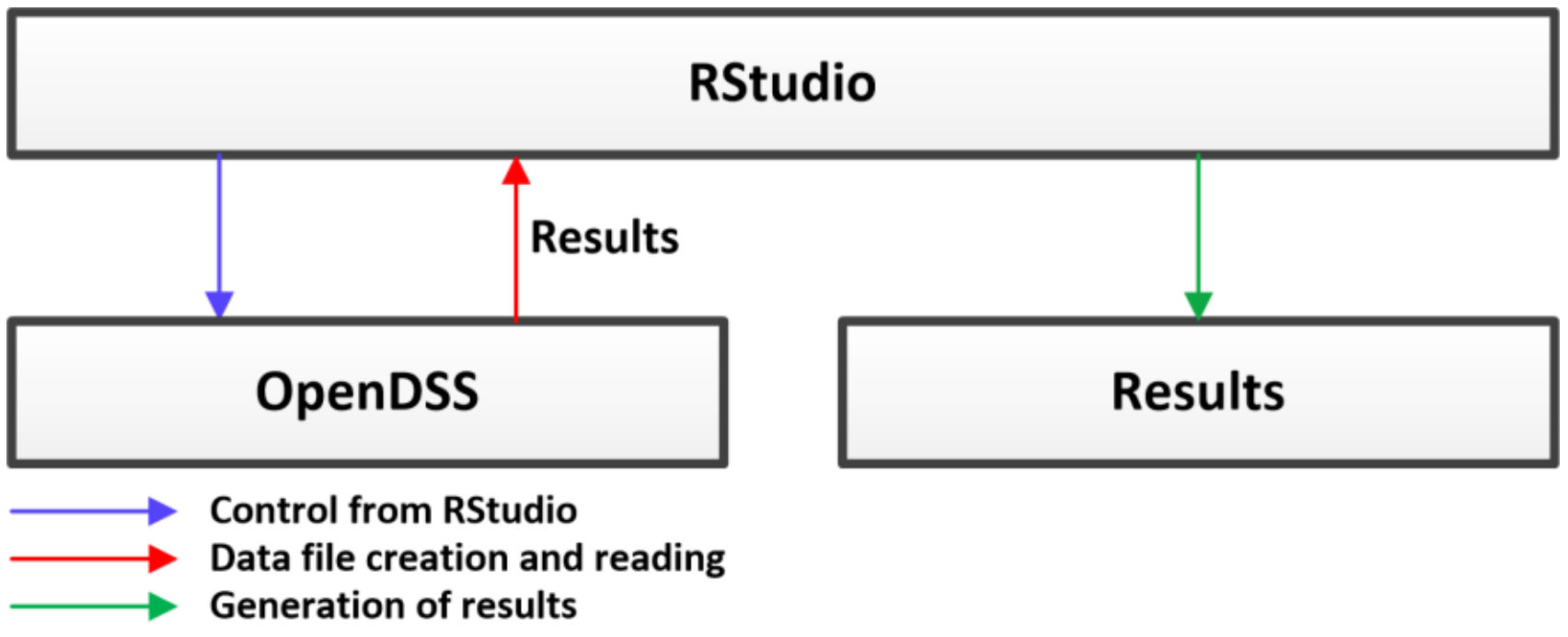

*Figure 30. Case study 3: Scheme of the RStudio-OpenDSS interaction.*

### 3.4.2 Time Mode Cases

This part of the case study analyzes the test system during 200 hours assuming loads vary with time in steps of 1 hour.

The three-phase loads are delta connected and modeled as constant PQ (model=1 in **OpenDSS**), with the following ratings:

- Load 1: Voltage=4.8 kV, Active power=285.0, Power factor=0.90
- Load 2: Voltage=4.8 kV, Active power=240.0, Power factor=0.89
- Load 3: Voltage=4.8 kV, Active power=192.0, Power factor=0.90.

The variation of the three loads is read from CSV files previously created. Figure 31 shows these shapes. The three factors are normalized prior to any simulation. A description of the four cases studied in this part follows.

**Case A1**. *Time mode without generation*: The first study analyzes the test system without generation or energy storage.

**Case A2**. *Time mode with generation*: The second study analyses the behavior of the test system after connecting a 300 kW photovoltaic generation parallel to Load 3. Figure 32 shows the photovoltaic generation factor. Note the zero value of the generation during the hours without sun. The photovoltaic generation factor is also normalized prior to any simulation.

**Case A3**. *Time mode with generation plus storage - Strategy 1*: The third study analyzes the behavior of the test system during 200 hours after connecting photovoltaic generation and energy storage in parallel to Load 3. The curves of yearly variation of load factors are those used with previous studies. The control curve of energy storage is depicted in Figure 33.

**Case A4**. *Time mode with generation plus storage - Strategy 2*: The new study analyzes the behavior of the test system during 200 hours after connecting photovoltaic generation and energy storage in parallel to Load 3, but using a different control strategy for energy storage. The new control strategy curve for energy storage is depicted in Figure 34. Figure 35 compares the two control strategies for energy storage.

In all cases, the CSV files generated by **OpenDSS** are read from **RStudio** for later manipulation and comparison with other studies.

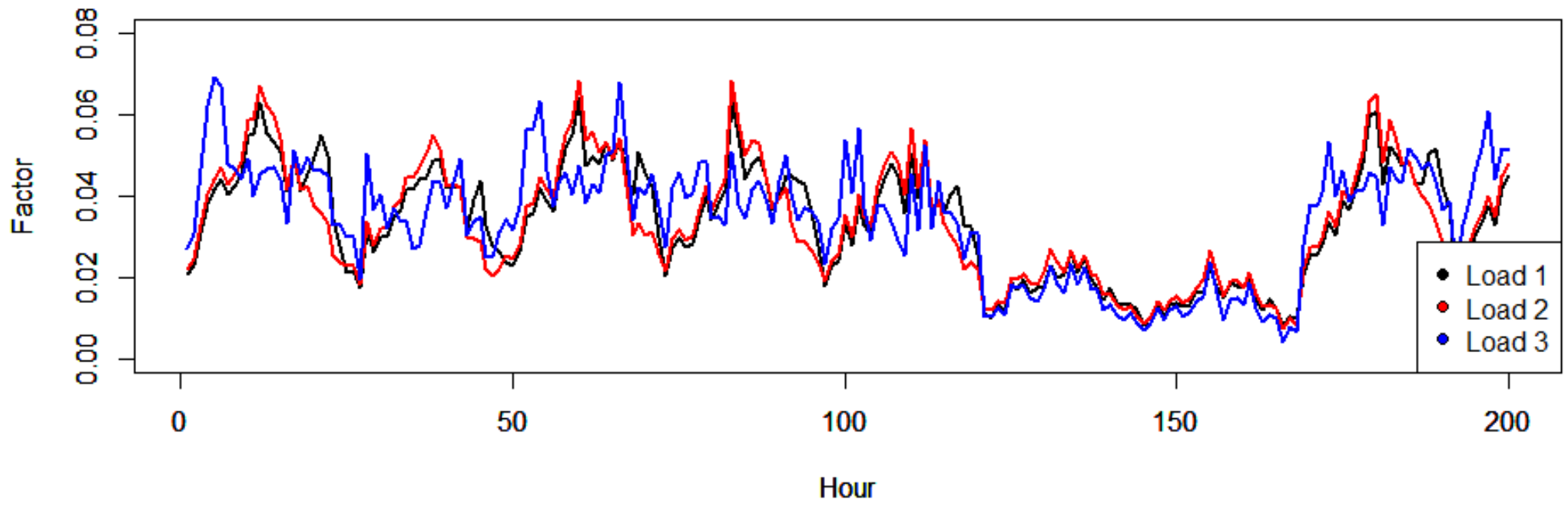

***Figure 31****. Case study 3: Comparison of load variation shapes.*

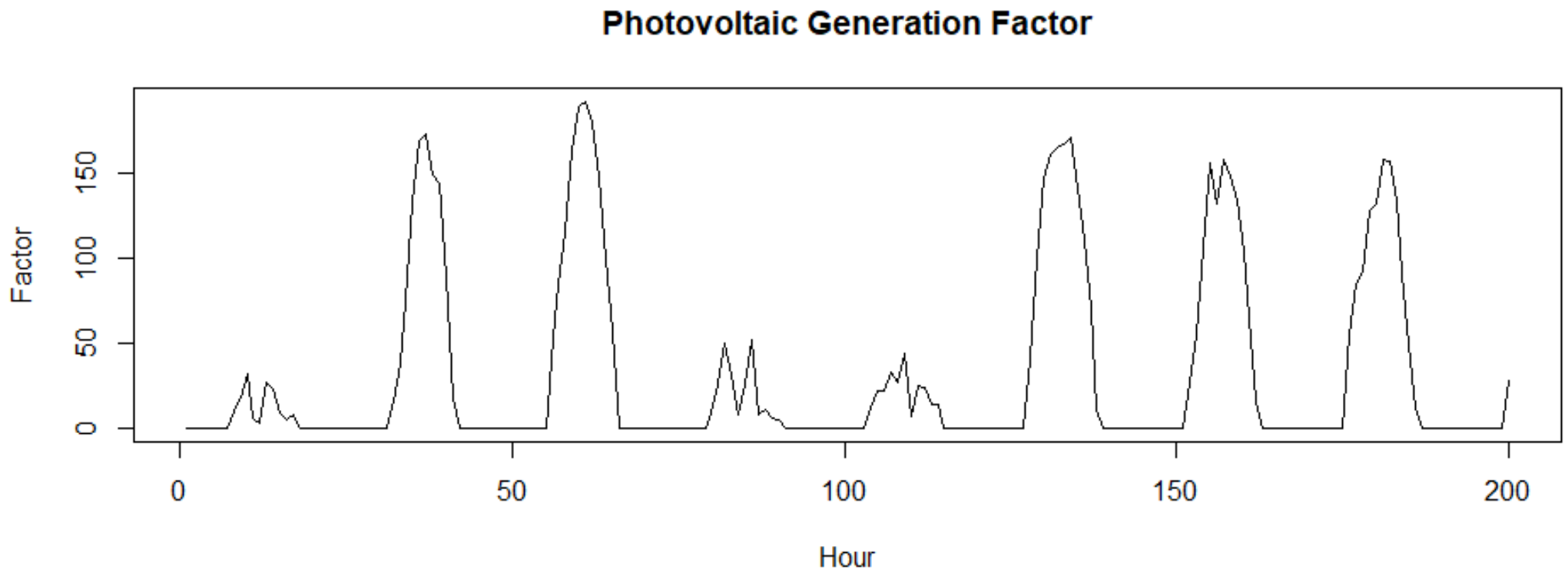

***Figure 32****. Case study 3A2: Photovoltaic generation.*

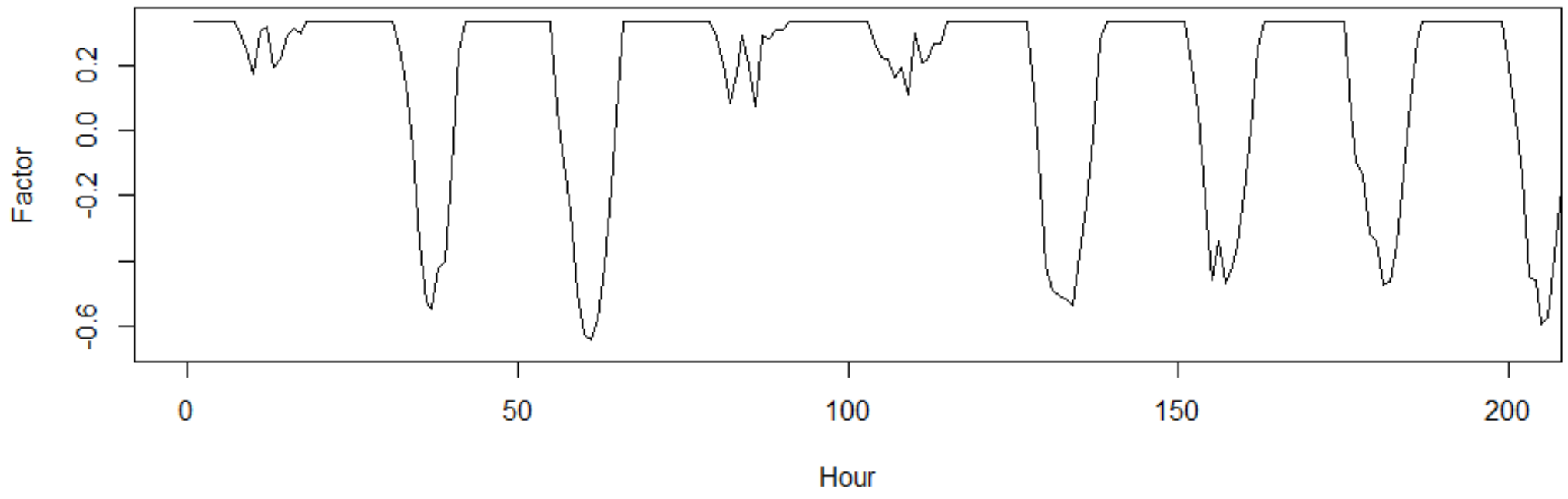

***Figure 33****. Case study 3A3: Storage control shape - Strategy 1.*

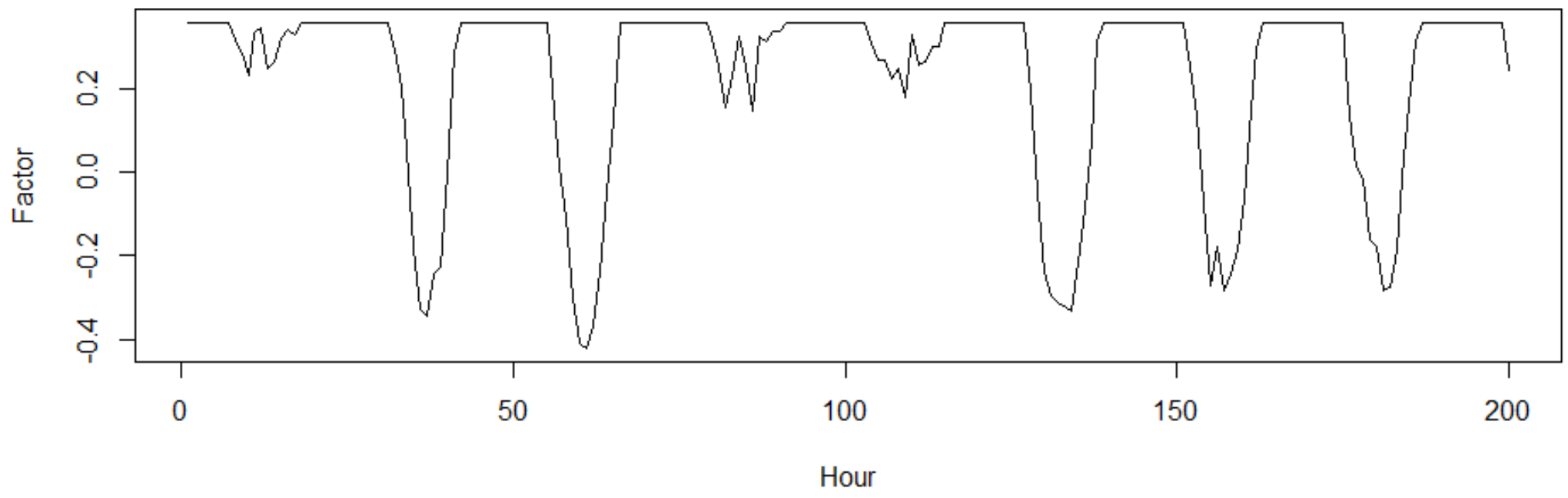

***Figure 34****. Case study 3A4: Storage control shape - Strategy 2.*

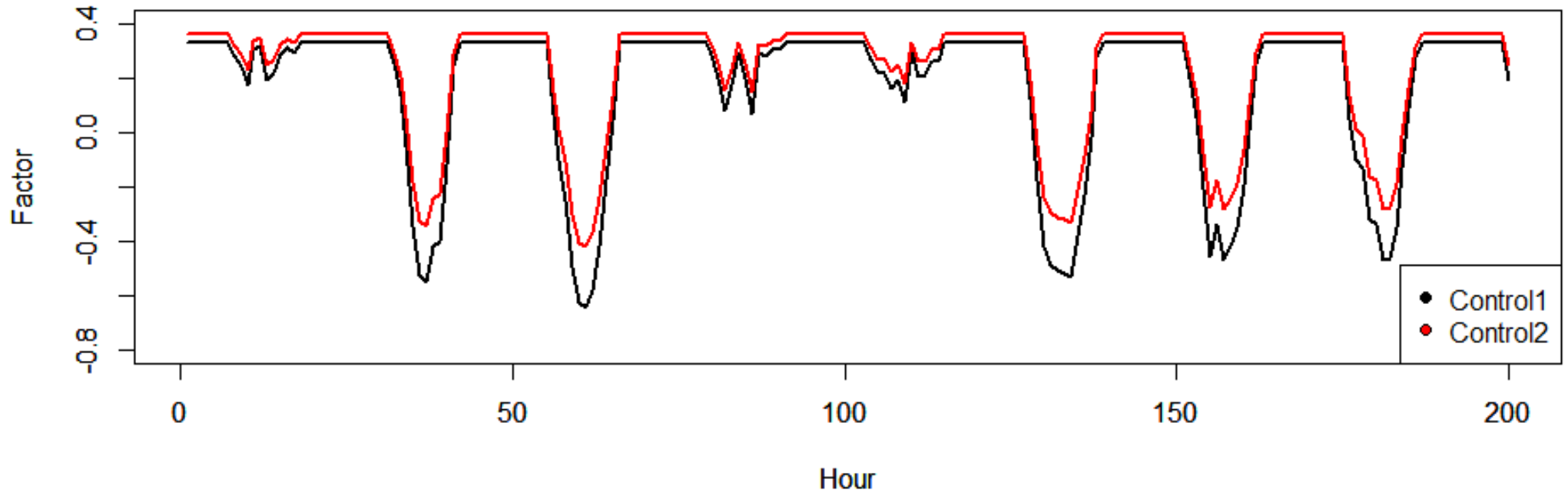

***Figure 35****. Comparison of storage control shapes.*

The following figures show a comparison of results derived from the four studies.

- Figure 36 shows the phase-to-ground voltage and the complex power measured at the Load 3 terminals.
- Figure 37 shows the complex power measured at the Line 1 sending terminals.
- Figure 38 shows the phase-to-ground voltage and the complex power measured at the photovoltaic generator terminals.
- Figure 39 shows some results derived from the two control strategies used with the energy storage unit.
- Figure 40 compares the power output from the photovoltaic generator and the energy storage unit with the two control strategies.
- Figure 41 compares the three-phase energy input (active and reactive) from the transmission system with the four studies.

Tables 3 and 4 compare results derived from the four studies. Some obvious conclusions are derived from this comparison: the energy required from the transmission system, as well as peak values of power, energy and losses decrease when generation (with or without storage) is connected. In addition, the comparison of results derived from the two control strategies used with the energy storage unit shows, as expected, that the selected control strategy matters, although differences are not significant.

***Table 3*** *- Maximum Power at the Substation Entrance*

| Case | kW | kvar | kVA |
|---|---|---|---|
| Without Generation | 558.56 | 649.51 | 856.66 |
| With Generation | 537.21 | 624.54 | 823.80 |
| With Generation and Storage 1 | 537.21 | 624.54 | 823.80 |
| With Generation and Storage 2 | 521.61 | 581.09 | 780.86 |

***Table 4*** *- Comparison of System Meters*

| | Without DG | With DG | With DG and Storage 1 | With DG and Storage 2 |
|---|---|---|---|---|
| kWh | 63525.36 | 53834.73 | 53592.86 | 53590.09 |
| kvarh | 34561.59 | 33894.55 | 33781.88 | 33792.56 |
| Peak.kW | 558.56 | 537.21 | 537.21 | 537.21 |
| Peak.kVA | 649.51 | 624.54 | 624.54 | 624.54 |
| Losses.kWh | 1492.66 | 1281.54 | 1222.17 | 1228.19 |
| Losses.kvarh | 3951.53 | 3141.41 | 3049.9 | 3035.75 |
| Peak.Losses.kW | 20.35 | 18.66 | 18.66 | 18.66 |

The improvements obtained after connecting generation and energy storage are not significant because no optimization procedure was implemented for selecting ratings (of both generation and storage) and control strategies of the energy storage unit. The capabilities of the **R** environment could be useful for such optimization procedure.

### 3.4.3 Monte Carlo Cases

This part analyzes the behavior of the test system assuming random loads and applying the Monte Carlo method. A description of the new four cases studied in this part follows.

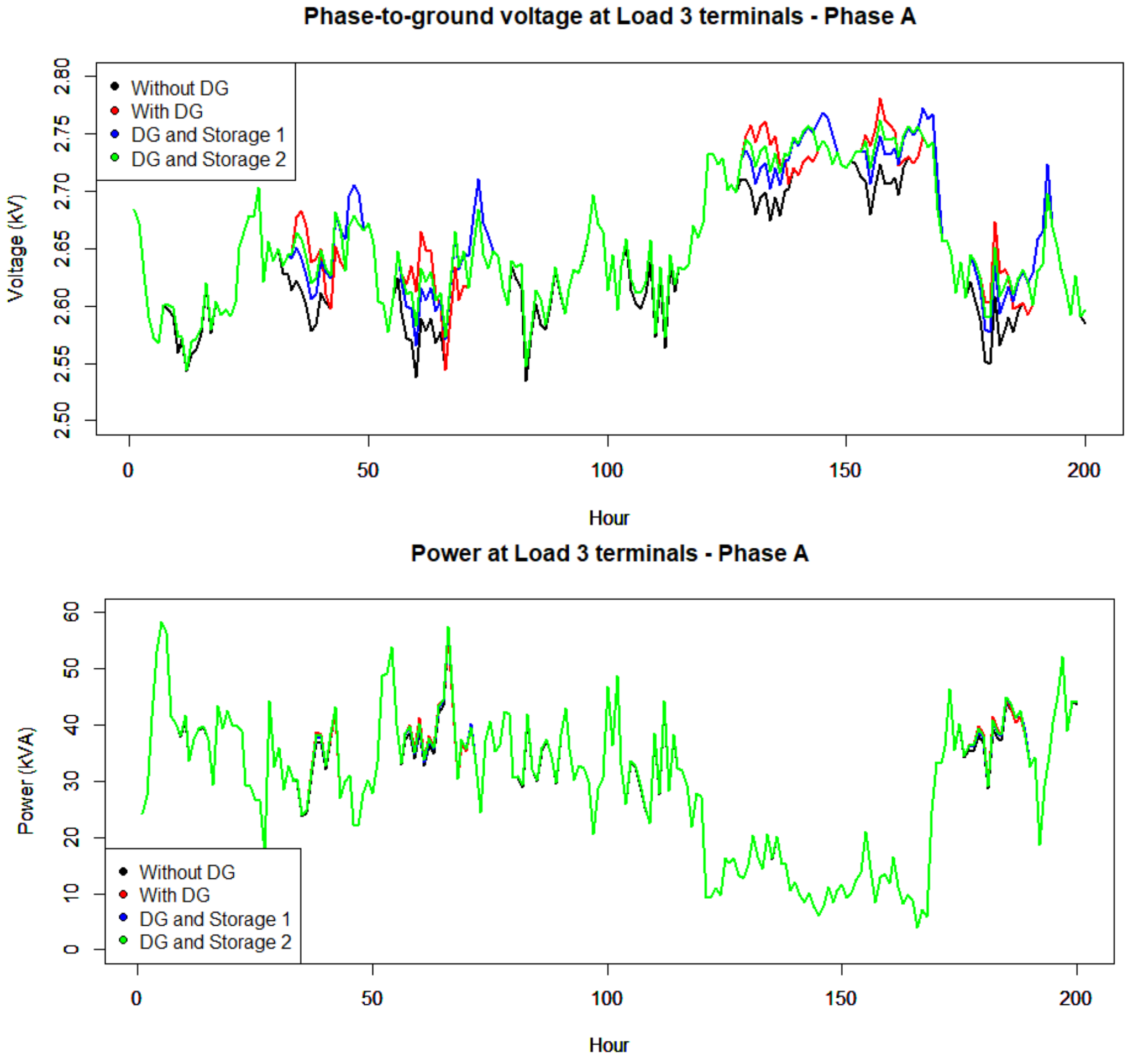

***Figure 36****. Comparison of results - Load 3 terminals.*

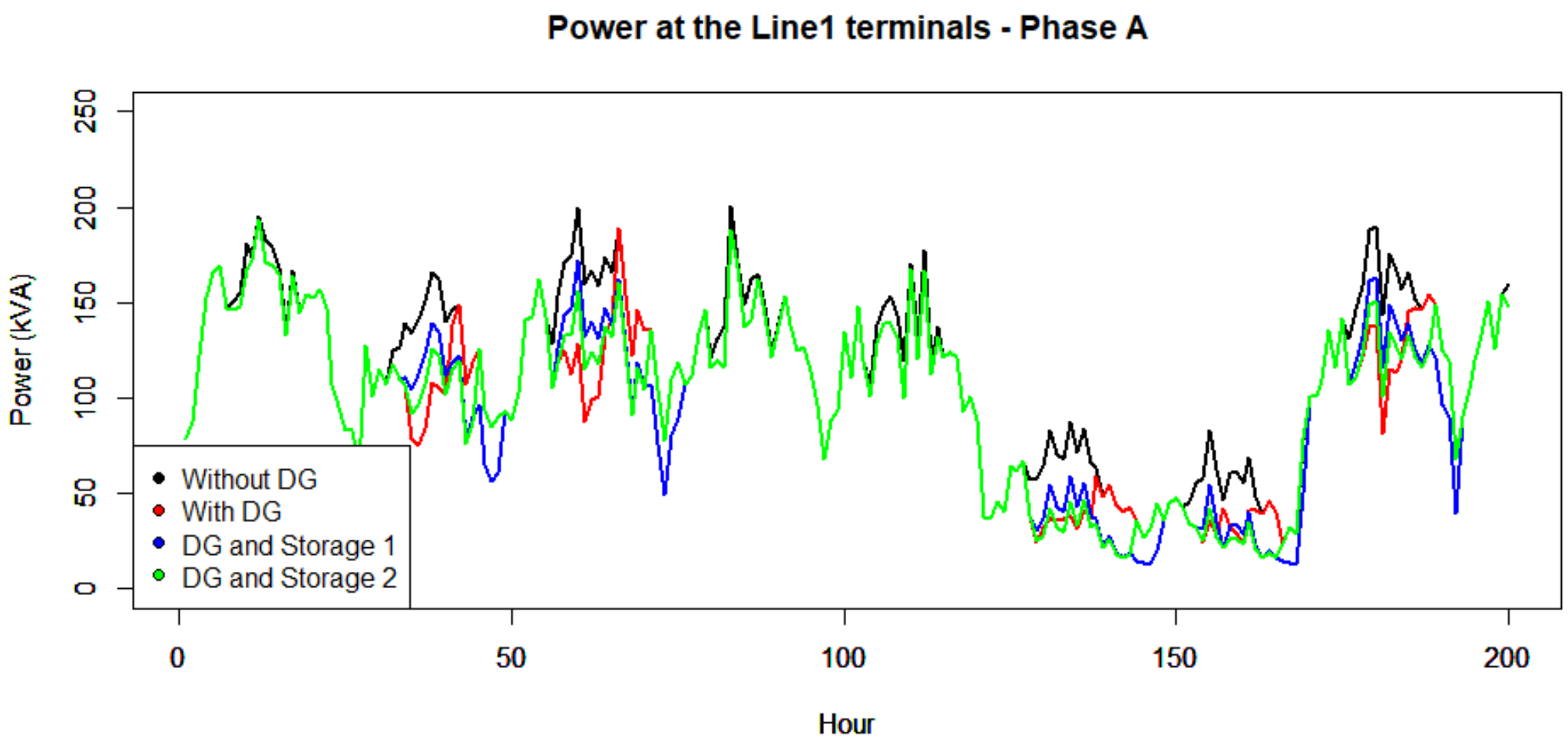

***Figure 37**. Comparison of results - Line 1 sending terminals.*

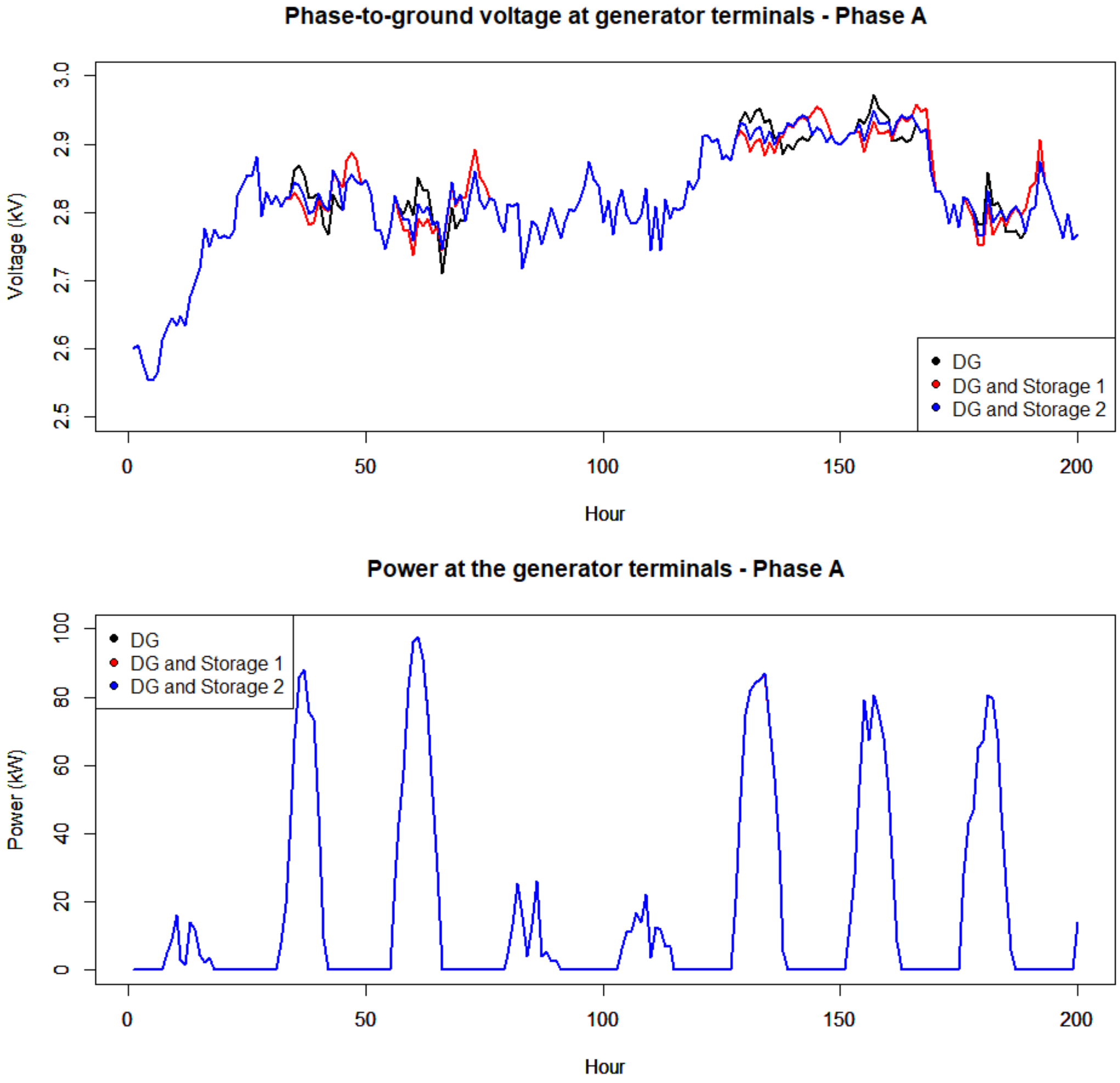

***Figure 38**. Comparison of results - Photovoltaic generator terminals.*

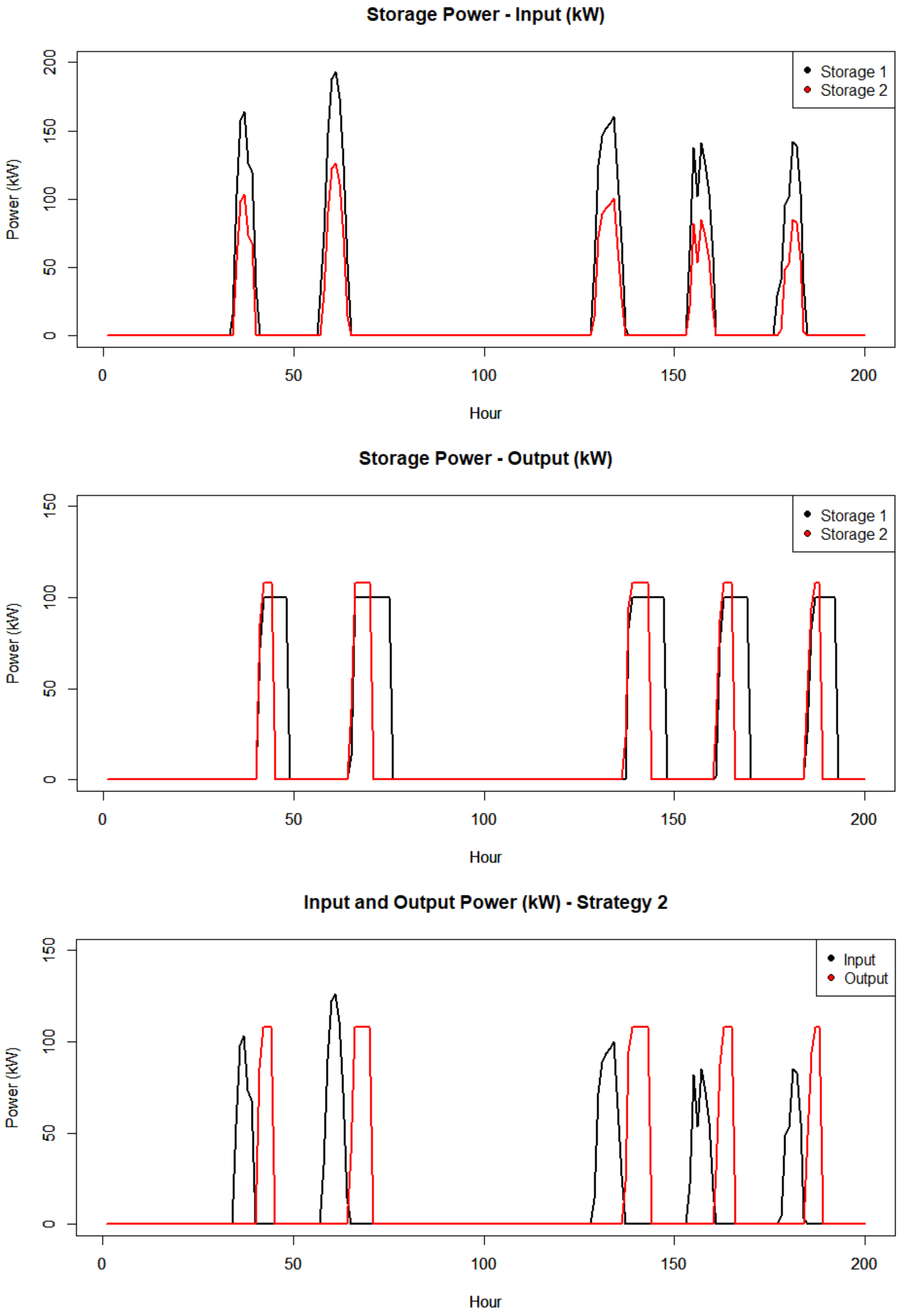

***Figure 39***. *Comparison of results - Storage. (Cont.)*

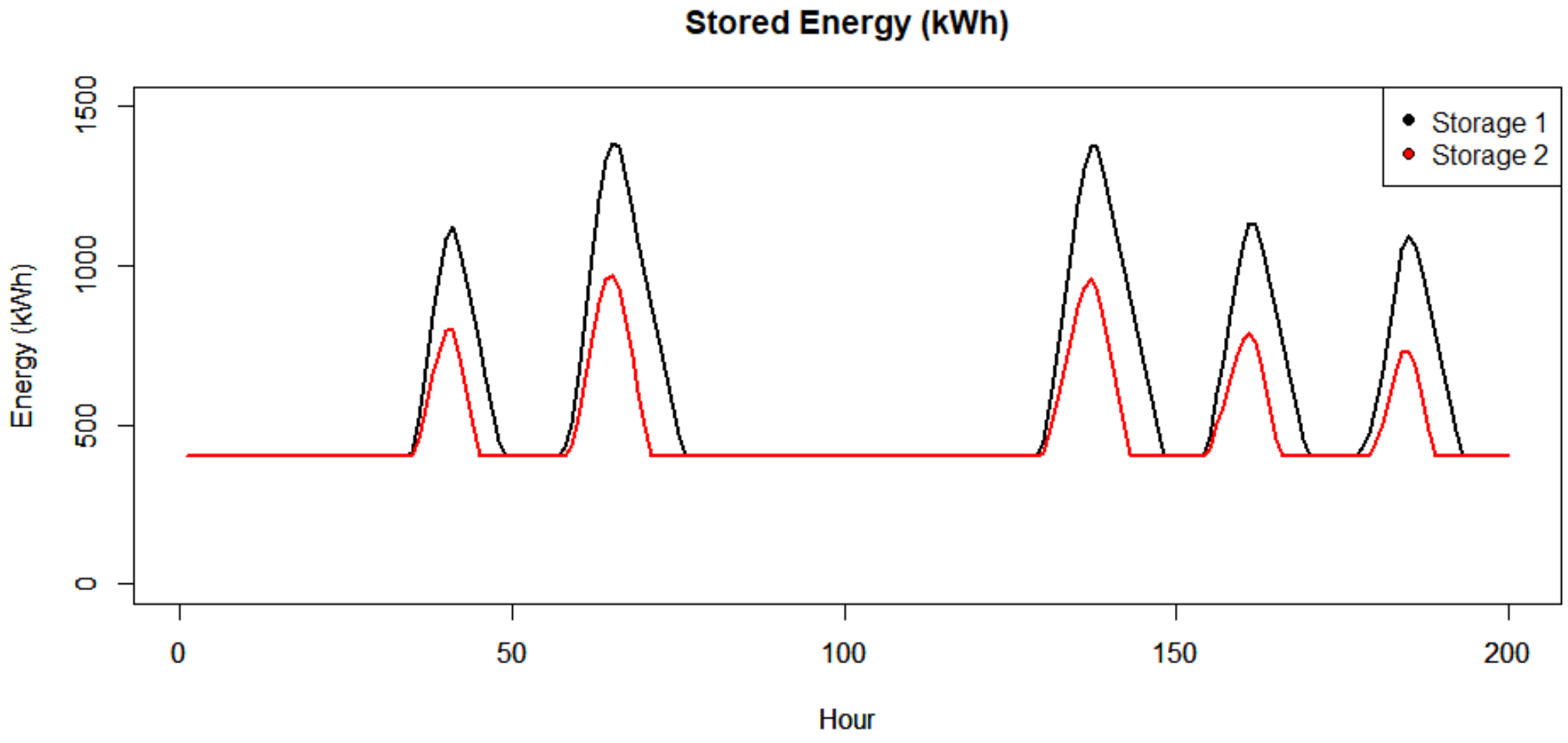

***Figure 39***. *Comparison of results - Storage.*

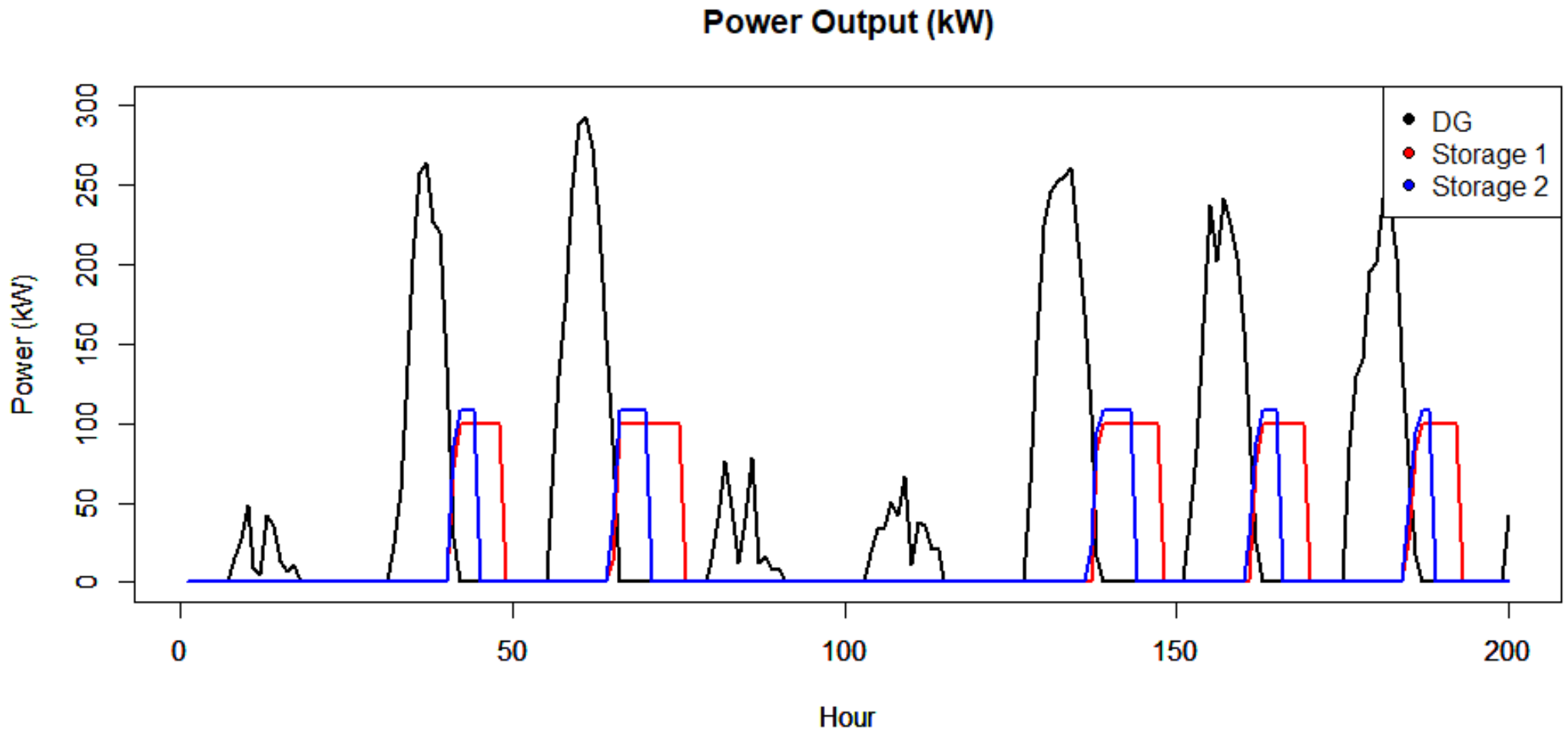

***Figure 40***. *Comparison of results - Power output of photovoltaic generation and storage.*

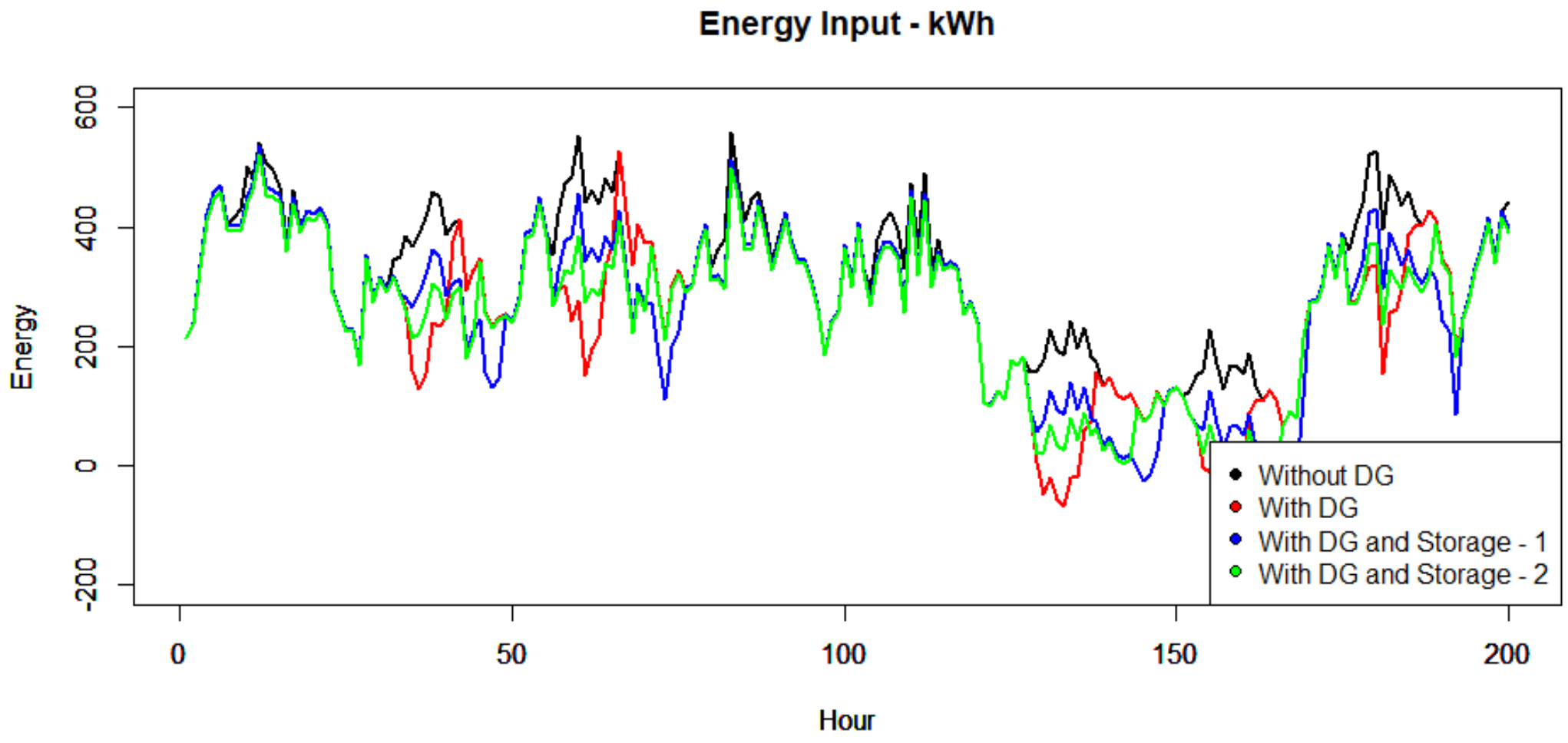

***Figure 41***. *Comparison of results - Energy input from the transmission system. (Cont.)*

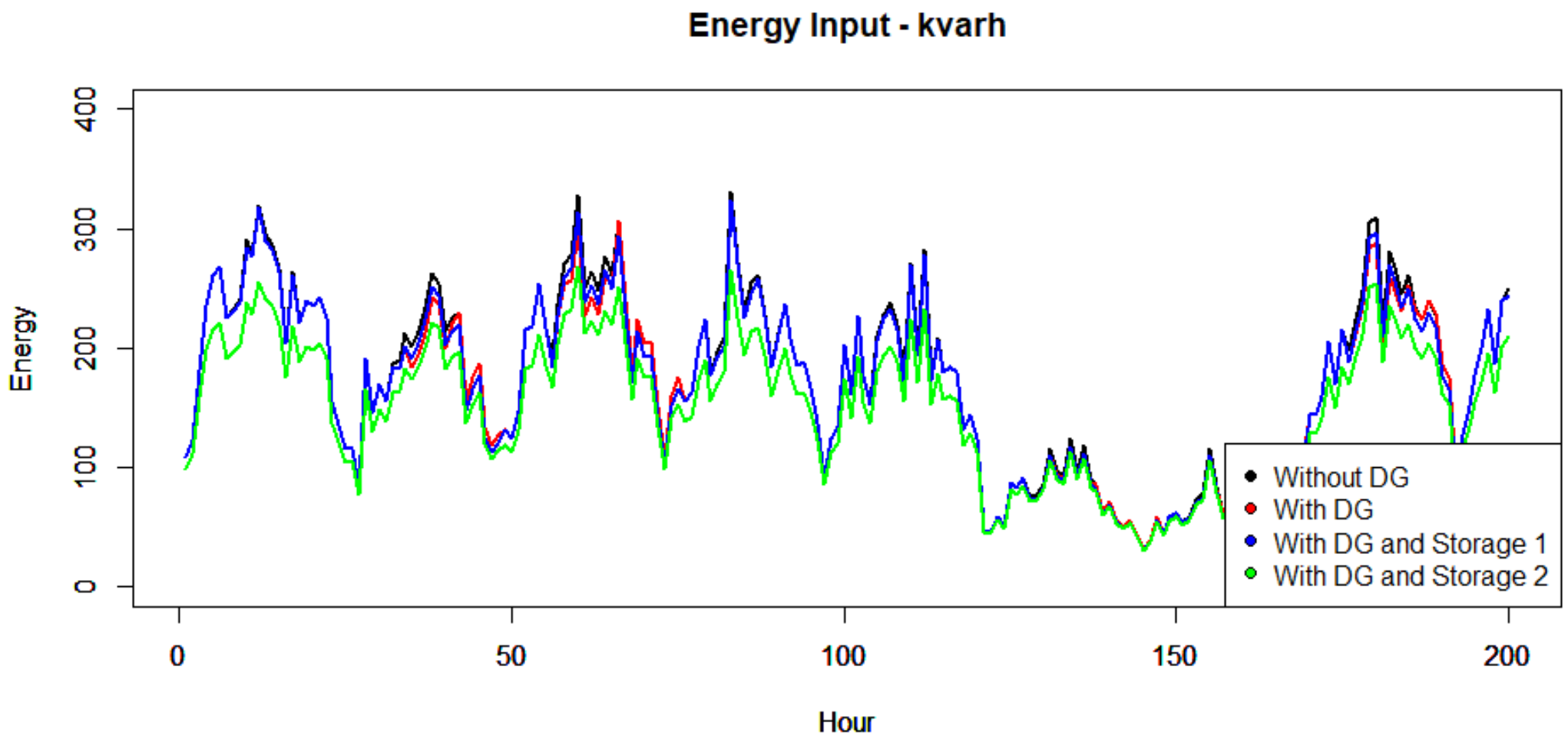

***Figure 41**. Comparison of results - Energy input from the transmission system.*

**Case B1**: This study analyzes the behavior of the test system assuming random loads. The load power values are randomly generated by **OpenDSS** assuming normal distributions with mean and standard deviation values specified in the input file. In this part, the three-phase loads are again delta connected and modeled as constant PQ (model=1 in **OpenDSS**), with the same ratings that were used in the first part:

- Load 1: Voltage=4.8 kV, Active power=285.0, Power factor=0.90
- Load 2: Voltage=4.8 kV, Active power=240.0, Power factor=0.89
- Load 3: Voltage=4.8 kV, Active power=192.0, Power factor=0.90.

Since the load values are now assumed random with a normal distribution, the power values for the three loads are calculated using a mean and a standard deviation of 50% and 5%, respectively.

**Case B2**: This new study analyzes the behavior of the test system assuming again random loads but with a generator in parallel to load 3. The load power values are again randomly generated by **OpenDSS** assuming normal distributions with mean and standard deviation values specified in the input file. Figure 42 depicts the factor of the generator power shape; it is obviously a constant generation. The rated power of the generator, which only injects active power to the system, is 300 kW.

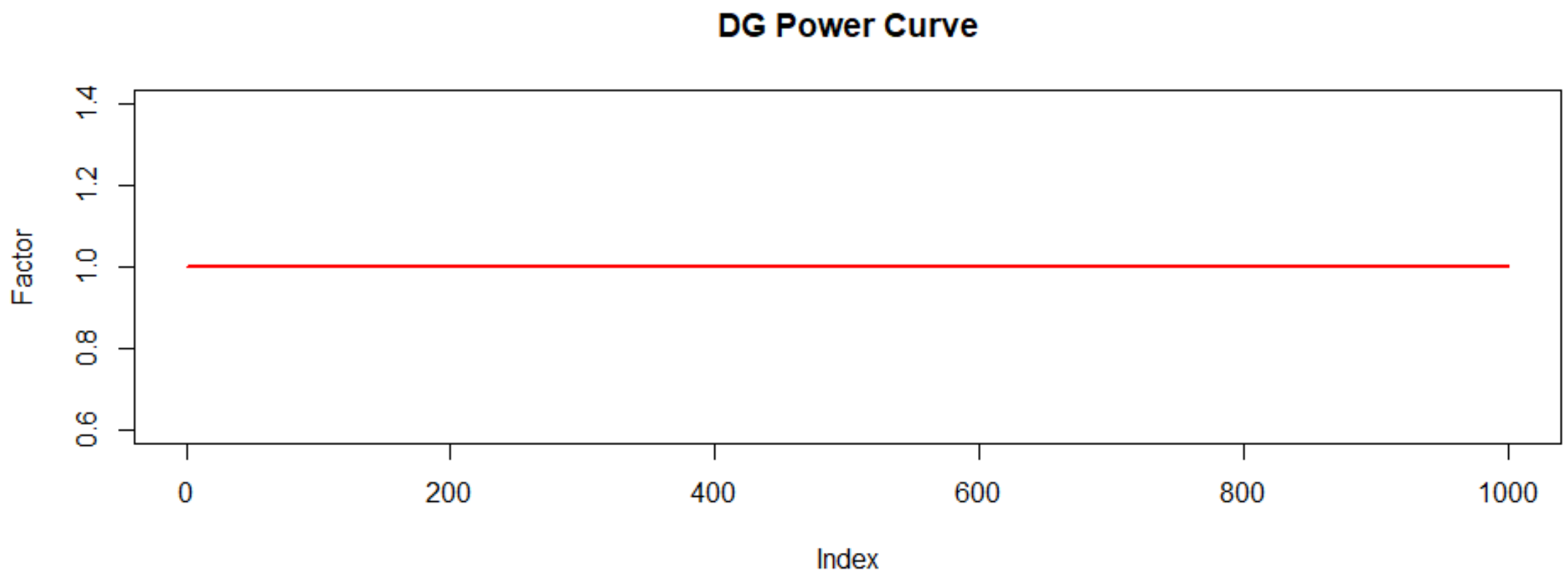

***Figure 42**. Case study 3B2: Monte Carlo simulation - Local generator factor.*

**Case B3**: This study is like the first study (i.e., the behavior of the test system is analyzed assuming random loads without generation) but now load power values are read from a CSV file.

**Case B4**: This new study analyzes the behavior of the test system assuming random loads but with a generator in parallel to load 3. The random values of power values are, as in the previous study, read from CSV file. The rated power of the generator, which only injects active power into the system, is 300 kW. Figure 43 shows the factor of the generator power shape.

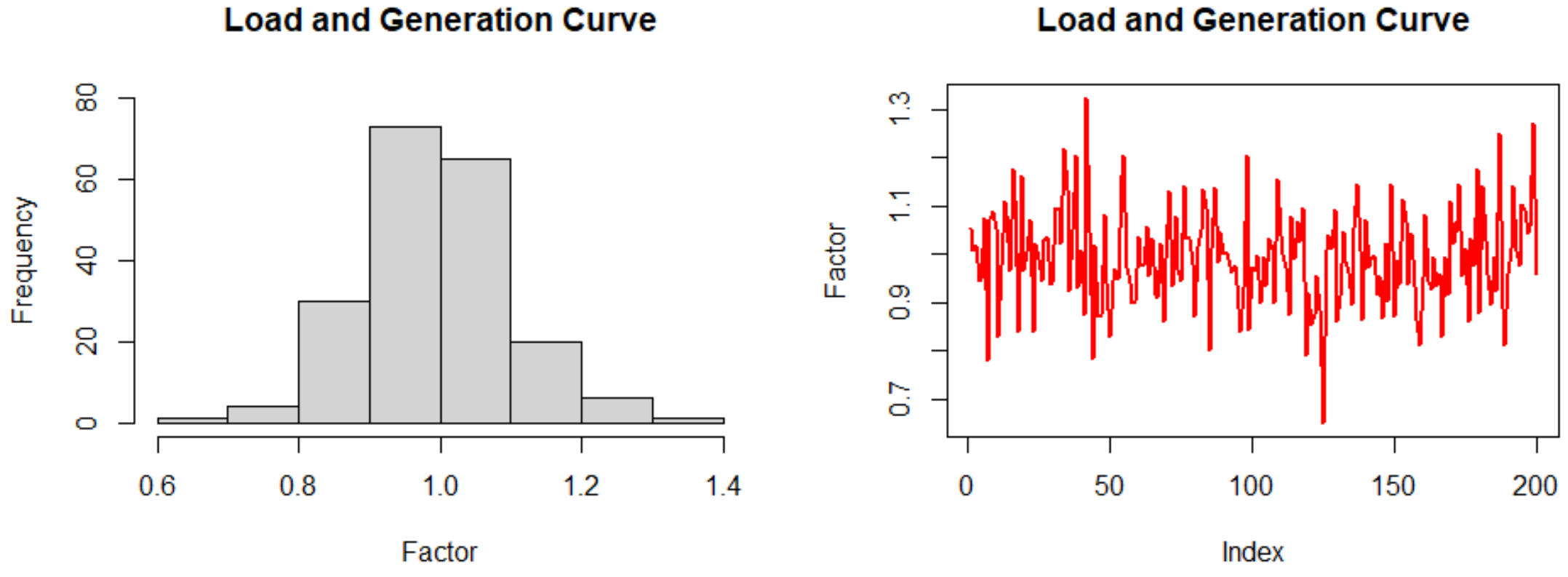

***Figure 43***. *Case study 3B4: Monte Carlo simulation - Local generator power factor.*

Tables 5 and 6 summarize some of the main results. Table 5 provides the mean and standard deviation values of active and reactive powers of the three loads. Observe that the four mean and standard deviation values corresponding to the same load are very close to each other. Table 6 provides the mean and standard deviation values of active and reactive powers of the three lines and the system source. It is evident the impact of the generation connected to the load node 3 (remember that in both Case 2 and Case 4 generator only inject active power): the mean value of the active power measured at the sending end of Line 2 and Line 3 is negative, and the mean value of the active power measured at the source terminals has been significantly reduced with respect the values measured when test system is without local generation.

The subsequent figures show a comparison of results derived from the four studies.

- Figure 44 compares the distribution of active and reactive powers obtained with the four studies and measured at the Load 1 terminals.
- Figure 45 compares the distribution of active and reactive powers obtained with the four studies and measured at the Line 1 sending terminals.
- Figure 46 compares the distribution of active and reactive powers obtained with the four studies and measured at the generator terminals.

It is evident the impact of the generation connected to load node 3 (see results related to Cases 2 and 4): the values of active powers measured at the sending end of lines 2 and 3 are negative, and the source values are significantly lowers than the cases without generation.

***Table 5***. *Comparison of results - Loads - Phase A*

| | Case 1 | | Case 2 | | Case 3 | | Case 4 | |
|---|---|---|---|---|---|---|---|---|
| | kW | kvar | kW | kvar | kW | kvar | kW | kvar |
| Load 1 - Mean | 47.65 | 23.09 | 47.41 | 22.96 | 47.30 | 22.92 | 47.40 | 22.96 |
| Load 1 - Std Dev | 4.54 | 2.20 | 4.81 | 2.33 | 4.37 | 2.12 | 4.53 | 2.20 |
| Load 2 - Mean | 39.73 | 20.36 | 40.01 | 20.51 | 39.67 | 20.33 | 39.92 | 20.46 |
| Load 2 - Std Dev | 3.87 | 1.98 | 4.10 | 2.10 | 3.53 | 1.80 | 3.82 | 1.96 |
| Load 3 - Mean | 31.56 | 15.29 | 31.93 | 15.47 | 31.53 | 15.28 | 31.94 | 15.48 |
| Load 3 - Std Dev | 3.09 | 1.50 | 3.30 | 1.60 | 2.71 | 1.31 | 3.06 | 1.48 |

*Table 6. Comparison of results - Lines - Phase A*

| | Case 1 | | Case 2 | | Case 3 | | Case 4 | |
|---|---|---|---|---|---|---|---|---|
| | kW | kvar | kW | kvar | kW | kvar | kW | kvar |
| Line 1 - Mean | 119.54 | 57.30 | 20.34 | 58.29 | 119.10 | 57.09 | 20.48 | 58.25 |
| Line 1 - Std Dev | 6.62 | 3.28 | 7.32 | 3.64 | 10.72 | 5.32 | 2.06 | 5.90 |
| Line 2 - Mean | 71.67 | 34.64 | -27.11 | 35.91 | 71.58 | 34.60 | -26.97 | 35.88 |
| Line 2 - Std Dev | 4.81 | 2.42 | 5.38 | 2.71 | 6.31 | 3.17 | 2.48 | 3.69 |
| Line 3 - Mean | 31.70 | 15.03 | -67.20 | 16.34 | 31.67 | 15.02 | -66.97 | 16.36 |
| Line 3 - Std Dev | 3.12 | 1.51 | 3.27 | 1.59 | 2.73 | 1.32 | 6.31 | 1.72 |
| Source - Mean | 121.90 | 66.61 | 20.84 | 60.27 | 138.42 | 28.65 | 20.97 | 60.23 |
| Source - Std Dev | 6.90 | 4.38 | 7.41 | 4.02 | 13.21 | 0.38 | 2.16 | 6.31 |

### 3.4.4 Discussion

Only a small percentage of **OpenDSS** capabilities has been used in this case study. Many other options can be very useful, mainly when analyzing large test systems; they could be used to detect, for instance, voltage exceptions (i.e., over- and undervoltages) or overloads. As already mentioned, **R** capabilities are especially useful when statistical analyses are required; this aspect has been illustrated in this case study although a very simple test system configuration has been simulated.

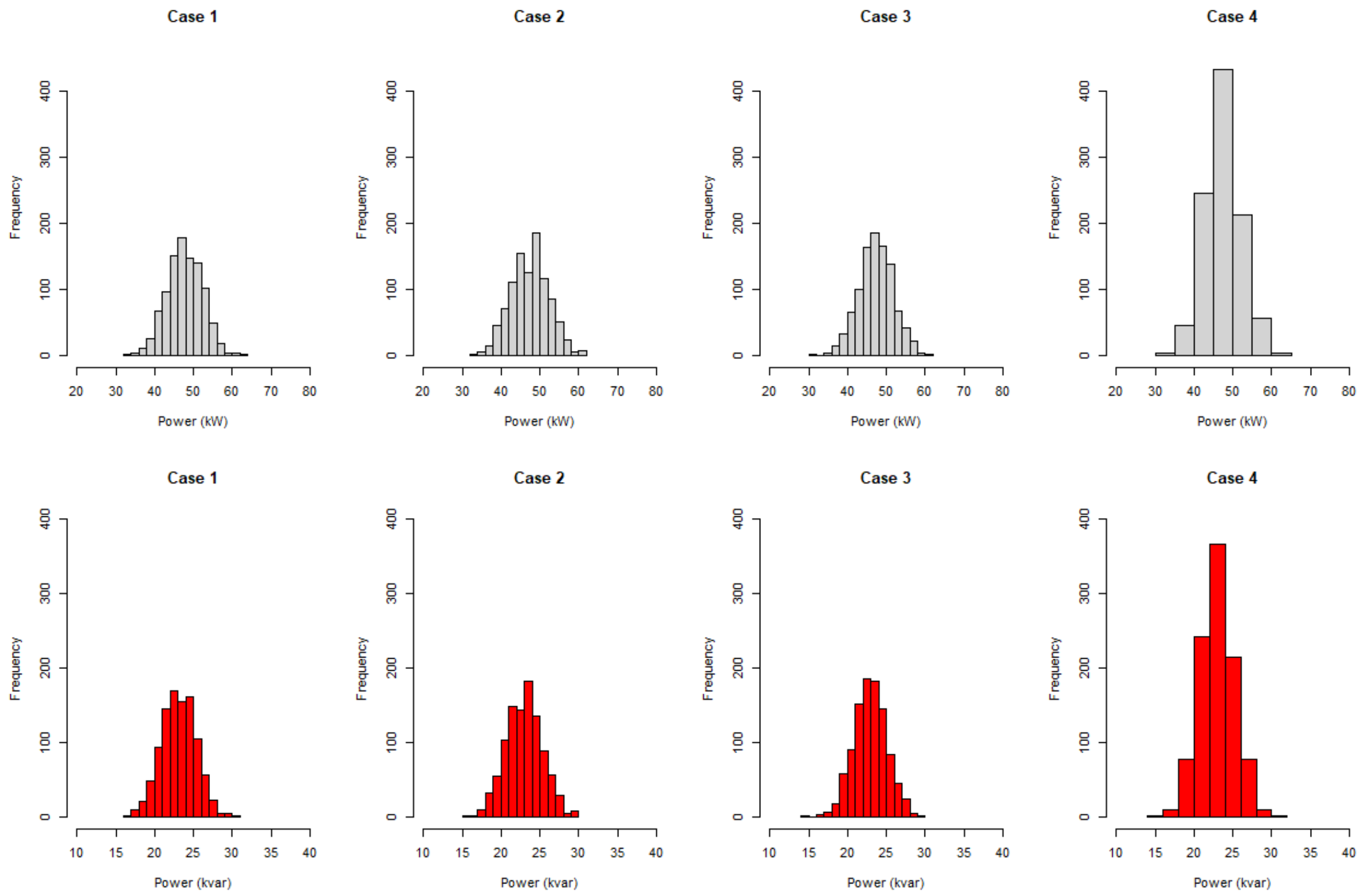

***Figure 44**. Monte Carlo simulation: Comparison of powers measured at Load 1 terminals.*

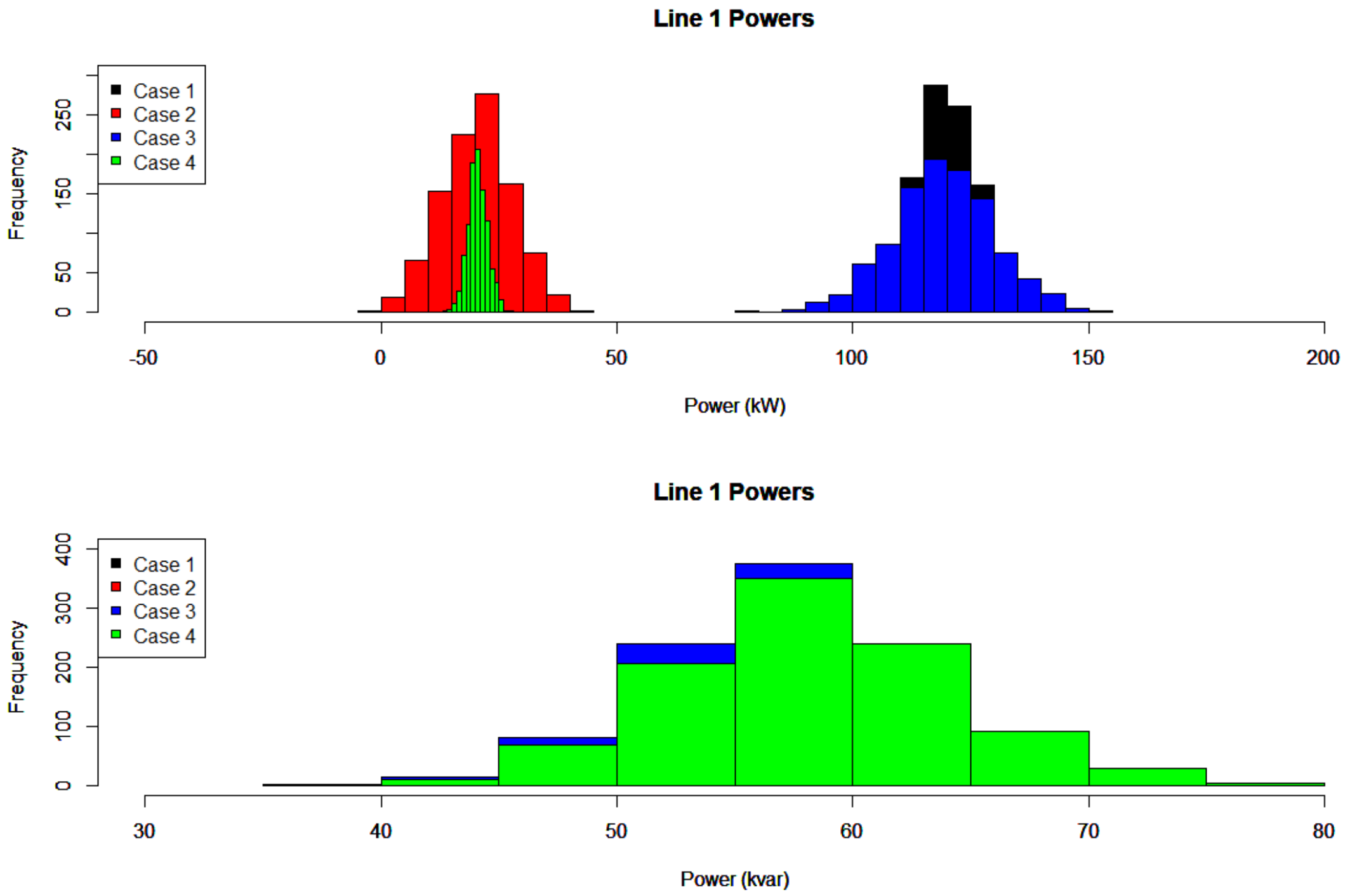

***Figure 45**. Monte Carlo simulation: Comparison of powers measured at Line 1 sending terminals.*

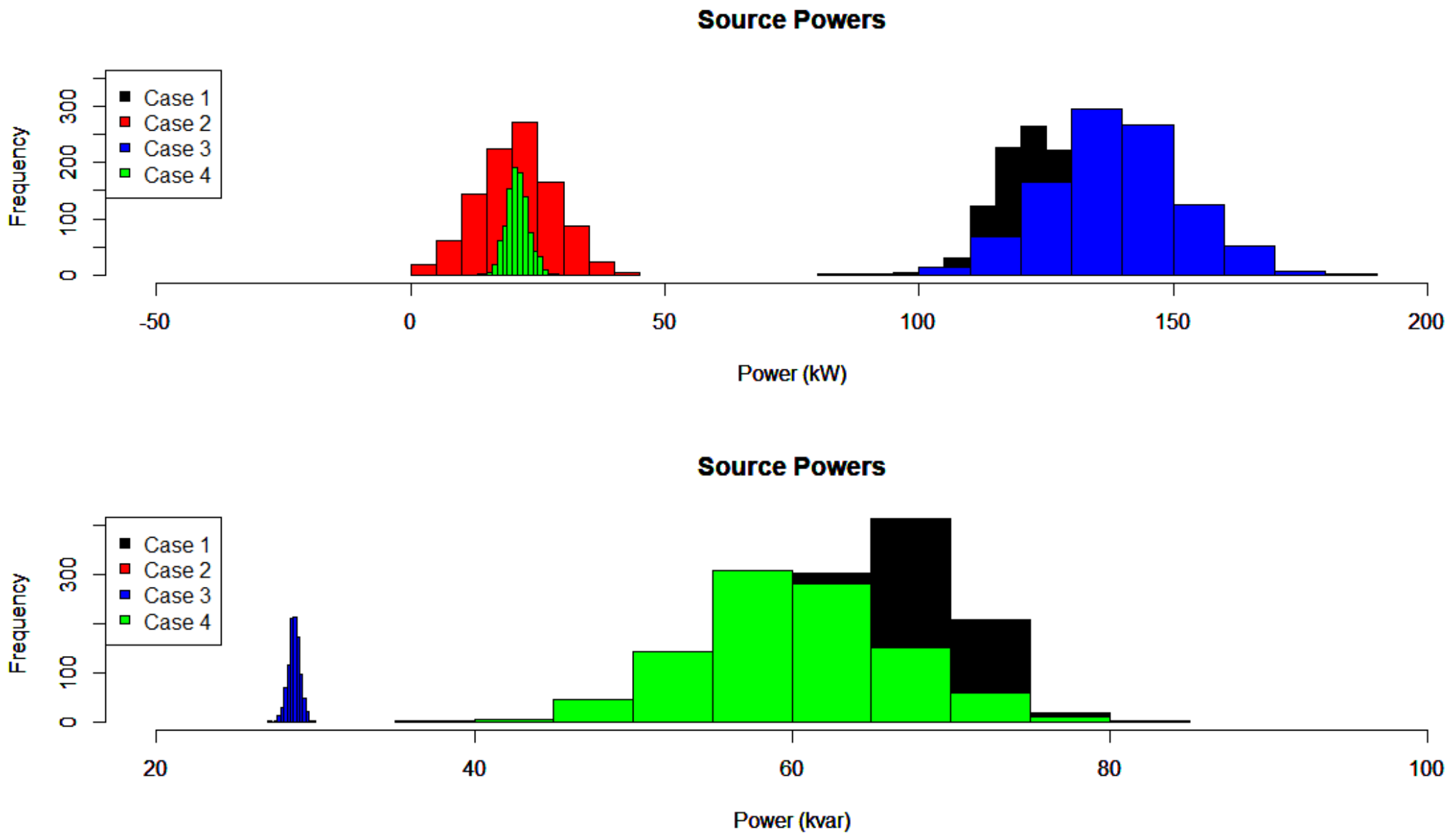

***Figure 46**. Monte Carlo simulation: Comparison of powers measured at the generator terminals.*

### 3.5 Case study 4 - Introduction to transient stability

#### 3.5.1 Introduction

The goal of this case study is to apply machine learning algorithms that could estimate the transient stability response of a very simple power system model. The study will be carried out by using a combination of capabilities of **OpenDSS** and **R**. **OpenDSS** is used here to simulate the transient response of the test system. Although **OpenDSS** is a very popular tool for analyzing the steady-state of distribution systems, it can also be applied in some transient studies. In this case study it will be applied to simulate the transient stability of a simple system configuration often used to introduce stability studies.
Figure 47 shows the scheme of the test system: a power plant composed of four synchronous generators and a step-up transformer is connected to a power system represented as an infinite bus by means of two parallel transmission lines represented as lossless reactances. Configuration and parameters of this test system come from the book by P. Kundur [41].
The parameters of the system components are as follows (see Figure 48):

- Power plant (four generators) with the following parameters: Rating (total) = 2200 MVA, kV = 24, X' (transient reactance) = 0.3 pu, Inertia Constant H = 3.5 MWs/MVA, Damping Coefficient D = 0.
- Step-up transformer: 24/345 kV, Xt = j0.15pu
- Parallel lines from transformer to infinite bus: Circuit 1 - X1 = j0.5 pu, Circuit 2 - X2 = j0.93 pu.

Per unit values come from using the following base values: Vbase = 24kV, Sbase = 2200MVA, Zbase (for 345 kV) = 345*345/2220 = 53.615 ohms.
The initial operating conditions are as follows:

- Infinite bus: V = 0.92 pu, Angle = 0.
- Power plant: Full load (i.e., 2220 MVA), P = 0.9, Q = 0.436.

It is assumed that the test system is running under balanced conditions prior to the fault.
The scenario to be analyzed is the transient stability of the power plant after a 3-phase bolted fault to ground located on Circuit 2 near transformer terminal. The fault is cleared by opening Circuit 2; see Figure 48.

### 3.5.2 Simulation of the test system

Figure 49 shows a scheme of the procedure implemented in **RStudio** to analyze the transient stability of the test system implemented in **OpenDSS** considering the possibility of generating within **RStudio** different parts of the input file, namely the fault model (fault type, location, duration) and the script needed to specify the initial conditions of the power plant (active and reactive power).
Two types of simulation results are presented below: (i) a three-phase fault cleared after 50 ms; (ii) a parametric study with different fault durations to illustrate the influence that the fault duration can have on the system response.

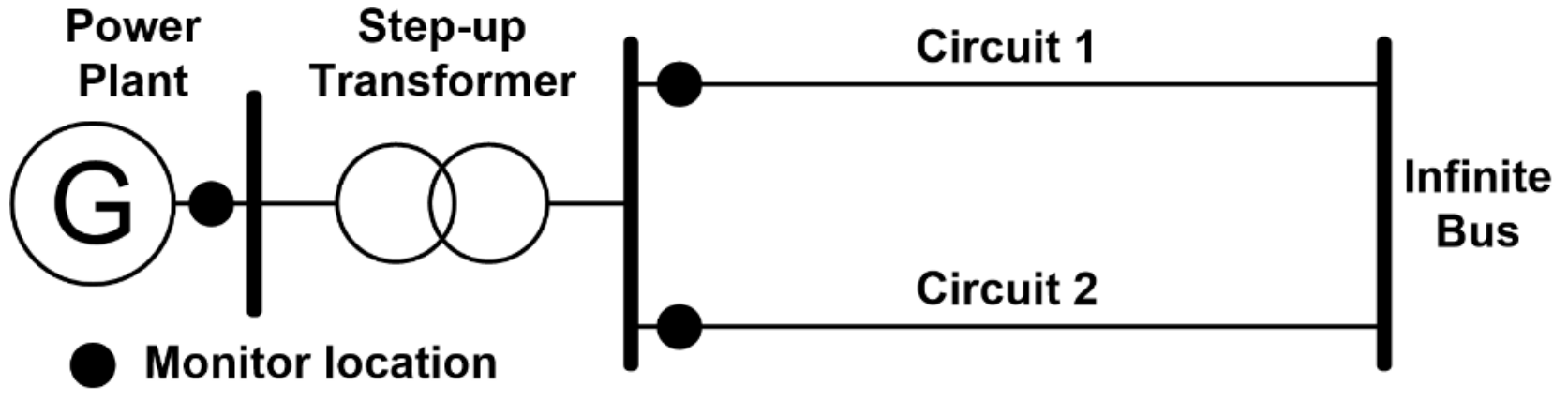

***Figure 47**. Case study 4: Test system configuration.*

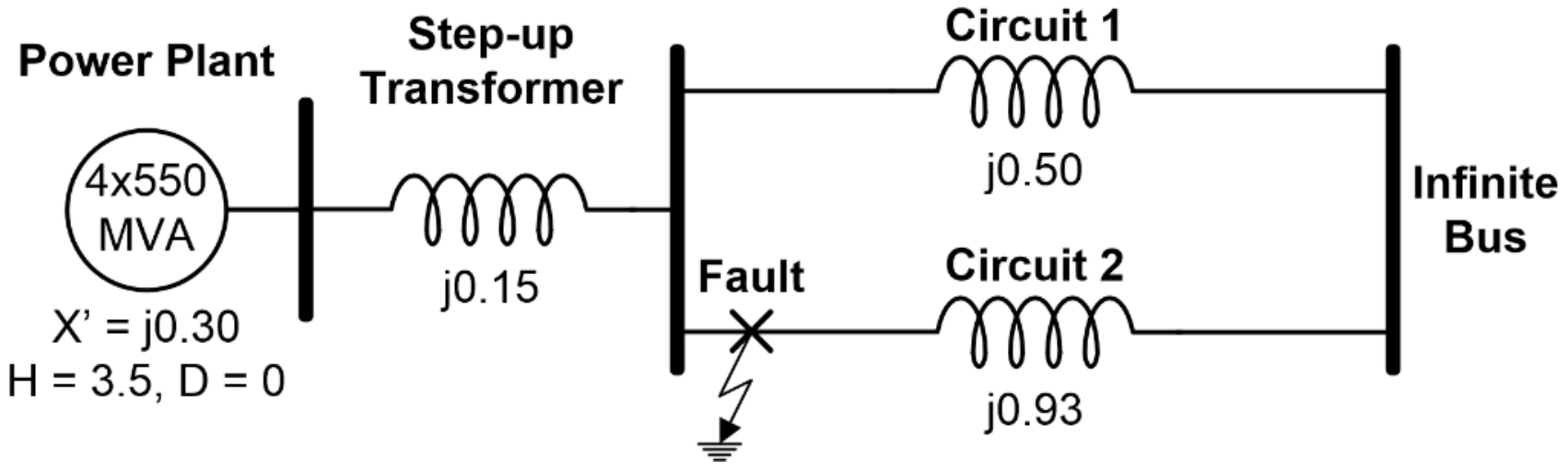

***Figure 48**. Case study 4: Test system model and parameters-All values in pu.*

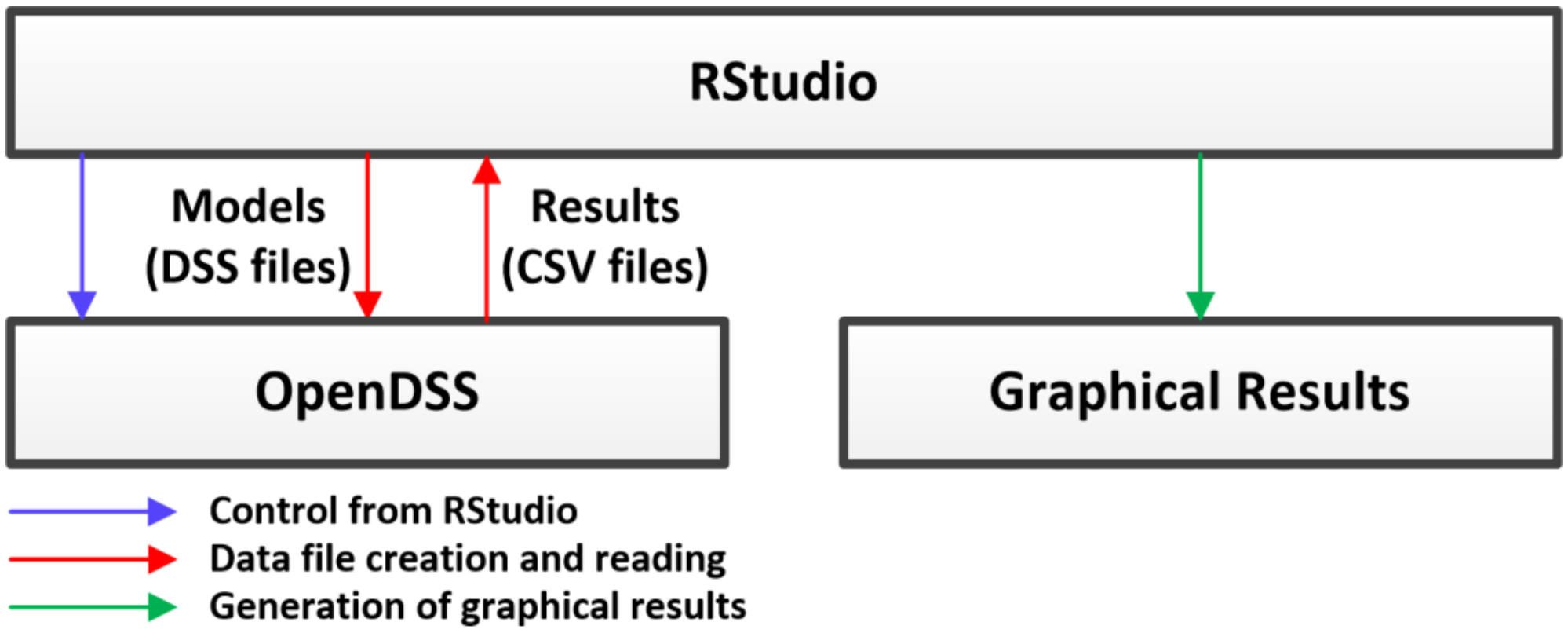

***Figure 49****. Case study 4: Scheme of the RStudio-OpenDSS interaction.*

The initial study is carried out using an input file in which the parameters of all power system components as well as the initial conditions and the fault characteristics (i.e., type, location, duration) are specified. In the second part, the fault model is inserted/redirected by using a text file generated within **RStudio** in which the characteristics of the fault (including the fact that the protection system clears the fault by separating Circuit 2 from the system) are specified.

#### *3.5.2.1 Initial study: Three-phase fault*

The present study is based on an input file released with the **OpenDSS** package. This file uses a three-phase model for the test system, and assumes it is running under balanced conditions.
After simulating the input file, several CSV files, in which the simulation results have been saved, are created. These files can be read using **R** capabilities and are further used to depict the results of concern.
Figure 50 shows the voltage values of the three phases and the active and reactive power measured at some system nodes (see Figure 47) prior to the fault. Remember that the test system is running under balanced conditions prior to the fault. Figure 51 depicts the transient response (i.e., active and reactive power generated by the power plant, speed deviation and angle of the generator rotor, voltages at the terminals of the two lines) prior to the fault, during the fault, and after fault clearing. It is obvious that the system is stable. Remember that the model used in this case study (see Figures 48) is an idealized representation of the test system.

#### *3.5.2.2 Parametric study*

The parametric study is aimed at collecting the behavior of the test system in front of a three-phase fault for which a variable duration will be assumed. System parameters and operating conditions prior to the fault remain the same as in previous studies.
Figure 52 shows the results derived from different fault durations when the generation plant is running at full load with a power factor of 0.9 (lagging). It is easy to deduce that the instability is caused when the duration of the fault is above certain value that depends on several parameters: generator load (i.e., active power injected to the power system at its terminals), some generator parameters (e.g., the values of the inertia constant, rotor damping, transient reactance). The fault location is another aspect to be considered, as well as the other fault characteristics (number of phases involved, duration).
The study is repeated with a different active power value, 1776 MW, which corresponds to a power factor of 0.8. Figure 53 shows that, since the generator active power is now lower, any fault duration value that was used in the previous study cannot cause instability.

```
## Results prior to the fault
```

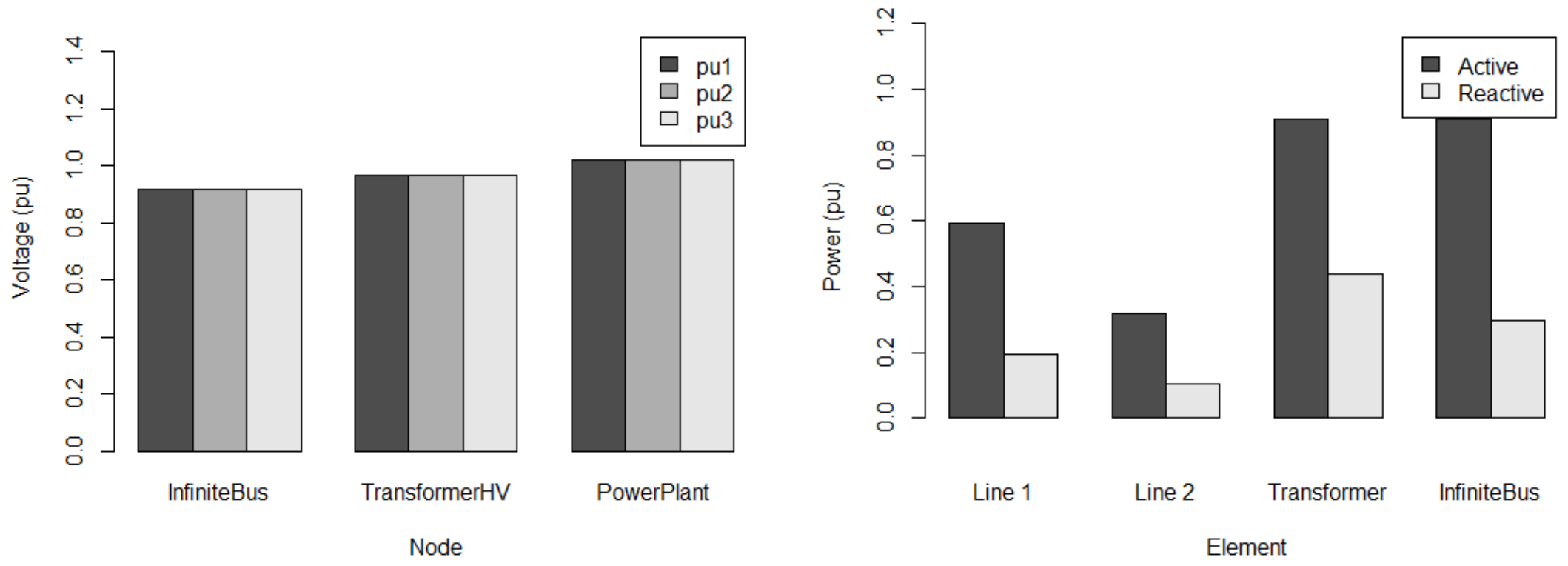

***Figure 50**. Case study 4: Initial scenario: Voltages and power prior to the fault.*

```
## Results after the fault
```

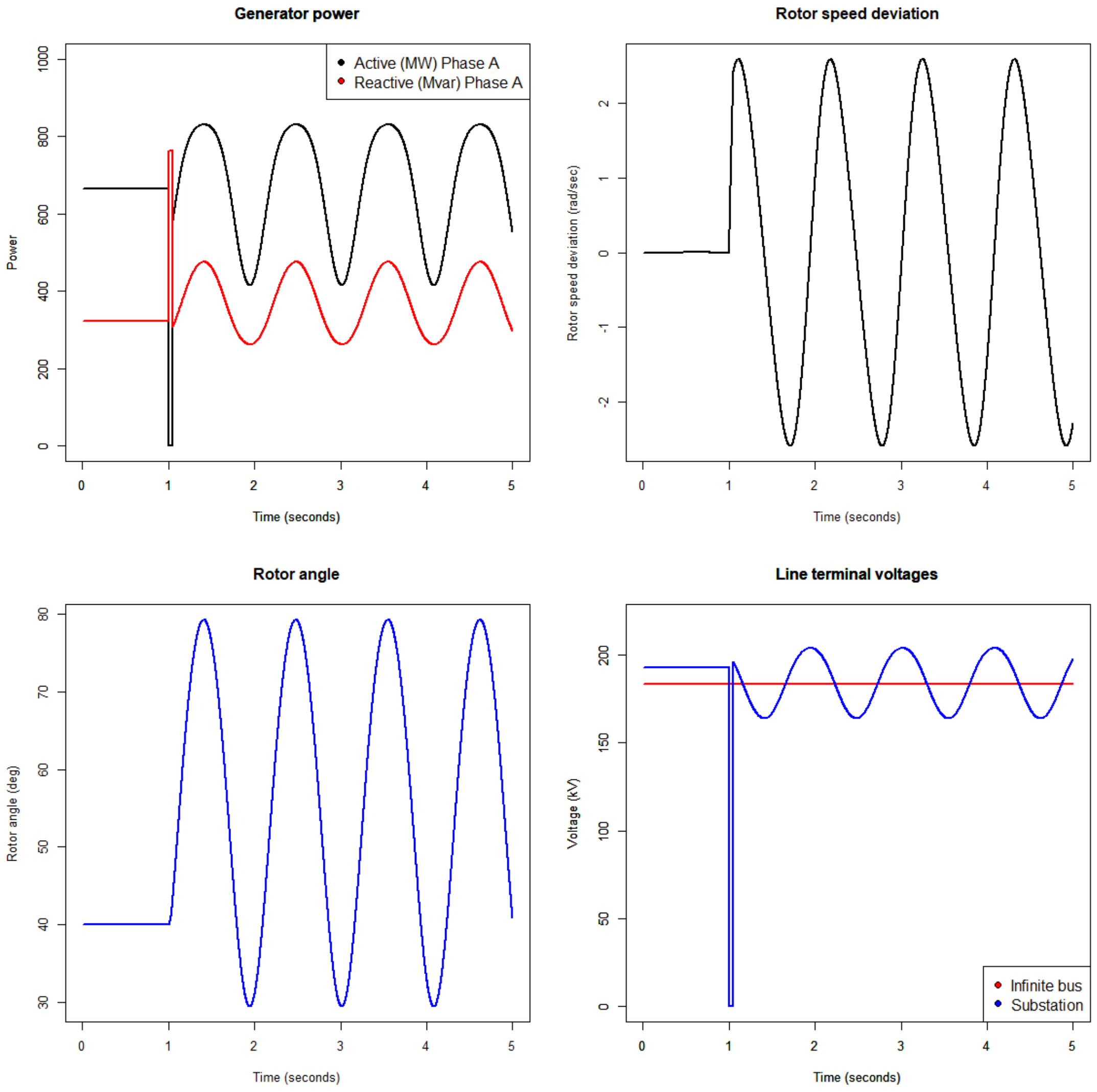

***Figure 51**. Case study 4: Initial scenario: Generator power, Rotor speed deviation, Rotor angle, Line terminal voltages prior, during and after the fault - Three-phase fault duration = 50 ms.*

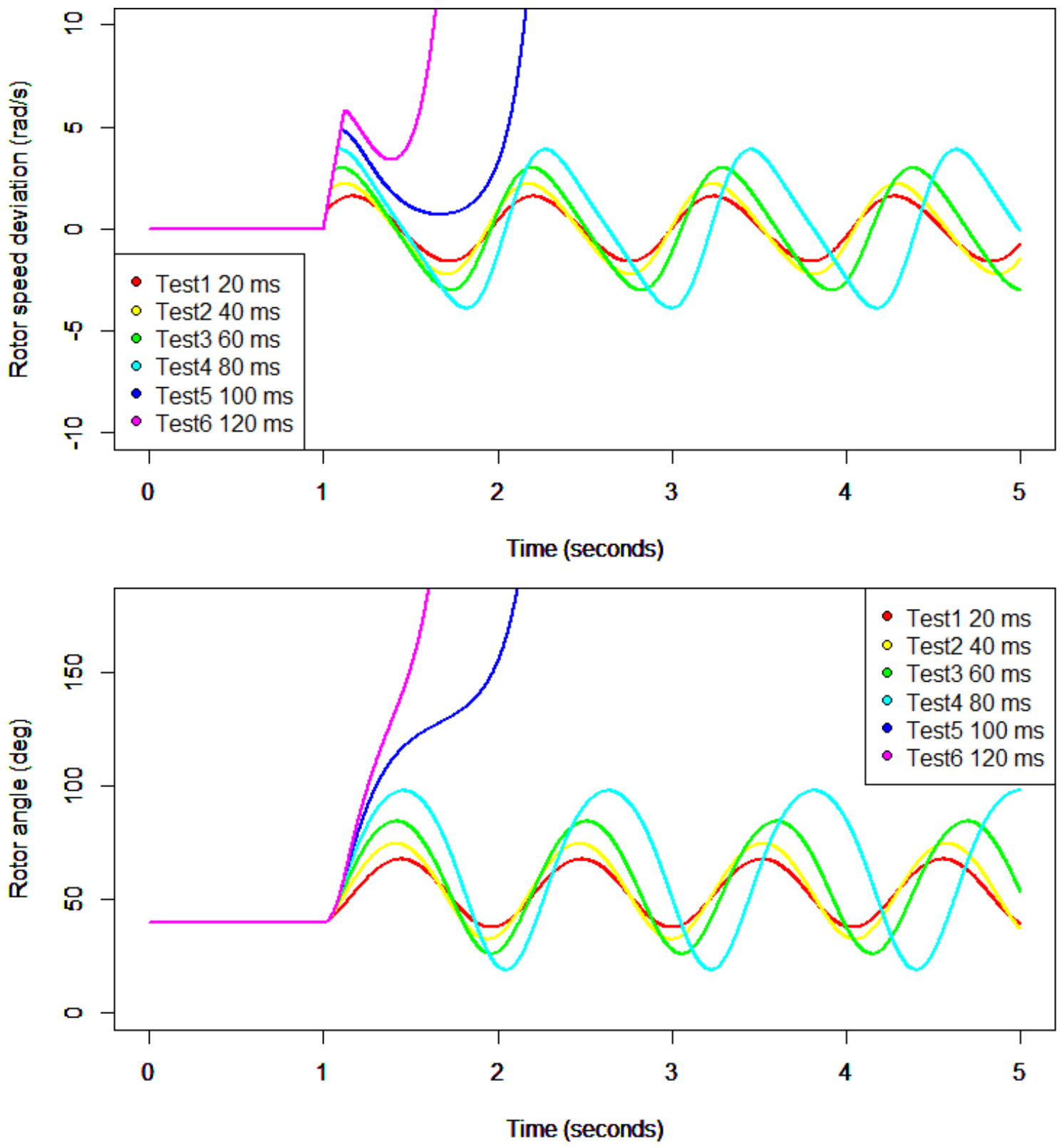

***Figure 52**. Case study 4: Parametric study - Generator power: 2220 MVA/1998 MW.*

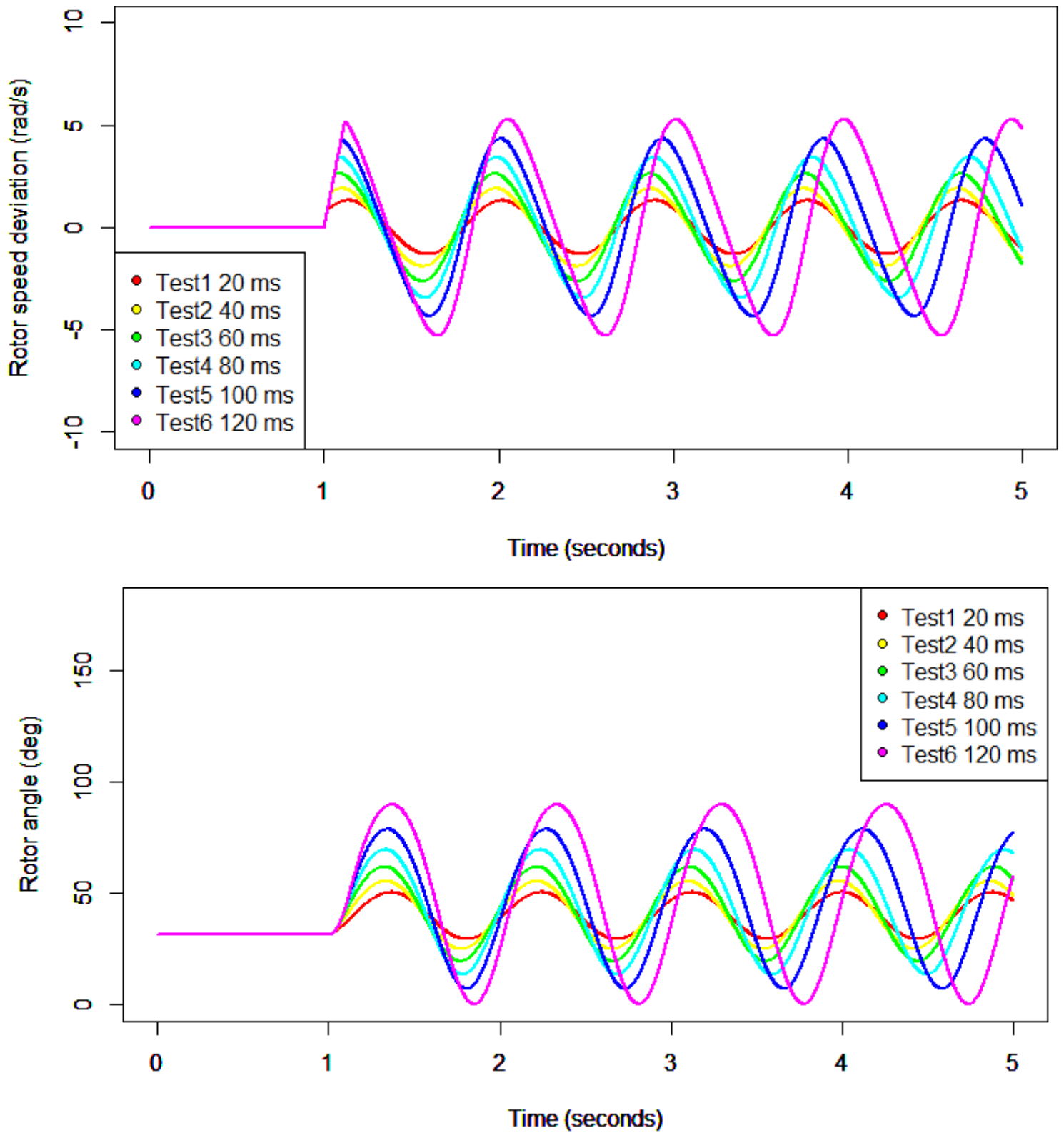

***Figure 53**. Case study 4: Parametric study - Generator power: 2220 MVA/1776 MW.*

### 3.5.3 Application of machine learning algorithms

The first step is to implement a parametric study with variable values of the fault duration and the active power injected by the power plant to the power system. The outcome should be a data frame that will be used to train ML algorithms. The data frame will have a third column with the stability result for every combination of the two values mentioned above.

Figure 54 depicts a flow chart with the procedure implemented in **RStudio** for generating the data frame that will be later used to train and test various ML approaches (ANN, SVM, kNN) in this case study.

The range of values considered for both parameters (fault duration, power plant generation) are as follows:

- The fault duration will vary between 70 and 250 ms.
- The power plant will always run at full load (i.e., 2220 MVA), but the power factor of the injected power will vary between 0.6 and 1.0. In all cases a leading power factor is assumed.

Figure 55 summarizes the results derived from simulations with **OpenDSS**; it shows how the various combinations of active power and fault duration values affect the stability of the test system. The figure clearly shows what combinations can cause a stable/unstable scenario.

Once the data frame with simulation results has been created, it is used to obtain data frames for training and testing.

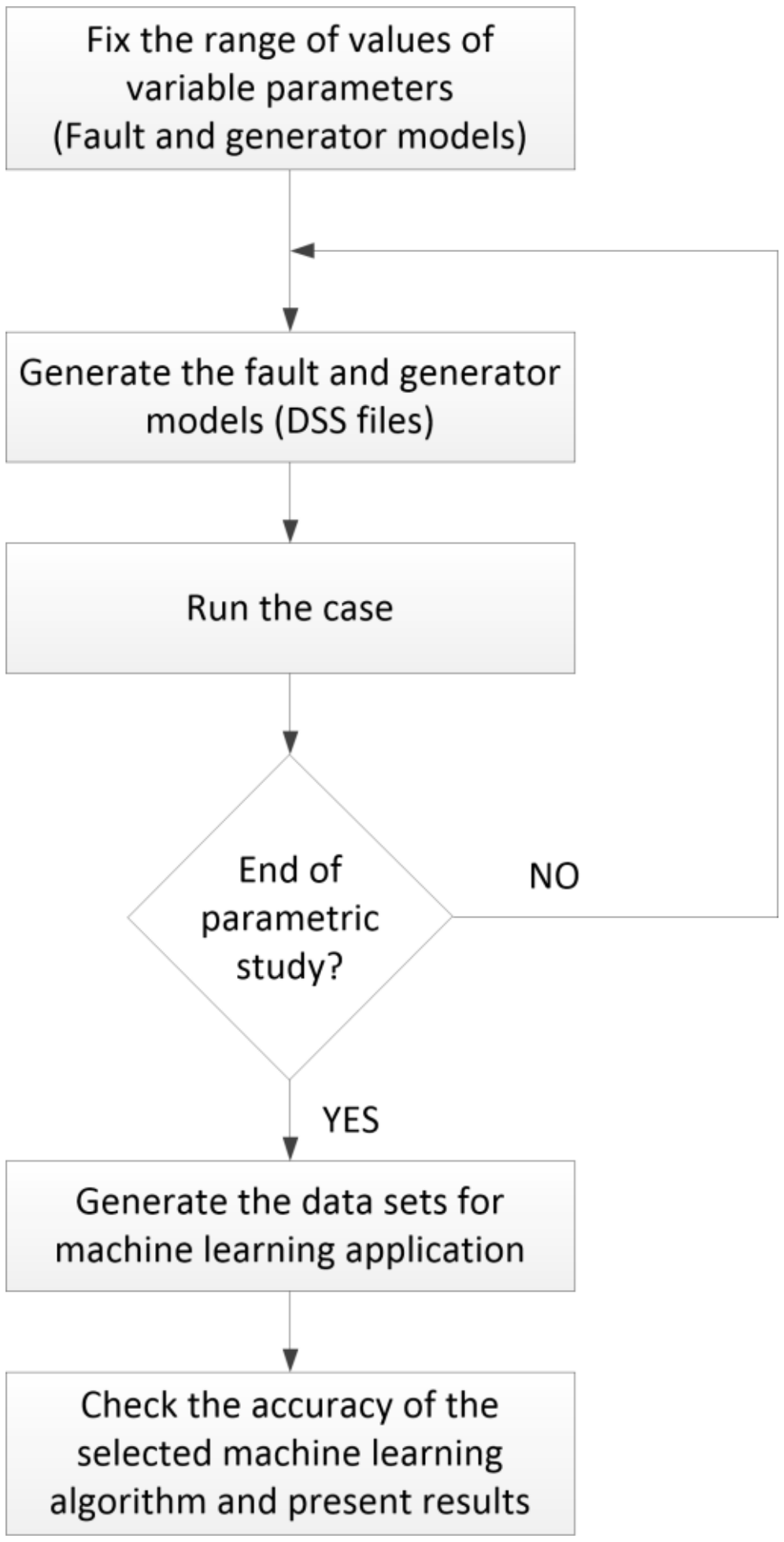

***Figure 54**. Case study 4: Procedure for the application of machine learning algorithms.*

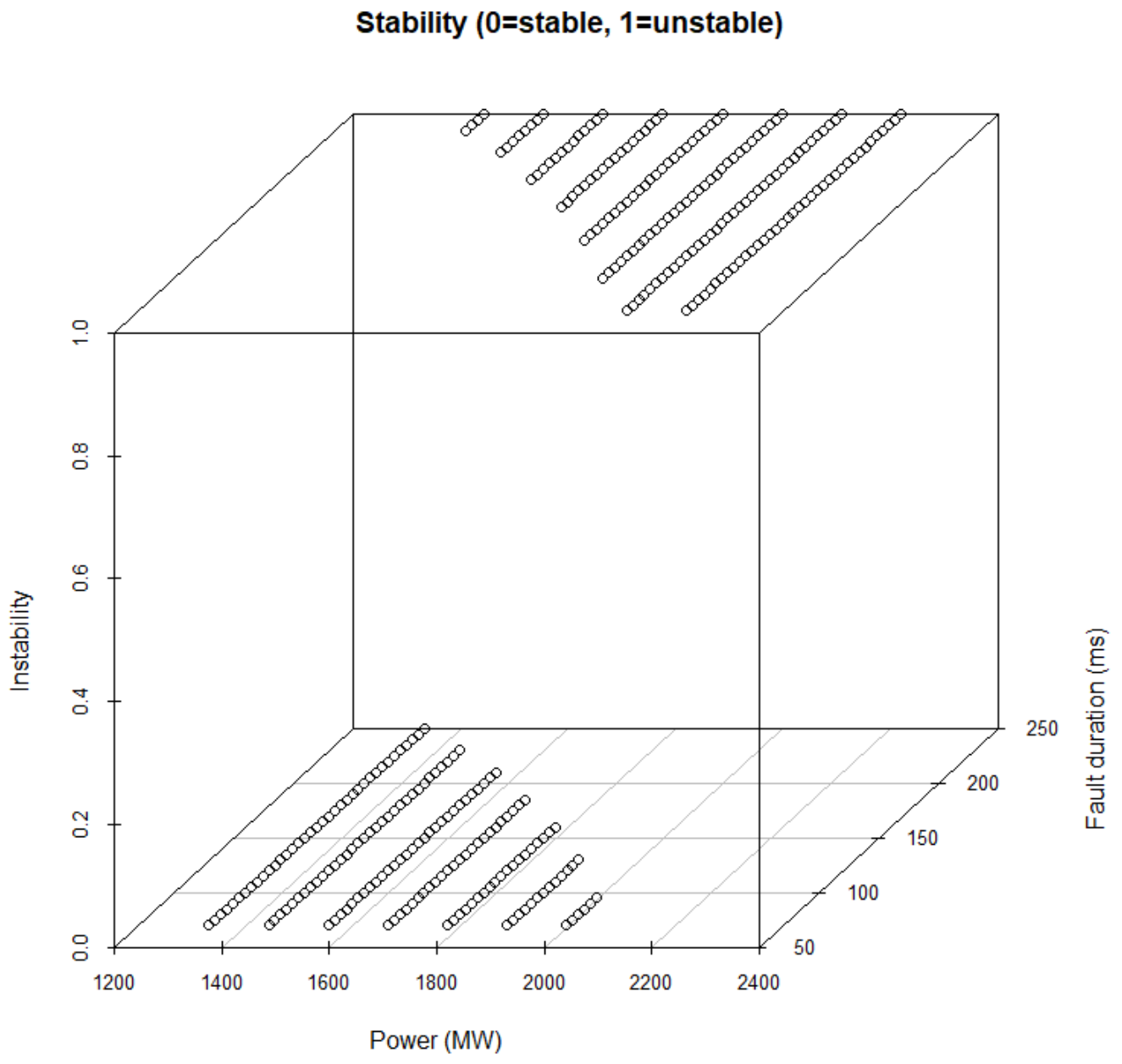

***Figure 55****. Case study 4: Data generated for training the neural network.*

80% and 20% of the results obtained in the previous parametric study are used to obtain the data frames to be used for training and testing, respectively. The combinations of active power and fault duration used for each data frame are randomly selected from the original data frame.
A neural network model can be obtained using several **R** capabilities. In this study the `neuralnet` library is used [42], [43].
First, a simple neural network with a single hidden neuron is considered; see Figure 56.

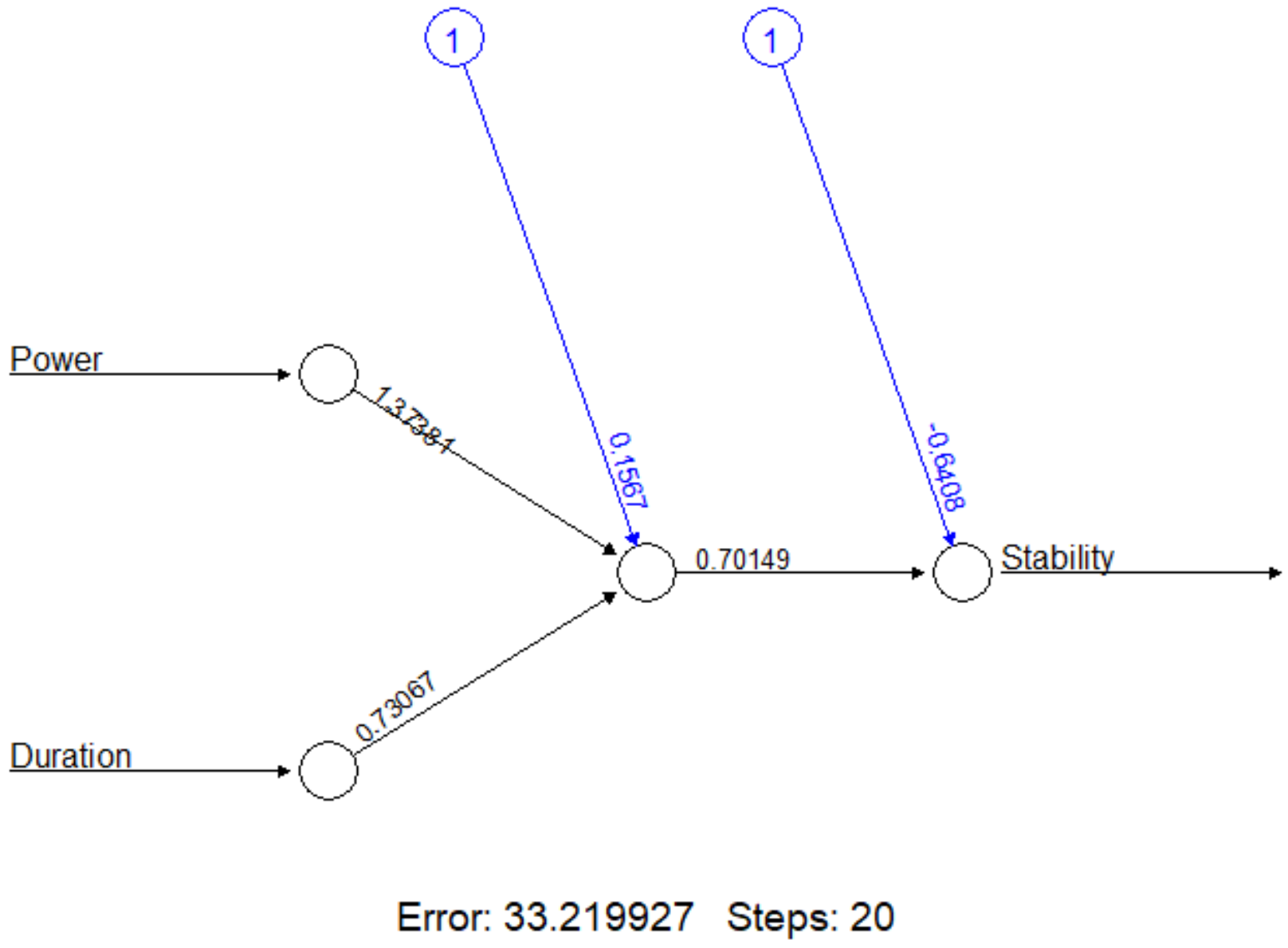

***Figure 56****. Case study 4: Artificial neural network with one hidden neuron.*

The accuracy of this ANN model is rather poor (52.24 %). A more sophisticated model with more hidden layers and neurons (2 layers and 8 neurons) is considered; see Figure 57. The accuracy is now 98.51 %.

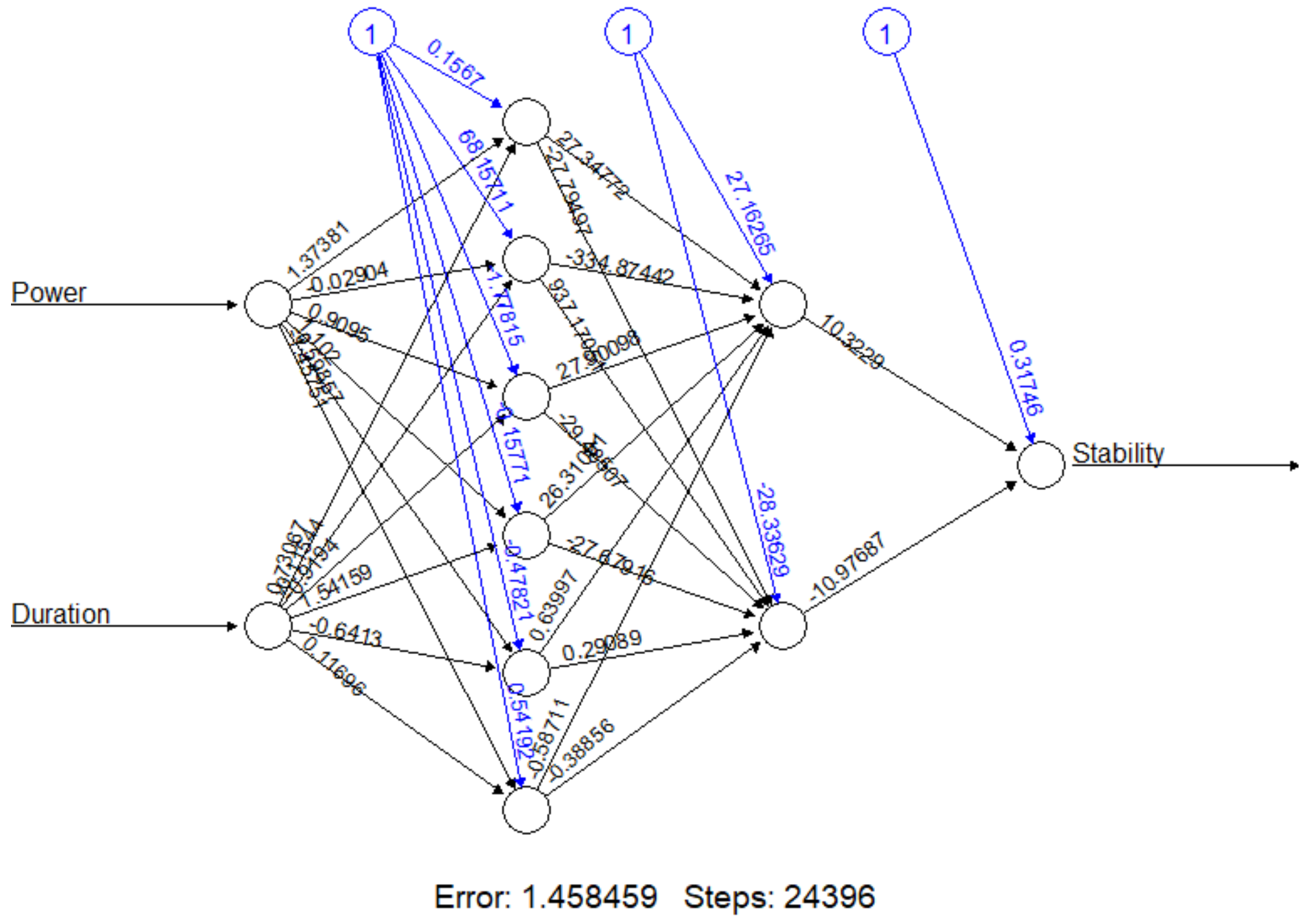

***Figure 57****. Case study 4: Artificial neural network with two hidden layers and eight hidden neurons.*

The performance of the neural networks is compared to that of other machine learning approaches (i.e. SVM and kNN).
The SVM algorithm is applied using again the `kernlab` library. The results are shown below:

```
## Prediction using support vector machine

##  Setting default kernel parameters

## agreement
## TRUE
##    1
```

The accuracy with the SVM reaches a value of 100 %.
The last study applies the kNN algorithm using again the `class` library and different numbers of neighbors. The results are shown below:

```
## Prediction using nearest neighbors

## Number of nearest neighbors = 1

## agreement
##      FALSE       TRUE
## 0.02985075 0.97014925

## Number of nearest neighbors = 2

## agreement
##      FALSE       TRUE
## 0.02985075 0.97014925

## Number of nearest neighbors = 3

## agreement
##      FALSE       TRUE
## 0.02985075 0.97014925
```

```
## Number of nearest neighbors = 4

## agreement
##      FALSE       TRUE
## 0.01492537 0.98507463
```

The maximum accuracy is 98.51 % and is obtained with 4 neighbors.
The procedure is repeated using randomly generated values for active power generation and fault duration; the data set obtained above is used for training. The test system response with the new combinations of values is shown below:

```
##     Power Duration Stability
## 1  1820.4   201.18         1
## 2  1820.4   140.81         0
## 3  1820.4    58.56         0
## 4  1820.4   221.22         1
## 5  1975.8   201.18         1
## 6  1975.8   140.81         1
## 7  1975.8    58.56         0
## 8  1975.8   221.22         1
## 9  2153.4   201.18         1
## 10 2153.4   140.81         1
## 11 2153.4    58.56         1
## 12 2153.4   221.22         1
## 13 1354.2   201.18         0
## 14 1354.2   140.81         0
## 15 1354.2    58.56         0
## 16 1354.2   221.22         0
```

Remember that 1 means instability and 0 stability.
The accuracy achieved with the application of the three approaches and the new data set are presented below:

- Artificial neural network with two hidden layers: 93.75%.
- Support vector machine: 100%.
- Nearest neighbor: 100% (with either 1, 2, 3 or 4 neighbors).

The power system configuration analyzed in this case study is a very simple one, usually employed to introduce the concept of stability in power system without renewable resources. Although a deeper study could be carried out by adding more parameters to the parametric study (i.e., the inertia constant and the transient reactance of generators), the procedure to be used would be like followed here. In addition, other studies can be carried out; for instance, considering other fault positions or other fault types (e.g., single-phase-to-ground faults).
As with the previous case studies, it is important to keep in mind the random nature of the results that can be derived with some machine learning algorithms. For this reason, the `set.seed` function must be considered to guarantee repeatable results every time **RStudio** is run.
According to the results derived from the application of three supervised machine learning approaches, it is evident that a high accuracy can be achieved. Remember that other **R** capabilities (i.e., other libraries) can be used with this case study and with which high accuracy can also be achieved. In addition, higher accuracy can be obtained with the `neuralnet` library by changing the configuration of the neural network; however, that should be made by using a trial-and-error approach since this library does not have a capability to optimize the neural network configuration.

## 4 Discussion

1. Although the simulation environment selected in this paper for studying power systems is based on **R**, it is evident that other simulation tools (e.g., **Python**, **MATLAB**) can be used for the same purpose. **Python** is probably the most popular approach at the time this paper is written. In fact, **OpenDSS** users can currently benefit from **PyDSS**, a high-level **Python** interface for **OpenDSS** [44], [45]. A good reason for selecting **R** is its capabilities for statistical studies and visualizations.

2. The links used to connect **RStudio** to either **TPBIG** or **OpenDSS** are of several types: (i) text files generated by **RStudio** that are inserted within the input files to be read by either **TPBIG** or **OpenDSS**; (ii) files with data generated by either **TPBIG** or **OpenDSS** that are later read and analyzed by **RStudio**; (iii) files with data generated by **RStudio** that are later read by **OpenDSS**. In all cases **RStudio** has been used to generate pure text files that are later inserted in the input file to be read by either **TPBIG** or **OpenDSS**. Remember that with **TPBIG** the connection to exchange data was unidirectional: **TPBIG** generates a file within a MODELS section that can be read by **RStudio**; therefore, all **TPBIG** input files must include a MODELS section that will pass the transient solution to a text file that can be read and interpreted from **RStudio**. As for **OpenDSS**, the connection uses CSV files, since both **RStudio** and **OpenDSS** can read this type of file. However, the connection is now bidirectional: **OpenDSS** generates CSV files that will be read by **RStudio**, and **RStudio** generates CSV files that will be read by **OpenDSS**.
3. There are many **R** capabilities that have not been used in this paper; for instance, parallel computing. **TPBIG** cannot be run using more than one core, although it has been used for applications in which a multicore environment could be useful; see, for instance, [27]. However, **OpenDSS** can be run using a multicore environment, so there is still room for new approaches/applications in which the combination **R-OpenDSS** could be run taking advantage of parallel computing.
4. There are many power system studies not covered in this paper for which any of the tool combinations used here could be very useful. One of the aspects to be considered is the fact that **RStudio** capabilities can be used to create text files that will be later inserted in the input file of either **TPBIG** or **OpenDSS**. These text files can represent power system component models not available as built-in capabilities.
5. Interaction between **R/Rstudio** and artificial intelligence platforms based on large language models (LLM), such as **ChatGPT** or **Copilot**, is now possible. Although an API secret key is always necessary, **R** users can now take advantage of many libraries (e.g., `askgpt` library for interacting with **ChatGPT** [46], Github-Copilot for interacting with **Copilot** [47]). There are other options; for instance, `ollamar` is an **R** package that provides an interface to **Ollama** (Omni-layer Learning Language Acquisition Model) [48], an open tool to locally run open-source LLMs, like **Gemma**, **LLaMa** or **Mistral**. **R** users can benefit from these tools to create **R** code, interpret the code, debug the code, turn **R** code into a tutorial, ask questions to AI platforms, and many more.
6. A new *integrated development environment* (IDE) for data science developed by Posit called **Positron** has already been released [49]. **Positron** unifies **Python** and **R** workflows with notebooks, scripts, interactive plotting, and AI assistance. At the time this document is being produced, only the beta version is available.

## 5 Conclusion

This paper has proposed a simulation environment based on **RStudio** and **R** capabilities for analyzing power systems. Since **R** is a tool specialized in statistical studies and with capabilities for generating high quality visualizations, the fields in which the proposed environment can be used should be mainly those requiring sophisticated statistical studies and high-quality visualizations.

The combination with **ATP** can be applied to studies of transient nature while the combination with **OpenDSS** can be applied to many other scenarios that also cover the transient behavior of power systems. Although the configurations of the test systems used in this paper are rather simple or small, the selected case studies cover enough number of applications to illustrate the scope of studies that can be carried out with the proposed environments using **RStudio** as a base.

The link between **RStudio** and the simulation tools specialized in power system studies (i.e., **TPBIG** and **OpenDSS**) is an important aspect that, as shown with the selected case studies, can be satisfactorily solved for automatically generating reports that can explain/detail the goal of the case study and the approaches chosen for its solution.

Finally, it is worth mentioning that the paper has emphasized the potential of the **R** environment for power system studies in which machine learning algorithms can be useful for predicting the behavior of the test system once a significant number of cases have been run for training purposes.

## 6 Acknowledgement

The authors want to express their gratitude to all those who make possible the development of the software tools used in this paper; without their work this contribution would not have been possible.